\input{psfig.sty}
\documentclass[galaxies,article,accept,moreauthor,pdftex,12pt,a4paper]{mdpi}

\usepackage{soul}
\usepackage{upgreek}
\usepackage{enumitem}
\usepackage{subfigure}
\usepackage{booktabs}
\usepackage{multirow}
\usepackage{longtable}
\usepackage{microtype}
\newcommand\myurl[1]{\changeurlcolor{black}\url{#1}\changeurlcolor{blue}}

\usepackage{subfigure}
\makeatletter
\renewcommand{\@thesubfigure}{\normalsize(\textbf{\alph{subfigure}})}
\makeatother

\firstpage{1}
\articlenumber{17}
\doinum{10.3390/galaxies5010017}
\pubvolume{5}
\pubyear{2017}
\copyrightyear{2017}
\externaleditor{Academic Editors: Jose Gaite and Antonaldo Diaferio}
\history{Received: {30 May 2016}; Accepted: {9 December 2016}; Published: %\hl
{16 February 2017}}
\usepackage[utf8]{inputenc}

 \theoremstyle{mdpi}
 \newcounter{thm}
 \newcounter{ex}
 \newcounter{re}
 \theoremstyle{mdpidefinition}

\def\aap{Astron. Astrophys.}
\def\aj{Astro. J.}
\def\apss{ Astrophys.  Space Sci. }
\def\apj{Astrophys. J.}
\def\apjl{Astrophys. J. Lett.}
\def\apjs{Astrophys. J. Supp.}
\def\jcap{J. Cosmo. Astrop. Phys}
\def\mnras{Mon. Not. Roy. Astro. Soc.}
\def\na{New Astron.}
\def\nar{New Astron. Rev.}
\def\nat{Nature}

\def\prd{Phys. Rev. D}
\def\physrep{Phys. Rept.}
\def\araa{Ann. Rev. Astron.  Astrophys.}

\usepackage{amsmath}

\def\nvphantom{\v@true\h@false\nph@nt}
\def\nhphantom{\v@false\h@true\nph@nt}
\def\nphantom{\v@true\h@true\nph@nt}
\def\nph@nt{\ifmmode\def\next{\mathpalette\nmathph@nt}%
  \else\let\next\nmakeph@nt\fi\next}
\def\nmakeph@nt#1{\setbox\z@\hbox{#1}\nfinph@nt}
\def\nmathph@nt#1#2{\setbox\z@\hbox{$\m@th#1{#2}$}\nfinph@nt}
\def\nfinph@nt{\setbox\tw@\null
  \ifv@ \ht\tw@\ht\z@ \dp\tw@\dp\z@\fi
  \ifh@ \wd\tw@-\wd\z@\fi \box\tw@}

\Title{Small Scale Problems of the $\Lambda$CDM Model: A~Short Review}

\Author{Antonino Del Popolo $^{1,2,3,%\dagger
}$* and Morgan Le Delliou $^{4,5,6%,\ddagger
}$}
\AuthorNames{Antonino Del Popolo and Morgan Le Delliou}

\address{$^{1}$ \quad Dipartimento di Fisica e Astronomia, University of Catania
, Viale Andrea Doria 6, {95125}
Catania, Italy
 \quad  \\
$^{2}$  \quad INFN Sezione di Catania,  Via S. Sofia 64, I-95123 Catania, Italy\quad  \\
$^{3}$ \quad International Institute of Physics, Universidade Federal do Rio Grande do Norte,
59012-970 Natal, Brazil\quad  \\
$^{4}$ \quad Instituto de Física Teorica, Universidade Estadual de S\~{a}o Paulo (IFT-UNESP),
 Rua Dr. Bento Teobaldo Ferraz~271, Bloco 2 - Barra Funda, 01140-070
S\~{a}o Paulo, SP Brazil; delliou@ift.unesp.br
 \\
$^{5}$ \quad Institute of Theoretical Physics
Physics Department, Lanzhou University
No. 222, South Tianshui Road, {Lanzhou 730000}, Gansu, China\\
$^6$ \quad Instituto de Astrofísica e Ciências do Espa\c co, Universidade de Lisboa, Faculdade de Ciências, Ed. C8,  \mbox{Campo Grande}, 1769-016 Lisboa, Portugal%\\
}

\corres{Correspondence: adelpopolo@oact.inaf.it%; Tel.: +x-xxx-xxx-xxxx
}

\firstnote{adelpopolo@oact.inaf.it%Current address: Affiliation 3
}
\abstract{The $\Lambda$CDM model, or concordance cosmology, as it is often called, is a paradigm at its maturity.
{%\bf
It} is clearly able to describe the universe at large scale, even if some issues remain open, such as the cosmological constant problem{%\bf
, the small-scale problems in galaxy formation}, or the unexplained anomalies in the CMB. %However,
$\Lambda$CDM clearly shows difficulty at small scales, which could be related to our scant understanding, from the nature of dark matter to that of gravity; or to the role of baryon physics, which is not well understood and implemented in simulation codes or in semi-analytic models.  At this stage, it is of fundamental importance to understand whether the
problems encountered by the $\Lambda$DCM model are a sign of its limits or a sign of our failures in getting the finer details right.
In the present paper, we will review the small-scale problems of the $\Lambda$CDM model, and we will discuss the proposed solutions and to what extent they are able to give us a theory accurately describing the phenomena in the complete range of scale of the observed universe.
}

\begin{document}

%%%%%%%%%%%%%%%%%%%%%%%%%%%%%%%%%%%%%%%%%%
%% Sections that are not mandatory are listed as such. The section titles given are for Articles. Review papers and other article types have a more flexible structure.

%% Only for the journal Gels: Please place the Experimental Section after the Conclusions

%%%%%%%%%%%%%%%%%%%%%%%%%%%%%%%%%%%%%%%%%%
%\setcounter{section}{-1} %% Remove this when starting to work on the template.

\section{Introduction}

%\cite{}

Despite the $\Lambda$CDM model being successful{%\bf
, according to the largest part of the cosmology community,}  in describing
%observations of the Universe, its
the formation and evolution of the large scale structure in the Universe,
%Spergel et al. 2003, Komatsu et al. 2011; DelPopolo 2007, 2013, 2014),
the state of the early Universe and the abundance of different forms of matter and energy \cite{spergel03,komatsu11,dp07,dp13,dp14a}, its predictive power --- already checked against new discoveries (e.g., lensing of the CMB \cite{smith,das}, B-mode polarisation \cite{hanson} the kinetic SZ effect) --- it presents several difficulties. Among the most famous, we recall the ``cosmological constant fine tuning problem'', and the ``cosmic coincidence problem'' \protect\cite{weinberg_co,ADP}.

The first problem is connected to the fact that most quantum field theories predict a huge cosmological constant from the energy of the quantum vacuum at present, more than 100 orders of magnitude too large, see Refs.  \protect\citep[][]{weinberg_co,martin_j,ADP}.
More precisely the theoretical expectations give
$\rho_{\Lambda} \simeq 10^{71} ~\rm GeV^4$, in contradiction with the cosmological upper bounds giving
$\rho_{\Lambda} \simeq 10^{-47} ~\rm GeV^4$ which gives rise to an extreme fine-tuning problem. It also entails fine tuning at Planck scale era, thus in the initial conditions of dark energy.
The second is connected to the reason why dark energy and dark matter energy densities are approximately equal nowadays (see Ref. \protect\citep[][]{sivanandam}).

Tensions of unknown origin are also present between the 2013 Planck parameters \cite{planck} and $\sigma_8$ obtained from cluster number counts and weak lensing, the actual value of the Hubble parameter, $H_0$, and SN IA data. The Planck 2015 data are still in tension with CFHTLenS weak lensing \cite{raveri} data, and~with the $\sigma_8$ growth rate \cite{maca}.

Another concern lies in that the large-angle fluctuations in the CMB show statistical anomalies (i.e., a quadrupole-octupole alignment \cite{schwa,copi,copi1,copi2,copi3},
%465-467,469, 471);
 a power hemispherical asymmetry \cite{eriksen,hansen,jaf,hot,planck1,akrami} and
% (458-463);
 a cold spot \cite{cruz,cruz1,cruz2}).
%(473-475)),
This collides with the idea that our universe should be a realisation of a statistically isotropic and Gaussian random field, which implies a statistical independence in the CMB multipoles. What is unclear is whether these anomalies are related to unknown systematics, if they are statistical effects \cite{bennet}, or a fingerprints of new physics.

The $\Lambda$CDM model also encounters problems in describing structures at small scales, e.g.,  \citep[][]{moore94,moore1,ostrik,boyl,boyl1,oh}. The main problems are/have been
\renewcommand{\theenumi}{\alph{enumi}}\begin{enumerate}[leftmargin=*,labelsep=5mm]
\item The cusp/core (CC) problem \cite{moore94,flores}, designating the discrepancy between the flat density profiles of dwarf galaxies (also coined dwarfs), Irregulars, and Low Surface Brightness galaxies (hereafter LSBs), and the cuspy profile predicted by dissipationless N-body simulations \cite{nfw,navarro10,Saburova:2014opa}, despite the fact that the observed galaxies are all of DM dominated types;
\item The ``missing satellite problem''
(MSP), coining the discrepancy between the number of predicted subhalos in N-body simulations {%\bf
\cite{Klypin:1999uc,moore1}} and those actually observed{%\bf
, further complicated by the ``Too Big To Fail'' (TBTF) problem, arising from the $\Lambda$CDM prediction of satellites that are too massive and too dense, compared to those observed, to hope for their destruction in the history of mass assembly up to today \cite{boyl,boyl1}};
\item The angular momentum catastrophe \cite{vanbuswa} labelling the angular momentum loss in Smooth Particle Hydrodynamics (SPH) simulations of galaxy formation that gives rise to dwarf galaxies' disks with different angular momentum distributions from those of cold dark
matter haloes, in addition to disc sizes that are much smaller in simulated galaxies compared with observed ones \cite{Cardone:2009jr};\label{enu:AMcat}
\item The problem of satellites planes, namely the alignment on thin planes of satellite galaxies of the MW and M31, a feature that proved difficult to explain in simulations of the $\Lambda$CDM paradigm~\cite{pawl}; \label{enu:satPlane}
\item The problem of re-obtaining the slope and scatter of the baryonic Tully-Fisher relation \mbox{($M_b \propto V_c^4$) \cite{mcgaugh}};\label{enu:TullyFisher}
%\item the problem of the close similarity of galaxies characteristics between voids and other environments \cite{peebl}.
\end{enumerate}
{%\bf
The $\Lambda$CDM model has some other issues discussed in Refs.~\cite{Kroupa:2004pt,Kroupa:2010hf,Kroupa:2012qj,Kroupa:2013yd,Kroupa:2014ria}, that we shall not consider in our short review.}
Problems \ref{enu:AMcat} and \ref{enu:TullyFisher} have actually been solved: baryonic models, as discussed in Section~\ref{sub:BsolCC}, have already been proposed to solve the problem posed by
SPH simulations baryons angular momentum  non conservation in collapse, leading them to typically only retain 10\%, and form~disks that are too small, compared to real galaxies \cite{NavarroBenz1991,NavarroSteinmetz1997,SomEtal99,NavarroSteinmetz2000b} (i.e., the ``angular momentum catastrophe'', AMC). Those solutions proceeded from feedback effects basically heating the gas , from~supernovae explosions~\cite{SomEtal99}, using clumps, in addition, to reproduce the correct  angular momentum distribution of baryons \cite{MallerDekel2002,SommerEtal2003,AbadiEtal2003} and selective outflows to obtain bulgeless disks \citep[][]{Governato2010} (see Section~\ref{sub:SNsolCC}), or from dynamical friction of baryonic clumps, able to explain all those features at once (see Section \ref{sub:DFclumps}, and Ref. \citep[][]{dpet14}).

The empirical optical luminosity-21 cm line width scaling in spiral galaxies, the Tully-Fisher relation \cite{TullyFisher}, that reflects the gas+luminous mass-rotation velocity scaling, the Baryonic Tully-Fisher Relation (BTFR) \citep[][]{Bell2000,Verheijen2001,Gurovich2004,McGaugh2005,Pfenniger2004,Begum2008,Stark2009,Trachternach2009,Gurovich2010}, have been used as a stumbling stone for the $\Lambda$CDM model by proponents of the MOND Modified Gravity Model (\citep[]{McGaugh2011}, discussed in Section \ref{sub:MTG}). However, this claim is less strong after {%\bf
some models and simulations have found a possible solution (see Section \ref{sec:btfr}).}

Problem \ref{enu:satPlane} is characterised in the MW by
\begin{itemize}[leftmargin=*,labelsep=5mm]
\item a highly flattened, planar distribution of the satellites in three-dimensional space,
\item a common orientation of the satellites orbits, and
\item an alignment of the satellites orbits within the distribution plane.
\end{itemize}and remains open:
if Ref.~\cite{Sawala2014} claimed a resolution in the EAGLE hydrodynamical simulations, the~review \cite{Pawlowski2015} concluded oppositely.

The present review will be restricted to the small-scale problems of the $\Lambda$CDM model (hereafter SSP$\Lambda$CDM) connected to the formation  of cusps, and to satellites. As we will see hereafter, these issues are strictly connected.

From one side, unified models have shown \cite{zolo,brooks,dpet14} that a mechanism which can transform cusps into cores conversely helps the solution of the MSP.
Several authors (e.g., \citep[][]{Mashchenko2006,Mashchenko2008,pen10}) noticed that the effects of a parent halo's tidal forces on a satellite depend fundamentally on the shape of the latter. A~cuspy profile allows the satellite to retain most of its structure, when entering the main halo. Inversely, for a cored profile, the tidal field of the main halo can easily strip the satellite from its gas and even destroy it in some cases \cite{pen10}. As a result, such satellite will not end up visible, either because it was destroyed or because it lacks the gas to make stars.

From the other side, the satellites most puzzling issue is now recognised as having shifted: rather~than the number of satellites, it is related to their inner mass density, specifically to their density profiles being flatter than those of N-body simulations.

In Sections {%\bf
\ref{sec:satPlanePb}, and \ref{sec:btfr}, we will summarise the discussions around, respectively, the satellite plane problem and the Baryonic Tully-Fisher Relation. In Section \ref{sec:ccp}, we will review the CC problem. We will shortly review the early discussions on that problem, then the solution related to the role of baryonic physics, as well as solutions dealing with changes of the dark matter properties or related to modified theories of gravity. In Sections} \ref{sec:MSP}, and \ref{sec:TBTF}, we review the MSP, and the Too-Big-To-Fail (TBTF) problem.
In~Section \ref{sec:UsolSSP},  we will discuss a unified baryonic solution to the quoted problems, and Section \ref{sec:concl} is devoted to conclusions.

{%\bf
\section{The Satellite Plane Problem}\label{sec:satPlanePb}

From the seventies on, it was noticed by several authors that the MW satellites are distributed on a planar structure, the so-called ``Magellanic Plane'' \cite{KunkelDemers76,LyndenBell76}, the~disc of satellites~\cite{Kroupa:2004pt,Metz:2008vp}, the MW ``Vast popular structure'' (VPOS) , and the ``great plane of andromeda'' (GPoA)~\cite{Pawlowski:2013cae,pawl,Pawlowski2015,Ibata:2013rh}. Recent~studies by Refs.~\cite{Kroupa:2004pt,Metz:2006zc,Pawlowski:2013cae} described the structure as a thin disc, having a height of 20 kpc, and 9 of the 11~classical dwarfs of the MW~\cite{Metz:2008vp} co-orbit within the structure. Other authors \citep{McConnachie:2005ck,Koch:2005kg,Metz:2006zc} found a similar distribution around M31, which was better evidenced by Refs.~\cite{Ibata:2013rh,Conn:2013iu}, who found the GPoA, \linebreak 
(see~\citep[][]{Shaya:2013xna,Tully:2015zfa,Bowden:2014lsa,Cautun:2014zxa,Libeskind:2005hs,Gillet2015ApJ...800...34G,Hammer:2013bga,Smith:2013ida}). In the local group (LG), similar alignments have been found in isolated dwarf galaxies~\cite{Pawlowski:2013kpa,Pawlowski:2013cae,Bellazzini:2013oea,Pawlowski:2014yxa}, as well as in more distant galaxies \cite{Galianni2010,Duc:2014bua,Paudel:2013bxa,Karachentsev:2014kaa}. According to several  authors \mbox{(e.g., \citep[][]{Kroupa:2004pt})}, this should be considered as a strong challenge for the $\Lambda$CDM model\footnote{%\bf
It was pointed out, however, for the first time by Kroupa et al. \cite{Kroupa:2004pt}, that the planar distribution of all known satellite galaxies in 2005 was in significant disagreement with the expected spatial distribution of dark-matter dominated satellite galaxies}. In~cosmological simulations the DM sub-haloes are isotropically distributed and then at odds with the planar VPOS. The VPOS and GPoA structures, studied from the point of view of space correlation led \linebreak Ref.~\cite{Metz:2006zc} to conclude on the tidal origin (i.e., tidal dwarfs) of a fraction of the MW and M31 \citep{BarnesHernquist92,Bournaud:2008}.\linebreak 
The structure of the observed dwarf galaxies could have been formed from encounters in the LG (e.g., a past encounter between the MW and  M31 \cite{Sawa:2004tx,Pawlowski:2012vz,Zhao:2013uya}, the tidal disruption of a \linebreak 
LMC-progenitor \cite{KunkelDemers76,LyndenBell76,Pawlowski:2011rv}, etc). The~tidal dwarf galaxies (TDG) scenario has shown to be able to reproduce structures such as the GPoA \cite{Hammer:2013bga}, or VPOS-like structure \cite{YANG:2010mk,Fouquet:2012tc}.

Several trials have been proposed to obtain the quoted structures in the $\Lambda$CDM model. Refs.~\mbox{\cite{Lake:2008zt,Li:2007mf}} proposed a scenario of accretion of groups of satellites, but this approach gives rise to properties of groups of dwarfs in disagreement with observations \cite{Metz:2009ys}, as the structures formed are too extended, and too thick ($\simeq$ 50 kpc), in comparison with the VPOS and GPoA. The VPOS structure was claimed to be reobtained in simulations such as that of Ref.~\cite{Lovell:2010ap}, but an analysis of the angular momenta directions (orbital poles), by Ref.~\cite{Pawlowski:2012vz} showed a disagreement with those of observations. This issue was also studied in the simulations by Refs.~\cite{Libeskind:2009mv,Deason:2011zv}. The first group of authors found that for at least 3 out of 11 satellite galaxies --- the latter representing about 35\% of the simulated satellite galaxy systems they consider in their $\Lambda$CDM simulations --- the orbital poles point along the short axis of the galaxy distribution. Similar alignment is also claimed by Ref.~\cite{Deason:2011zv}. Futher improved studies by Ref.~\cite{Pawlowski:2013cae} found the distribution of orbital poles of the 11 brightest MW satellites to be much more concentrated, displaying a good alignment for $\geq$ 6 of the orbital poles, which seems to be difficult to explain for $\Lambda$CDM. Another way to solve the puzzle is to claim that, since the GPoA-like structures are rare \cite{Bahl:2013zda}, there is no contradiction between the $\Lambda$CDM Millennium II simulations and observations, as VPOS in MW are as frequent in the Millennium II simulation as in observations \cite{Wang:2012bt}. Finally, the~paradigm of cold mode accretion streams \cite{Goerdt:2013gza} could naturally explain the GPoA, as, in it, host~galaxies acquire not only their gas but also their satellites along cold streams: therefore GPoA would represent indirect observational evidence of the paradigm.

Such claims were criticized by several authors (see \citep[][]{Pawlowski2015}), which point the finger at the choice of the parameters used for the comparison.

More recent simulations (e.g., \citep[][]{Sawala:2015cdf}) claim that the problem of reproducing those  structures is merely due to neglecting the effects of baryons, the comparison being made with N-body only simulations, while when taking those effects into account, such structures are no-longer problems.

Ref.~\cite{Pawlowski2015} rejected such claims, since the scale on which baryon physics can act is much smaller than that of the satellite systems (some hundreds of kpc)

To summarize, the issue is still open, and from our point of view, should a solution be found, it is highly improbable that it would be connected to baryonic physics.

\section{The Baryonic Tully-Fisher Relation}\label{sec:btfr}

The Tully-Fisher relation was discovered to empirically correlate spiral galaxies luminosities with their HI line-width \cite{TullyFisher}, that is their stellar mass with their rotation velocities \citep{McGaugh:2000sr,Bell2000,TorresFlores:2011uc}. This~correlation is well fitted by a power law with small scatter for late-types of high mass and velocities \cite{McGaugh:2000sr,Bell2000}, while it no longer follows a single power law at low mass and velocities because cold gas starts to contribute significantly. It has therefore been generalised by using baryonic mass instead of just stellar mass, and the resulting Baryonic Tully-Fisher Relation (BTFR) fits well with a single power law over several orders of magnitude \cite{Stark2009} and with small scatter for selected high-quality data \cite{McGaugh2011,Lelli:2015wst}.

The concordance cosmology framework understands the BTFR as an imprint of the finite age of the Universe on the halo mass-rotation velocity relation (see, e.g., \citep[][]{Mo:1997vb,Steinmetz:1998gr}), that translates into a constant density contrast, and thus implies  linear scaling of virial radius with velocity, i.e., from~geometry,  $M\propto V^3$. The problems for the $\Lambda$CDM model coming from the scatter and slope of the BTFR highlighted by \citep[]{Lelli:2015wst} (and references therein) hinge on that relation. The link with the BTFR is made by assuming proportionality of galaxies baryonic mass and rotation velocity with their host haloes total virial mass and velocity, respectively. The latter assumption is not trivially satisfied since the observations used for the BTFR clearly provide a very limited baryonic mass and velocity probe compared with equivalent haloes in simulations \cite{White:1991mr}. In fact, so few halo baryons assemble in actual galaxies that the correlation with virial mass is unclear, while, as the disk sizes remain so small compared to the estimated virial sizes of galaxy haloes,  scaling between their two characteristic velocities appears unrealistic. In~addition, baryonic feedback mechanisms determining the galaxies' baryonic masses seems too variable to ensure similar galactic baryon fractions for all haloes. The tightness of the observed BTFR is therefore difficult to explain in the $\Lambda$CDM model, contrary to the fundamental acceleration scale included in the MOdified Newtonian Dynamic (MOND) proposition \cite{McGaugh2011}.
\linebreak Indeed, \cite{Lelli:2015wst} claim that observation of a select sample of disk galaxies, assuming constant stellar mass to light ratio, displays a significantly lower scatter than in the $\Lambda$CDM simulations of Ref.~\cite{Dutton:2012jh}, and that the residuals correlations with the radius or surface brightness of galaxies are not following the $\Lambda$CDM semi-analytic predictions of Ref.~\cite{Desmond:2015nja}, a conclusion which puts the $\Lambda$CDM model at odds with~observations.

For many years, numerical galaxy formation simulations were unable to produce morphologically realistic galaxies, let alone reproduce the BTFR (see, e.g.,  \citep[][]{Navarro:1999fr,Scannapieco:2011yd}, and references therein), and even semi-analytic models, with empirical inputs, had a hard time predicting it correctly (e.g., \citep[][]{Cole:2000ex}). Recently, however, this state of affairs has evolved with improvements in modeling of the baryonic feedback mechanisms allowing production of realistic rotation disk galaxies \cite{Guedes:2011ux,Brook:2012fm,McCarthy:2012wr,Aumer:2013gpa,Marinacci2014}.  The small statistics of these models did not allow for predictions on the BTFR, while attempts at reproductions led to controversy on the impact of those baryonic processes on the dark halo, some claiming drastic feedback was needed to obtain the BTFR \cite{Dutton:2008zf,chan} while others reported no need for such feedback (see, e.g.,~\mbox{\citep[][]{Vogelsberger:2014pda,Schaller2015}).}

Recently, the combination of the large EAGLE simulation programme \cite{Schaye:2014tpa,Crain:2015poa} --- calibrated on small scales on observed galaxy stellar mass function and present radius but not on the BTFR, with multiple realisations of Local Group-like galaxies in smaller volume, i.e. the APOSTLE \linebreak project \cite{Fattahi2016MNRAS.457..844F,Sawala:2015cdf} --- claimed to have successfully reproduced  the BTFR over four decades \cite{Sales:2016dmm}, even reproducing its break-down at the faint end, as was indeed observed \cite{Geha:2006mc,Trachternach2009}.

This, after the other claim of successful model from Ref.~\cite[]{DiCintio2015} (a semi-empirical model coupling observed Halo Abundance Matching baryon mass fractions with $\Lambda$CDM haloes and claiming to generate a realistic BTFR), seems to have found a path to solve the problem.}

\section{The Cusp/Core Problem }\label{sec:ccp}

\begin{figure}[H]
\centering
%\begin{center}
\includegraphics[width=7cm]{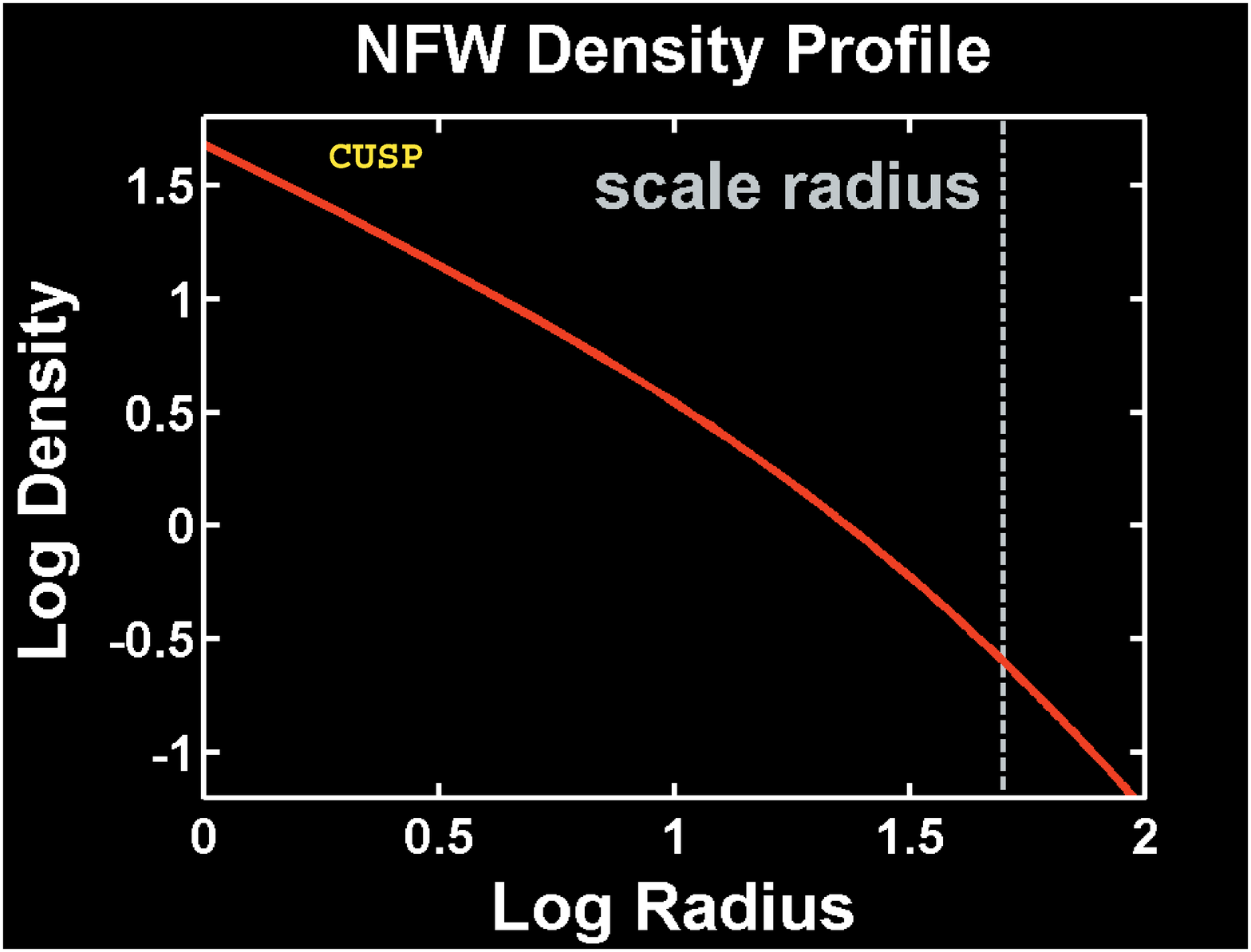}
\includegraphics[width=7cm]{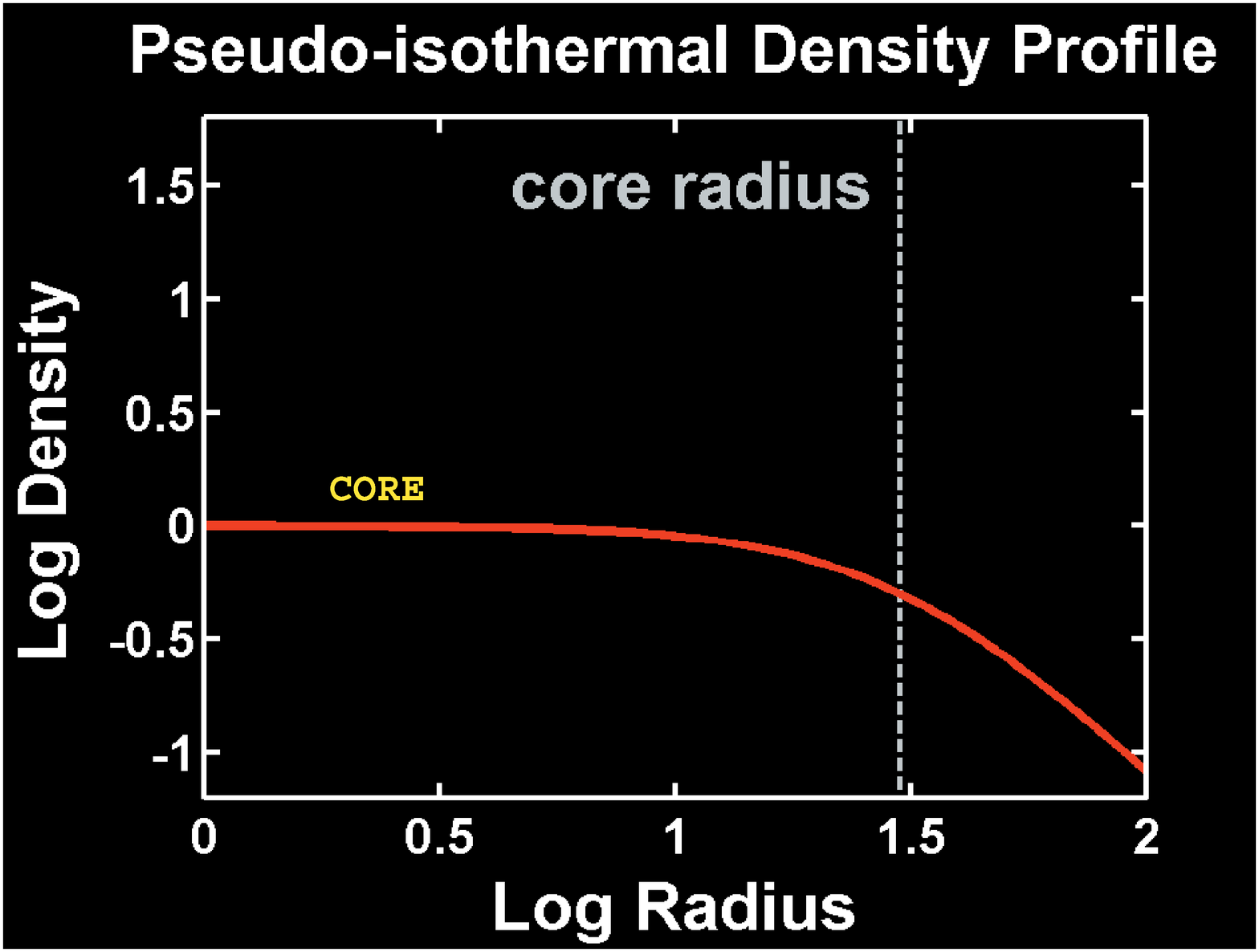}
%\end{center}
%\includegraphics[width=10cm]{cusp2.eps}
%\resizebox{6.3cm}{!}{\includegraphics{cusp1.eps}}
%\resizebox{6.3cm}{!}{\includegraphics{cusp2.eps}}
\caption{Schematic representation of the CC problem. The left panel represents the cuspy NFW profile. The~dotted vertical line is the scale radius of the profile. The~right panel represents the case of a cored profile. The~dotted vertical line is the core radius of the profile.}
\label{fig:spectra}
\end{figure}

Flores \& Primack \protect\cite{flores} and Moore \protect\cite{moore94} ruled out cuspy profiles from DDO galaxies' rotation curves, and showed them to be well approximated by cored (or pseudo-) isothermal density profiles. The~problem then lies in the cuspy profiles produced in dissipationless simulations of the CDM model (see~Figure \ref{fig:spectra}).

\begin{figure}[H]
\centering
\includegraphics[width=14cm]{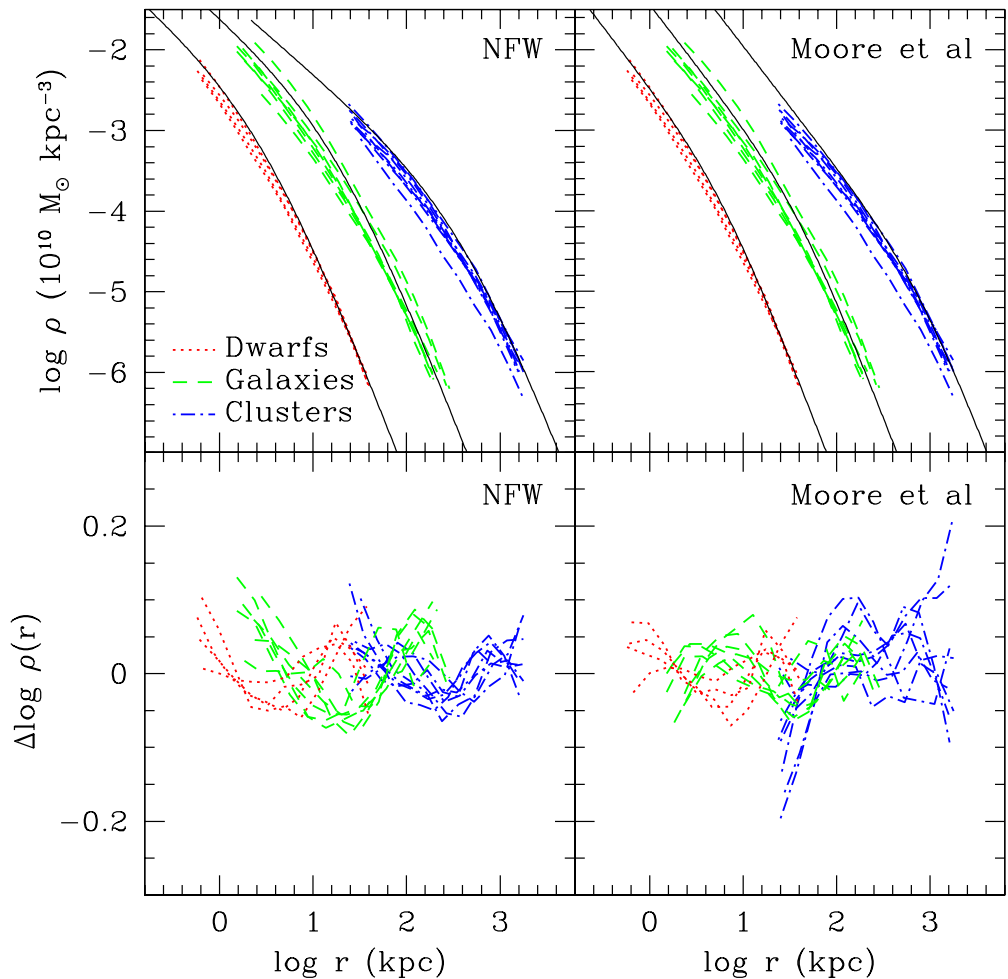}
\vspace{-6pt}
\caption{Density profiles comparison.  The top left corner is a comparison of the NFW profile (solid~line) with  Ref. \cite{navarro04}'s simulations, that are well fitted by the Einasto profile \cite{navarro10}, in the cases of dwarfs (red line), galaxies (green line), and clusters (blue line). The top right corner is the same as top left, but for the Moore profile. The bottom left and right corners display the residuals \mbox{(Figure reproduced from~\citep[][]{navarro04}}).
%{\bf Right panel panel}:
%Density estimates have been multiplied by
% $r^2$ in order to emphasize details in the comparison. Radii have been scaled to$ r_{-2}$, the radius where the logarithmic slope has the ‘isothermal’ value, $−2$. For comparison, we also show the NFW and
%M99 profiles, which are fixed in these scaled
%units. This scaling makes clear that the inner profiles curve inwards more gradually than NFW, and are substantially shallower than predicted by M99. The
%bottom panels show residuals from the best fits (i.e. with the radial scaling free) to the profiles using various fitting formulae (Section 3.2). Note that the Einasto
%formula fits all profiles well, especially in the inner regions. The shape parameter, α, varies significantly from halo to halo, indicating that the profiles are not
%strictly self-similar: no simple physical rescaling can match one halo on to another. The NFW formula is also able to reproduce the inner profiles quite well,
%although the slight mismatch in profile shapes leads to deviations that increase inwards and are maximal at the innermost resolved point. The steeply cusped
%Moore profile gives the poorest fits
}
\label{fig:NFWmoore}
\end{figure}

{The dissipationless N-body simulations of Navarro, Frenk, \& White \protect\cite{nfw,nfw1} }, then showed that DM profiles are cuspy, with inner density $\rho \propto r^\alpha$, with inner density power index $\alpha=-1$, that they are universal in dissipationless simulations, that is independent from the cosmology, and from the scale (coined NFW profiles).
An even steeper profile predicted by \cite{Moore1998,Fukushige2001} gave $\alpha=-1.5$, while other
authors found that the inner slope is dependent on the object considered, and/or its mass~\citep{Jing2000,Ricotti2003,Ricotti2004,Ricotti2007,DelPopolo2010,Cardone2011b,DelPopolo2011,DelPopolo2013d,
DiCintio2014}. More recent N-body dissipationless simulations tend to agree on the fact that a profile flattening towards the
centre to a minimum value of $\simeq\!\!-0.8$  \citep{Stadel2009}, namely the Einasto profile, seems to give a better
fit to simulations \citep[]{Gao2008} (see Figure \ref{fig:NFWmoore}). The problem lies in the fact that the smallest value predicted by dissipationless N-body simulations is larger than the values
obtained respectively  \mbox{by observations \citep{Burkert1995,deBlok2003,Swaters2003,KuziodeNaray2011,Oh2011a,oh}}, in SPH simulations
\citep{Governato2010,Governato2012}, or in semi-analytical models
\citep{DelPopolo2009,Cardone2012,DelPopolo2012a,DelPopolo2012b, dpc2012,DelPopolo2014a,Popolo:1999im}.

That discrepancy has been fervently debated for two decades. Early HI observations of LSBs led to contradicting results. McGaugh \& de Blok \cite{mcgaugh_deblok} found a discrepancy between their Rotation curves (RCs) and the NFW halo, as did Cote et al. \cite{cote}. Conversely, van den Bosch \& Swaters \cite{vdbswater} could not exclude steep profiles in several of the objects they studied. Similar contradictions were obtained using H$\alpha$ observations: de Blok \& Bosma \cite{deblok_bosma} found evidence against steep profiles, de Blok et al. \cite{deBlok2003} measured inner slopes $\alpha = -0.2 \pm 0.2$, and similarly Spekkens et al. \cite{spekkens} obtained  $\alpha= -0.22 \pm 0.08$, while Hayashi et al. \cite{hayashi} showed   RCs   in agreement with cuspy profiles. High resolution observations usually agreed on flat profiles (\citep[][]{blaise,kuzio08,kuzio09,weldrake,trachte}, etc.).

\begin{figure}[H]
\begin{center}
\phantom{e}\hspace{-1.5cm}
\includegraphics[width=15cm]{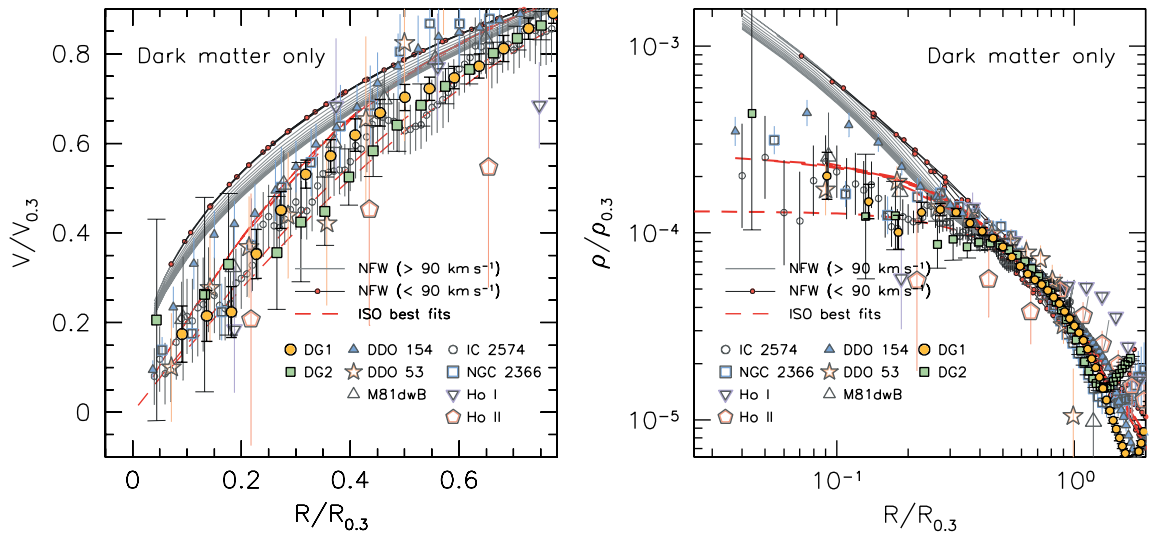}\hspace{2cm}
\end{center}
\vspace{-6pt}
\caption[]{
{%%\bf
Left panel}: Comparison of the RCs from (\textbf{a}) 7 dwarf galaxies from THINGS;
%\phantom{Figure3. Left panel: Comparison of the RCs from }
(\textbf{b}) the two galaxies DG1 and DG2 simulated by \cite{Governato2010},
%\phantom{Figure3. Left panel: Comparison of the RCs}
to the NFW (solid lines), and pseudo-isothermal (ISO) profiles (red dashed line). The NFW RCs  with $V_{200}$ in the range 10--90 km/s are highlighted by small red dots. The rotation velocity $V$ is scaled to $V_{0.3}$, namely the value of  $V$ at
$R_{0.3}$, representing the distance at which $\frac{d \log V}{d \log R}=0.3$.
%\phantom{Figure3.:}{%%\bf
Right panel%}
: same as the left, but for the density profiles \mbox{(figure reproduced from~\citep[][]{oh}}).
}\label{fig:RCdenNFWiso}
%}
\end{figure}

Gentile et al.~\protect\cite{gentile,gentile1,gentile2} decomposed the total rotation curves of some spiral galaxies in stellar, gaseous, and dark matter components.
% (see Figure 15).
Fitting the density with various models they found that constant density core models are preferred over cuspy profiles. Similar results were obtained by \mbox{Oh et al. \protect\cite{oh}} using 7 dwarf galaxies from THINGS (The HI Nearby Galaxy Survey) galaxies. The~comparison of the RCs (and the density profiles) with the NFW, and pseudo-isothermal (ISO) profiles is plotted in Figure~\ref{fig:RCdenNFWiso}. The plot clearly shows that the ISO profile is a better fit to the RCs.

Similar results were also presented using the LITTLE THINGS galaxies \cite{Oh2015}.
\pagebreak

%
%Dwarf galaxies are dark matter (DM) dominated, and have a low baryon fraction \citep{deBlok1997}. They have been widely used because of their simple dynamical structure, at least disc galaxies without %bulges.
%

These studies were mainly related to dwarf or low surface brightness galaxies. In the case of high-surface brightness objects, or large galaxies, determining their inner density structure is more complicated.  Therefore, stating the nature of the inner density profile for all galaxies, cored or cuspy, is not so obvious nowadays.
While, according to \cite{Spano2008}, high-surface brightness galaxies are cored, other authors (e.g., \citep[][]{Simon2005,deBlok2008,dpc2012,DelPopolo2013d,Martinsson2013}) conclude differently. The THINGS  sample tends to be better described by isothermal profiles (ISO) for low luminosity galaxies,
$M_B>-19$, while  for $M_B<-19$, cuspy or cored profiles describe them equally well.

The situation is even more confusing as dwarfs do not always have flat slopes, as seen in
%In the case of NGC 2976, 4605, 5949, 5693, 6689, the authors showed that the profiles range from 0 (NGC2976) to -1.28 (NGC5963).
\mbox{Simon et al. \protect\cite{Simon2005}}. They studied the low mass spirals NGC2976, NGC6689, NGC5949, NGC4605, and NGC5963, where they found a large scatter in the inner slope $\alpha$: for NGC2976, they obtained $\alpha \simeq 0.01$, compatible with a cored profile, while for NGC5963 they got a cuspy one, $\alpha \simeq 1.28$. The~other three galaxies had $\alpha \simeq 0.80$ (NGC6689), $\alpha \simeq 0.88$ (NGC5949), and $\alpha \simeq 0.88$ (NGC4605).

\begin{figure}[H]
\hspace{-0.5cm}
\centering
%\begin{center}
\includegraphics[width=15cm]{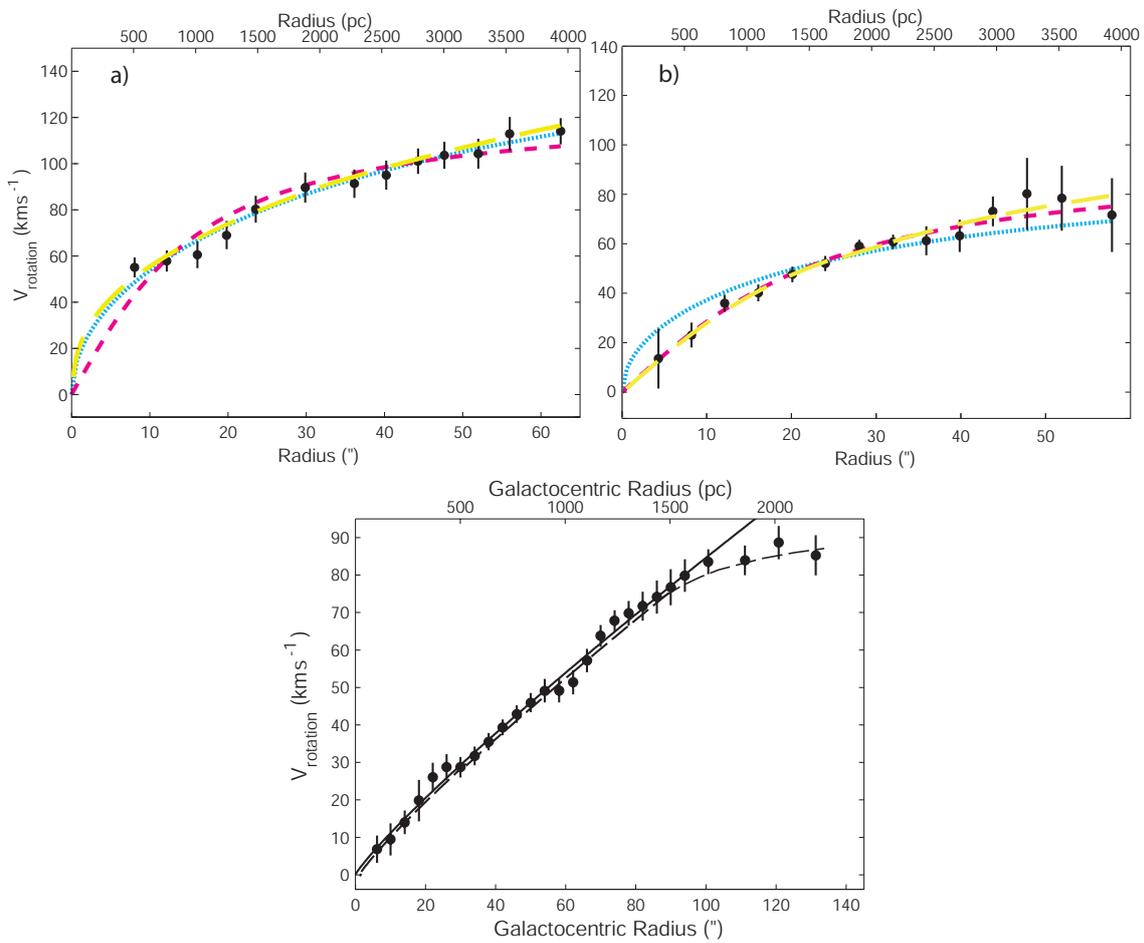}
%\end{center}
%\psfig{file=v_m1.eps,width=13.0cm}
%\psfig{file=DMO_DMB.eps,width=10.0cm}
\caption[]{Top left panel: DM halo RC of NGC5963 (black points with error-bars) computed by \cite{Simon2005}. The dotted cyan represents the RC obtained from a fitted NFW profile, the short-dashed magenta, from~the pseudo-isothermal profile fit (ISO), and the yellow dashed line, from the model of \cite{DelPopolo2012a}.
%\\ \phantom{Figure 4. T}
Top~right panel: same as the top left panel, for NGC5949.
%\\ \phantom{Figure 4. T}
Bottom panel: DM halo RC of NGC2976 (black points with error-bars) obtained by \cite{Simon2005}. The solid line is the RC computed from a power-law fit to the corresponding density (slope $\simeq 0.01$), and the dashed line, the RC obtained by the model of~\citep[][]{DelPopolo2012a} (from whom this figure is adapted).
%Minimum disk rotation curve of NGC 2976. Black circles with error-bars represent the rotation velocities relative
%to the minimum disk obtained by S03. The solid line is a power-law fit to the rotation curve which corresponds to a density profile of
%$\rho \propto   r^{-0.27}$, obtained by S03. The dashed line plots the result of the model of the present paper. (b) Maximum disk rotation curve of
%NGC 2976. Similarly to panel (a), but in this case $\rho \propto r^{-0.01}$.
}\label{fig:RCobs1}
%}
\end{figure}

In Figure \ref{fig:RCobs1}, the top left panel plots the DM halo RC of NGC5963 (black dots with error-bars), together with the RC obtained from a fitted NFW profile (cyan line), from a fitted pseudo-isothermal profile (ISO, short-dashed magenta line), and from the model of \citep[][]{DelPopolo2009} (yellow dashed line) taking account of baryonic physics. The top right, and bottom panels display the same data for the cases of  NGC5949, and NGC2976,
 respectively, with the same kind of fitted profiles for NGC5949, and just a flat power law (black line) and the same model \citep[][]{DelPopolo2009} (dashed line) for NGC2976. The cuspy density profile of NGC5963 is well approximated by an NFW profile, which reflects in the RCs. NGC5949's
RC is fitted equally well by RCs from an ISO or an NFW profile, while NGC2976 displays a very flat inner density ($\alpha \simeq 0.01$. All the three RCs are well approximated by the \cite{DelPopolo2012a} model.

In other terms, if a large part of dwarfs are well described by cored profiles, others are not.

Different results have even been obtained using similar techniques for the same object. For~example, the dark matter profile inner slope of NGC2976 is bracketed between $-0.17<\alpha<-0.01$, according to \cite{Simon2003}, while \cite{Adams2012} got $\alpha=-0.90 \pm 0.15$, and \citep{Adams2014} found, considering tracers being stars, $\alpha=-0.53 \pm 0.14$, or gas, $\alpha=-0.30 \pm 0.18$.

Somehow in agreement with the previous discussion, \cite{Oman2015} found that the shapes of observed rotation velocity of galaxies display a much larger variation than in simulations.

The discussion above highlights that the determination of the inner slope of galaxies, even for dwarfs, is no easy task. Moreover, it points out that the CC problem must be defined in terms of the inner mass of galaxies rather than of the inner slope of RCs or density profile. The result from the studies discussed above, and of several others, shows the existence of a range of profiles, and that,  \vphantom{even with}
 even with the improvements of nowadays kinematic maps, there is no agreement on the exact dark matter slopes distribution based on morphologies \citep{Simon2005,oh,Adams2014}.

The situation is even more flagrant going to larger masses (e.g., spiral galaxies) dominated by stars, or especially to smaller masses (e.g., dwarf spheroidals (dSphs)) where biases that enter in the system modelling
%\citep{Battaglia2013}
lead to opposite results.
%\footnote{See Section~\ref{sect:results} for a wider discussion.}

Several techniques have been used to understand and evaluate this problem on dSphs. The spherical Jeans equation gives results highly dependent on the assumptions,
since mass and anisotropy of the stellar orbits are degenerate in such model \citep{Evans2009}.
Maximum likelihood in the parameter space approach applied to Jeans modelling \citep{Wolf2012,Hayashi2012,Richardson2013} is plagued by similar such
degeneracies. Schwarzschild modelling has been applied to, e.g., Sculptor and Fornax dSphs, finding cored profiles~\citep{Jardel2012,Breddels2013,Jardel2013b,Jardel2013a}. Methods based on multiple stellar populations concluded that
Fornax (slope measured at $\simeq$ 1 kpc) and Sculptor (slope measured at $\simeq$ 500 pc) have a cored profile
\citep{Battaglia2008,Walker2011,Agnello2012,Amorisco2012}. However, a cusp is found in Draco using a Schwarzschild model \citep{Jardel2013a}. This latter results show that in
reality there is no accepted conclusion on a unique inner structures of dSphs.

A similar problem is also present in galaxy clusters. Sand et al. \protect\cite{sand1} combining weak lensing, strong lensing, and velocity dispersion studies of the stars of the BCG (Brightest Central Galaxy) found that out {%\bf
of} the clusters MACS1206, MS2137-23, RXJ1133, A383, A1201, A963, only RXJ1133 had a profile compatible with the NFW model, and similar studies of Newman et al.~\protect\cite{newman} (for A611), \mbox{Newman et al.}~\protect\cite{newman1} (for A383), and Newman et al.~\protect\cite{newman2}
 (for MS2137, A963, A383, A611, A2537, A2667, A2390) (see also \citep[][]{dpclust}) also found flatter profiles than other studies.
For example,  \mbox{Donnaruma et al.}~\protect\cite{donnaruma} found a cuspy profile for A611 combining strong lensing and X-ray observations, among other discrepancies from Newman's \protect\cite{newman2}, which covered seven relaxed, massive clusters with flat and cuspy profiles and an average slope $\alpha=0.50 \pm 0.1$.

In general gravitational lensing yields conflicting estimates: they sometimes agree with numerical simulations \protect\cite{dahle,gavazzi2,donnaruma} but can also find much shallower slopes ($-$0.5) \protect\cite{sand2,sand1,newman,newman1,newman2,DelPopolo:2002nd,DelPopolo:2005sq}. X-ray analyses have similarly led to a wide range of slope values, from $-$0.6 \protect\cite{ettori} to $-$1.2 \protect\cite{Lewis2003} till $-$1.9 \protect\cite{arabadjis1}, but can also agree with the NFW profile \protect\cite{schmidt}.
%; 34 Chandra X-ray observatory Clusters) Newman et al. 2012\protect\cite{newman2}

While early observations obtained conflicting results concerning the inner structure of the density profiles, high resolution observations, on average, agree on profiles flatter than the NFW's. At the same time, the new observations show a diversity in the inner structure from galaxy to galaxy, as also shown in \cite{Oman2015} simulations.

\begin{figure}[H]
\begin{center}
\includegraphics[width=7cm]{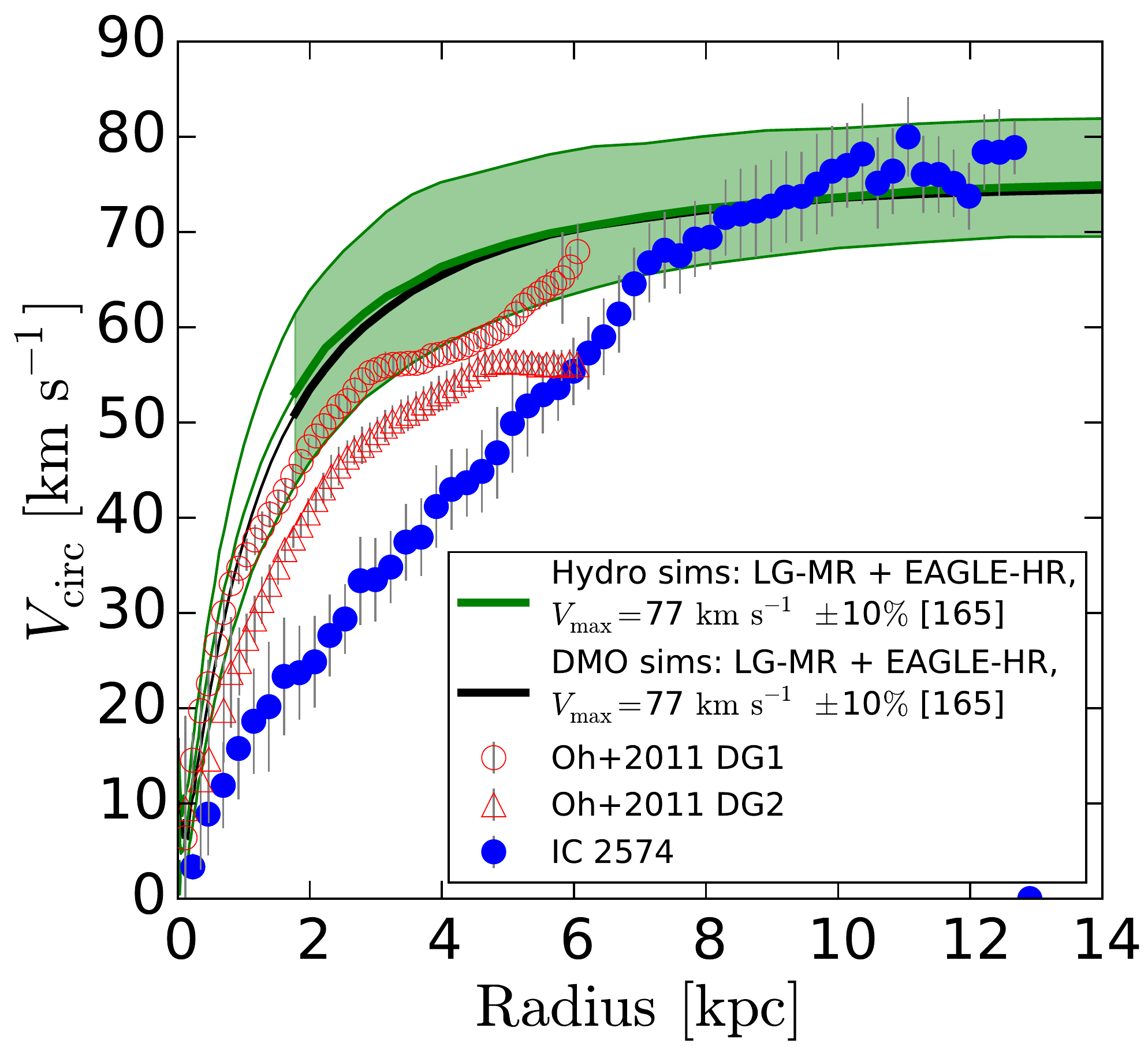}
\end{center}
\vspace{-12pt}
\caption{Comparison of the observed IC2574 RC (filled circles) with the two simulated galaxies DG1, and DG2 of \citep{Governato2010}.
The green line, and shaded region, represent respectively  the median rotation velocity, and scatter, of the galaxies simulated by \citep[][]{Oman2015} (from which this figure is reproduced).
%The green line shows the median circular velocity curve of all galaxies from our LG-MR and EAGLE-HR simulations (see Section 2.1)
%with $Vmax = 77 km/s  \pm 10$ per cent, matching the value of Vmax of IC 2574. The shaded area indicates the 10th–90th percentile range at each
%radius. The lines become thinner and the shading stops inside the average convergence radius, computed following the prescription of Power et al. (2003). The numbers in square brackets in the legend are the %numbers of galaxies/haloes that contribute to that velocity bin. The solid black line is the median circular velocity profile of haloes of the same Vmax, identified in
%our DMO simulations.
}
\label{fig:RCsim1}
\end{figure}

Even if \cite{Oman2015}'s simulation results (shaded green band in Figure \ref{fig:RCsim1}) are in agreement with the RC of galaxy IC2574 at radius $> 6$ kpc, their behaviour in the inner parts is completely different. This~discrepancy points out that convergence in the inner slope of RC between simulations does not mean that they are correctly describing the whole behaviour of the RC.
In fact, the deficit of the mass in the inner part of the profile better characterises the CC problem \cite{Oman2015}. The key issue is not the shape of the density profile but the excess amount of DM predicted by CDM in the central kpcs of the galaxy. The tension is already evident at scales at which the circular velocity reaches its asymptotic value \cite{kuzio_spekkens}.

Unfortunately, almost all the observation papers in the literature estimate
the inner slope of galaxies through $\alpha$, except for \cite{Walker2011} who measured the inner slope
through $\Gamma \equiv \Delta\log{M}/\Delta\log{r}<3-\alpha$ for Fornax and Sculptor. The latter slope, an integrated quantity, is
more easily evaluated than the local quantity $\alpha$, but only provides some constraints, not a precise value for the inner
slope.

%Moreover, to our knowledge, nowadays this quantity is only estimated for the two quoted galaxies.

%So, in summary, since in literature all the estimates of the inner slopes are done with $\alpha$, we will
%use this quantity to make a comparison with theory.
%This is also what was done in a recent paper \cite{Oh2015}, who determined the inner slopes of the LITTLE THINGS
%galaxies, as customary, determining $\alpha$ and compared it with Di Cintio theoretical model (see the following).

On the other side of the mass spectrum, a clear determination of the inner structure of dSphs, cored or cuspy, is very important, as objects with smaller masses are more likely to display a similar inner profile than that of dissipationless
N-body simulations predictions (cuspy).

\subsection{Early Solutions}

Many solutions have been proposed to solve the CC problem and in general the SSP$\Lambda$CDMs. A decade ago some authors (e.g., Refs. \protect\citep[][]{vandenbosch,vdbswater}) turned against observations, claiming that the inconsistencies could be due to poor resolution or to an improper account of systematic effects.  Non-circular motions, beam smearing, off-centring, slit-misplacement, which tend to systematically lower slopes, were charged with the discrepancies.

In HI observations, the finite beam size produces a smear out of HI emission, giving rise to larger disks. The effect depends on the size of the beam, the HI distribution, the inclination angle of the galaxy, and the intrinsic velocity gradients. The problem can be solved with high spatial resolution observations, $<$1 kpc (see the following).

In H$\alpha$ observations a slit misplacement may lead to missing the dynamic centre of the galaxy, with the result of having flatter profiles. The problem can be solved in different ways \mbox{(e.g., 3d spectroscopy~\citep[][]{kuzio08,Spano2008}).}
The gas is usually assumed to move on circular orbits, so non-circular motions produce an underestimation of the slope. Those motions are, however, of the order of a few km/s, as shown by~\cite{trachte}. Nowadays high resolution observations can distinguish cored and cuspy haloes by deriving their asymptotic inner slopes from rotation curve data~\protect\cite{KuziodeNaray2011}.

%According to Ref. \protect\cite{hayashi}, was the following, error bars were large enough so that the cores are favored but cusps can usually not be ruled out.
Several authors \protect\cite{power,navarro04,hayashi} suggested ways to reconcile simulations with  observations, claiming their simulations were in good agreement with observations, since they become progressively shallower from the virial radius inwards, and that the discrepancy came from projection effects.
Indeed, DM haloes are 3-D, but the practice is to compare spherically averaged circular velocity of DM~haloes with the rotation speeds of gaseous disks.
%Comparing rotation speeds of gaseous disks to spherically averaged
%circular velocity of DM haloes, one should expect differences.
 In other words, the observational disagreement would be with the fitting formulae, rather than with simulated haloes \protect\cite{hayashi}.

Nowadays, this proposal is easily rejected: high resolution DM-only simulations have a minimum inner slopes $ \simeq$$-0.8$ \protect\cite{Stadel2009}, while the inner slope of galaxies observed with high resolution techniques are much smaller.

Another possibility took seriously the failure of the CDM model, claiming the problems were with the simulations \protect\cite{deBlok2003,deblok3,borriello}. However, modern simulations {do not} suffer from their past problems: lack of resolution, relaxation, and over-merging. Already then, convergence tests performed by~\protect\cite{diemand} showed that N-body simulations correctly determine the CDM density profiles. The~N-body simulations used in the past were
dissipationless, meaning they only took account of DM, while~baryons are not negligible in the inner regions of galaxies (inner kpc), and dominate over DM in the central 10 kpc of clusters \protect\cite{sand1,newman,newman1,newman2}. Nowadays, high resolution cosmological hydrodynamic simulations are available and we will discuss their important role in the next sections, dealing with the baryon  solutions of the SSP$\Lambda$CDM model.
%SSP$\Lambda$CDM model.

Pushing further, the validity  of the CDM paradigm was questioned, so it was speculated that other forms of DM (warm \protect\cite{som_dol}, fuzzy \protect\cite{hu}, repulsive \protect\cite{goodman}, fluid \protect\cite{peebles2000}, annihilating \protect\cite{kap}, decaying~\protect\cite{cen1}, or self-interacting \protect\cite{sperg_ste}) could solve the SSP$\Lambda$CDM.
More radical alternative solutions modified the spectrum at small scales (e.g.,\citep[][]{Zentner2003}), or gravity (e.g., $f(R)$  \citep{Buchdahl1970,Starobinsky1980}, $f(T)$ theories -- see \citep[][]{Bengochea2009,Linder2010,Dent2011,Zheng2011}, or the Modified Newton Dynamics -- MOND, \citep[][]{Milgrom1983b,Milgrom1983a})\footnote{
$f(R)$, and $f(T)$ theories are types of modified gravity theories, generalisation of Einstein's General Relativity.
First proposed in 1970 by Buchdahl~\protect\cite{Buchdahl1970}, and turned into an active research field by Starobinsky~\protect\cite{Starobinsky1980}, $f(R)$ theories are defined by a different function of the Ricci scalar in their Lagrangian \protect\cite{capozziello_de}. Inspired by the Teleparallel Equivalent of GR, the $f(T)$ theories have been introduced to explain Universe acceleration without using dark energy (see Ref.  \protect\citep[][]{Bengochea2009}). Finally, the Modified Newtonian Dynamics, was introduced in 1983 by Milgrom~\protect\cite{Milgrom1983b,Milgrom1983a} as a way to model rotation curves of galaxies.
}. In what follows, we will denote those paradigm changing solutions, whereas for DM, initial conditions or gravity, as ``cosmological solutions''.

\label{sub:BsolCC}\subsection{Baryonic Solutions to the CC Problem}

As discussed above, the CC problem could be solved with ``cosmological solutions'' that do not preserve the $\Lambda$CDM model, also known as the concordance model. Such modification could, however, alter the successful predictions of the concordance model, that explains many of the observations and features of our Universe.

Thus, before throwing away such a model, it would be wise
to verify if some piece of poorly understood or neglected local physics could be connected to the small scale problems.

$\Lambda$CDM solutions of the CC problem are based on ``astrophysical solutions'', for which some mechanism,  ``heating'' the DM, would produce an inner flatter density profile, such as
\begin{enumerate}[leftmargin=*,labelsep=5mm]
\item the effect of a rotating bar;
\item transferring angular momentum (e.g., from baryons to DM) through dynamical \mbox{friction \protect\cite{ElZant2001,ElZant2004,DelPopolo2009};}
\item AGN feedback, gas bulk motions generated by supernova (SN) explosions~\protect\cite{Mashchenko2006,Mashchenko2008,Governato2010};
\item the presence of a central black hole giving rise to a shallower cusp, as claimed by some authors~\protect\cite{delliou_b,delliou_b1,delliou_b2};
\item the role of angular momentum in structure formation.
\end{enumerate}

Several authors used spherical~\protect\cite{nusser,hiotelis,delliou,ascasibar,williams,dpl,dpl1,dpl2,dpl3,dpl4,dpl5,DelPopolo:2008mp}, or elliptic~\protect\cite{Popolo:2002mp} infall models\footnote{General Relativity secondary infall models have been presented by a group around Mimoso and Le Delliou~\protect\cite{mimoso,mimoso1,delliou_she,delliou_she1}.}, arriving to the conclusion that the larger the angular momentum of a proto-structure, the flatter its inner density profile, and finding agreement with the rotation curves of dwarfs~\protect\cite{williams}.
%, and showing that the specific angular momentum acquired in the semy-analitical models is larger than that of simulations.
% (see Figure 54).
Since the angular momentum acquired by a structure is anticorrelated with the peak height\footnote{The peak height of a proto-structure is defined as $\nu=\delta(0)/\sigma$, where $\delta(0)$ is the central peak overdensity, and $\sigma$ is mass variance  (see \citep[][]{DelPopolo1996}). $\nu$ is larger for more massive objects.}, the density profile of dwarfs is flatter than that of giant galaxies.

El-Zant et al.~\protect\cite{ElZant2001,ElZant2004} showed how clumps of baryons lose energy, transferred  through dynamical friction to the DM component of the system, flattening or erasing the natural DM cusp, both in dwarf galaxies and clusters of galaxies.
% (see Figure 55).
Other authors studied adiabatic contraction of DM haloes~\protect\cite{blumenthal,gnedin}\footnote{Calculated through iterative techniques (e.g., \citep[][]{spedi}).},  that conversely produces a steepening of the density profiles.

Del Popolo  (see also \citep[][]{DPK2009}) took simultaneously into account these effects \protect\cite{DelPopolo2009}:
\begin{itemize}[leftmargin=*,labelsep=5mm]
\item ordered angular momentum acquired by the proto-structure through tidal torques;
\item random angular momentum;
\item energy and angular momentum exchange between baryons and DM through dynamical friction;
\item and adiabatic contraction.
\end{itemize}

\begin{figure}[H]
\centering
%%\begin{center}
\includegraphics[width=8cm]{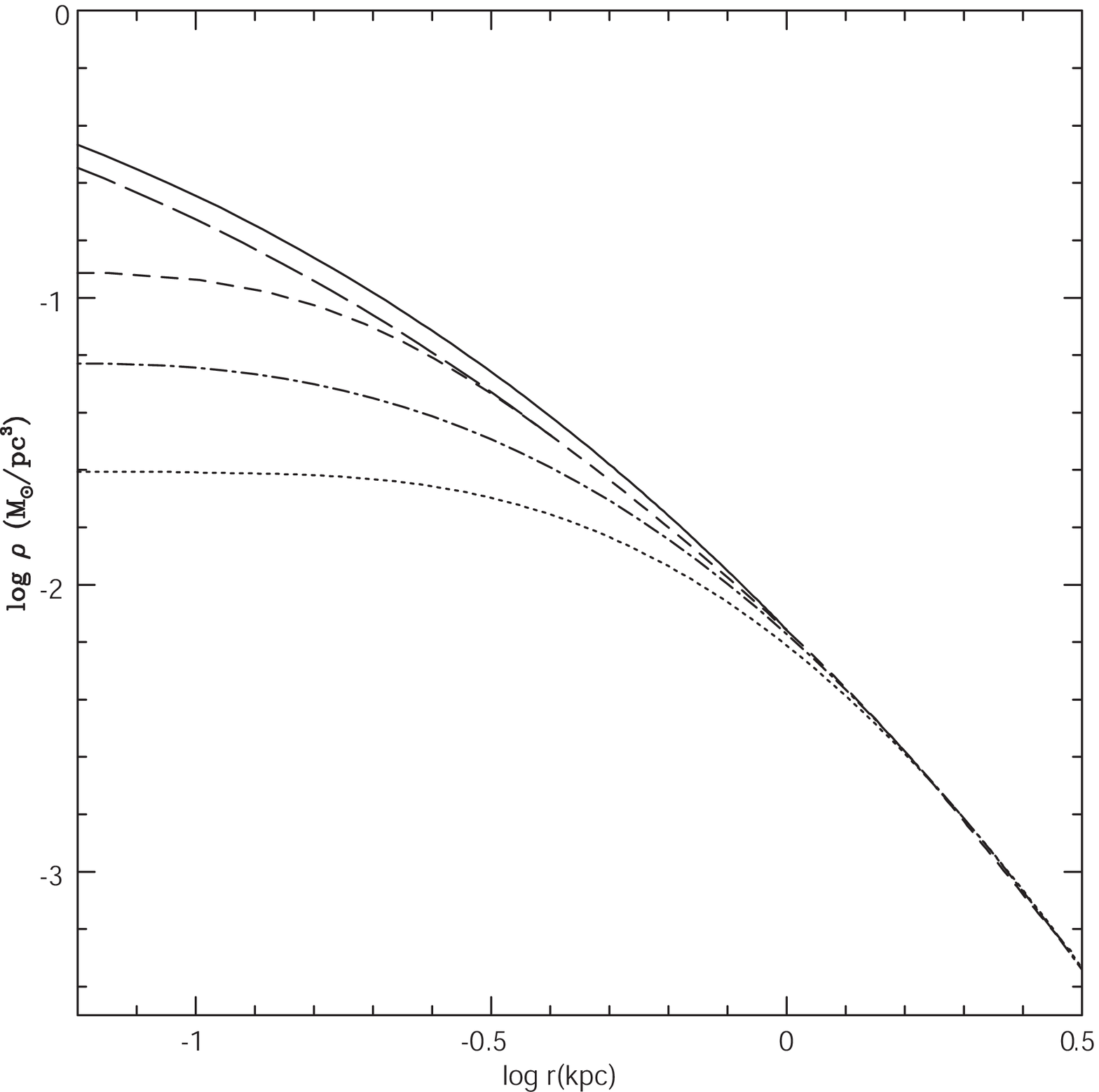}
\includegraphics[width=6.8cm]{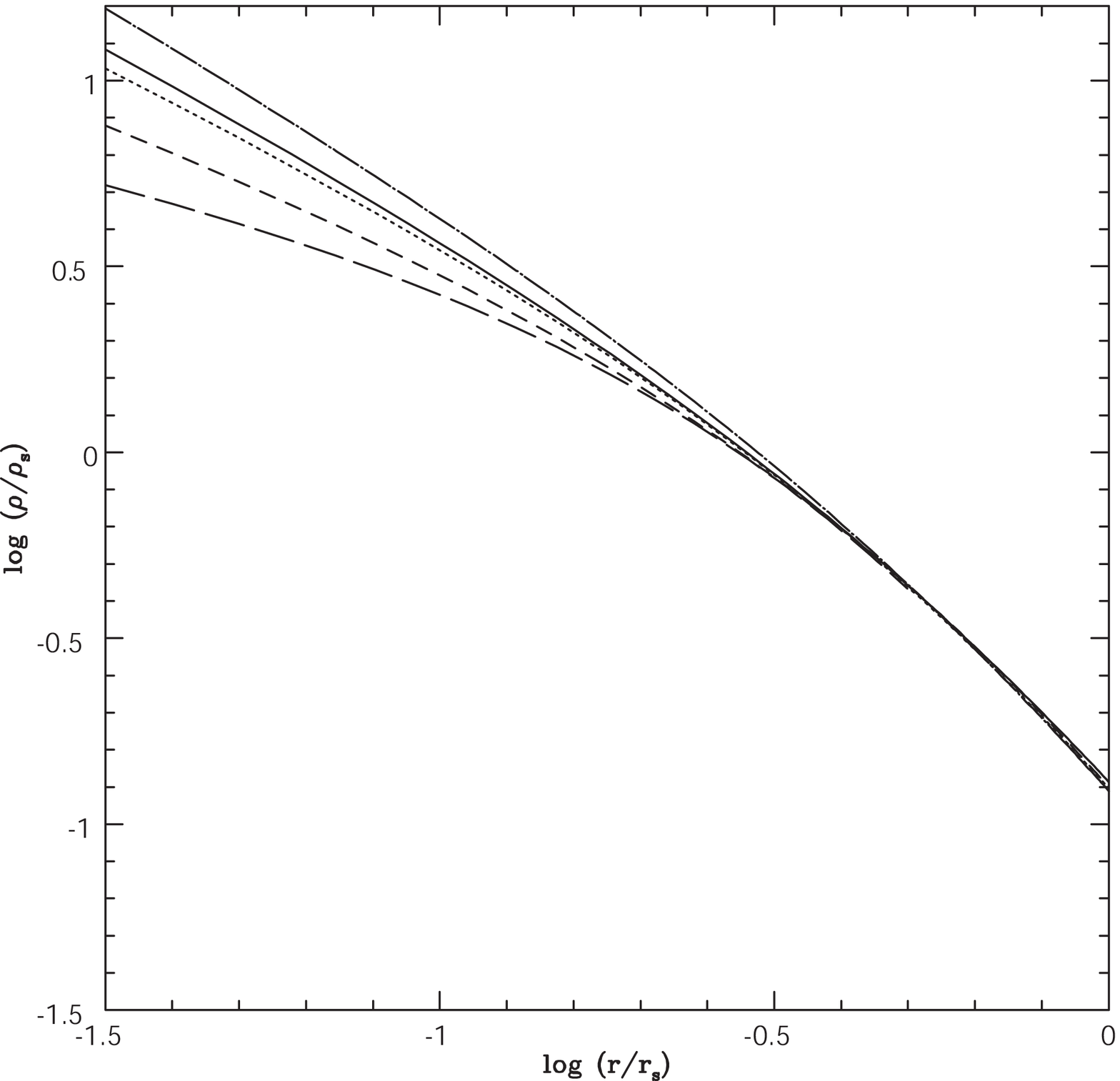}
%%\end{center}
\caption{Left panel: evolution of the density profile of a $10^9 M_{\odot}$ halo  \cite{DelPopolo2009} (see the discussion on the DFBC scenario). The profile at $z=10, 3, 2, 1$, and 0, are represented by the solid line,
long-dashed line, short-dashed line, dot-dashed line, and dotted line, respectively.
%\\\textcolor{white}{\bf Figure 6.}
Right panel: evolution of the density profile of a $10^{14} M_{\odot}$ halo of \cite{DelPopolo2009} (see the discussion on the DFBC scenario).
The dot-dashed line represents the total mass density profile (DM+Baryons) at $z=0$, while the DM profile at $z = 3, 1.5,$ $1$~and  0 is represented by the solid line,
dotted-line, short-dashed line, and long-dashed line, respectively (figure reproduced from \citep[][]{DelPopolo2009}).
}
\label{fig:DFevolDen}
\end{figure}

Angular momentum, and dynamical friction in galaxies and clusters not only tend to flatten their density profiles but also to change their global structure \cite{dpclusta,dpclustb,dpclustc,DelPopolo:2003iu}.

Ref.~\protect\cite{DelPopolo2009} showed that comparing dissipationless simulations with real structures containing baryons is not correct: the role of baryons in the inner part of the proto-structure is not negligible, explaining the discrepancy between N-body simulations and observed density profiles. In Figure \ref{fig:DFevolDen}, the evolution of haloes of $10^9$, and $10^{14} M_{\odot}$ are shown. Ref.~\protect\cite{DelPopolo2009} inscribed itself in the ``Dynamical Friction from Baryonic Clumps'' scenario (hereafter DFBC scenario) discussed in the following.

 In Del Popolo \protect\cite{DelPopolo2012a}, the model was applied to dwarfs galaxies showing the influence of the formation history, the content of baryons, and the environment on their density profiles.

%Differently from N-body simulations, the previous semi-analytic models does not take into account mergers. The commonness of spiral galaxies and
%their fragile discs, testifies that the role of mergers at least in some case is not so important (Toth \& Ostriker 1992\protect\cite{toth_ostr}). As %shown by Le Delliou (2008)\protect\cite{delliou_merg}
%mergers are an intermittent mass inflow accretion.

A different process was for the first time proposed by Navarro et al. \cite{Navarro1996a}, based on supernovae feedback (see the following), that was  able to flatten the DM profile.

Currently, the most favoured astrophysical solutions are
\begin{enumerate}[leftmargin=*,labelsep=5mm]
\item ``supernovae feedback flattening'' (SNFF) of the cusp
\citep{Navarro1996a,Gelato1999,Read2005,Mashchenko2006,Mashchenko2008,Governato2010,Governato2012}, and
\item ``dynamical friction from baryonic clumps'' (DFBC)
\citep[][]{ElZant2001,ElZant2004,Ma2004,Nipoti2004,RomanoDiaz2008,RomanoDiaz2009,DelPopolo2009,Cole2011,Inoue2011,Nipoti2015}.
\end{enumerate}

Before discussing those mechanisms, it is useful to recall how baryons and DM can ``interact''.

Smoothly distributed baryons with DM give rise to the adiabatic contraction (AC) of DM and baryons collapse towards the galactic centre, which steepens the DM
profile and thus increases the
central density of the structure. However, the effects of AC can be counteracted if energy is transferred from baryons to DM. This can happen if
\begin{enumerate}[leftmargin=*,labelsep=5mm]
\item The orbital energy of incoming clumps is transferred to DM through dynamical friction. \\As a consequence, DM particles move
to the outskirts of the galaxy, and the density DM profile is flattened
\citep{ElZant2001,ElZant2004,Ma2004,Nipoti2004,RomanoDiaz2008,RomanoDiaz2009,DelPopolo2009,Cole2011,Inoue2011,
Nipoti2015} (see also the review of \citep[][]{Pontzen2014}).
\item {\it Internal} energy sources in the galaxy \citep{Pontzen2014} heat up the DM particles.\\ If baryonic matter is expelled from, or even moved in, the galaxy (bulk motions produced by supernovae explosions,~\citep[][]{Mashchenko2006,Mashchenko2008}), this produces a temporary flattening of the gravitational potential, moving DM particles outwards, and flattening the cusp.
\end{enumerate}

\label{sub:SNsolCC}\subsection{Supernovae Feedback Flattening}

Since the suggestion from Ref.~\cite{flores}, stressed in many following works, of the importance of baryons in solving the CC problem, the first mechanism envisaged was connected to supernovae feedback \citep{Navarro1996a,Gelato1999,Read2005,Mashchenko2006,Mashchenko2008,Governato2010,Governato2012,Teyssier2013,Sawala:2015cdf}.
% (AGGIUNGERE ALTRI)

Ref.~\cite{Navarro1996a} showed that the sudden expulsion of baryons into the halo in a single event could flatten
the profile. The process is most efficient for galaxies with shallow potential, such as dwarfs. However, Ref.~\cite{Gnedin2002} showed that, while a single explosive event did not move sufficient energy to form a
core, repeated moderately violent explosions could reach the goal (however see \citep[][]{GarrisonKimmel2013} for a different point of view).

Gelato and Sommer-Larsen \cite{Gelato1999} studied more in detail the \cite{Navarro1996a} scenario, trying to reproduce DDO154's RC starting from NFW profiles: they tried to reproduce a gas outflow event by abruptly changing the disk potential. They found that it was necessary to expel at least 75\% of the disk mass in order to reproduce the RC .

Read and Gilmore \cite{Read2005}
%(MNRAS356,107)
showed that repeated outflows followed by gas re-accretion, could give rise to a core even in larger galaxies.

Refs.~\cite{Mashchenko2006,Mashchenko2008} showed that random bulk motions of gas driven by SN explosions in primordial galaxies could form a core. Similar results were obtained in  \cite{Governato2010} simulations. Refs.~\cite{Oh2011a,oh} compared the average slope of THINGS dwarfs with the simulations by \cite{Governato2010}. Governato et al.~\cite{Governato2012} ran simulations to study galaxies larger than in
\cite{Governato2010}, and compared the results with observations. They found, for~$M_*> 10^7 M_{\odot}$ galaxies, a correlation between the stellar mass $M_{*}$ and the inner slope.

%{\bf \cite{Governato2010} resolving "clumps" due to star formation, showed that outflows from explosion of supernovae
%will reduce the inner DM density. \footnote{These clumps have density similar to that of molecular clouds.}}

Governato's papers used the code GASOLINE \citep{Wadsley2004}, a N-Body+SPH code to simulate galaxies.
By means of the ``zoom'' technique \citep{Katz1993}, the resolution for gas particles was brought down to
$M_{\rm p,gas} = 3 \times 10^3 M_{\odot}$, while the DM particles had $M_{\rm p,DM} = 1.6 \times 10^4 M_{\odot}$, and
the softening retained at 86 kpc. The authors performed a two runs with different star forming thresholds: one in which stars formed if the hydrogen density was $>$$100/$cm$^3$
(High Threshold run, HT), and another with hydrogen density threshold $>$$0.1/$cm$^3$
(Low Threshold run, LT). These simulations, similarly to \cite{DiCintio2014}, implemented the blast wave SN feedback mechanism
\citep{Stinson2006}, and/or early stellar feedback \citep{Stinson2013}. There, the interstellar medium (ISM) received $10^{51}$ ergs of energy from $>$$8$$M_{\odot}$ stars.
%Even metals are allowed to diffuse between the particles of gas \citep{Shen2010}.
Energy ejected from SN ejected energy was coupled with the coefficient $\epsilon_{\rm esf}$ to the ISM. The MaGICC simulations \citep{Stinson2013} adopted the fiducial $\epsilon_{\rm esf}=0.1$.

{
%\color{blue}
Similarly, \cite{Pontzen2012} showed that final result of cusp flattening was generated by combining bursty star formation together with supernovae feedback, resulting in fast oscillations of the inner (1 kpc) galaxy potential, and expanding gas bubbles. This process only starts after the galactic centre accumulated cold gas density reaches $>$100/cm$^3$ and stars form\footnote{Consistent with \cite{Governato2010} assumptions; the threshold $>$10/cm$^3$ \citep{Ceverino2009} marks the limit for bulk gas flows.}.
Smaller densities (e.g., 0.1/cm$^3$) do not produce any visible changes in the DM inner density profiles. Governato et al. \cite{Governato2012}, repeated the calculation for larger mass galaxies (see Figure \ref{fig:DenHydro}, top left and right panels). Teyssier et al. \cite{Teyssier2013} used the adaptive mesh refinement code RAMSES together with a new SN feedback scheme, finding results in agreement with \citep[][]{Pontzen2012} (see Figure \ref{fig:DenHydro}, bottom left panel), which showed that $M_*> 10^7 M_{\odot}$ galaxies have a flat inner profile. Onorbe et al. \cite{Onorbe2015} found similar results, but for $M_*> 10^6 M_{\odot}$ galaxies (Figure \ref{fig:DenHydro} bottom right\label{ftn:}~panel)\footnote{\label{ftn:coreMassLimit} In general, the supernovae feedback mechanism is not able to transform cusps into cores in galaxies with  $M_*< 10^6 M_{\odot}$.}.

This view was recently criticised by \cite{Oman2015}. According to their simulations, in systems having
$V_{\rm max}<60$ km/s, baryons have little effect on the rotational curve even in the inner regions of the~galaxy. }

{
%\color{red}
According to  \cite{Oman2015}, the cores formed in \cite{Pontzen2012}  are fundamentally related to an ad hoc choice of parameters, while in their own simulations had no evidence of core formation.

\begin{figure}[H]
%\begin{center}
\hspace{2cm}
\includegraphics[width=12cm]{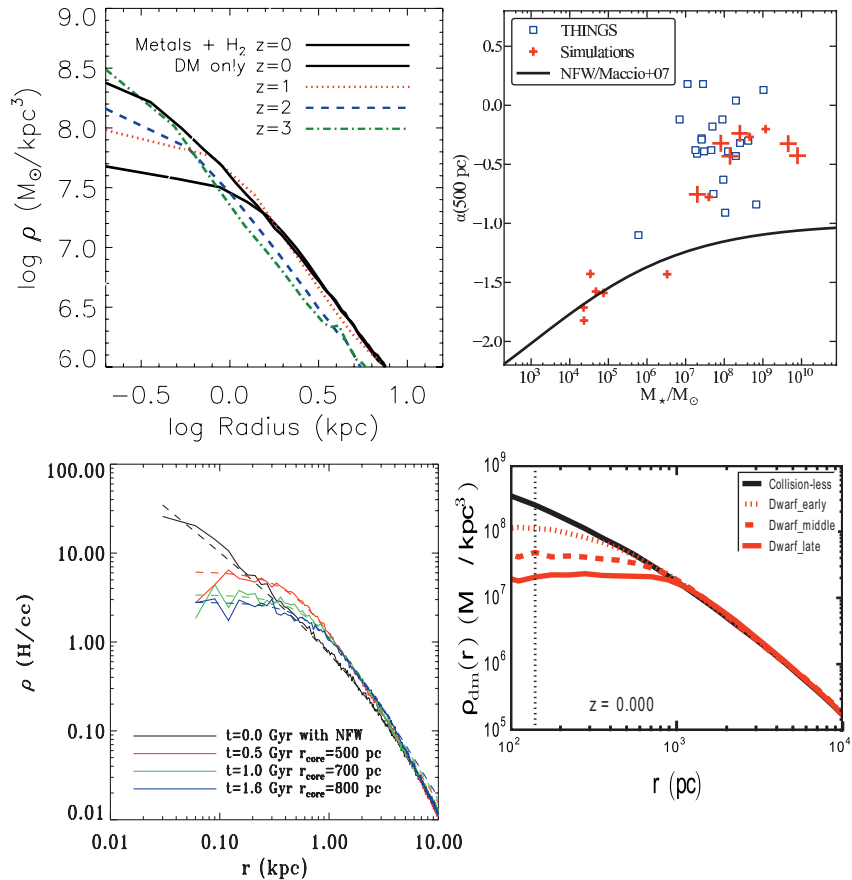}
%\includegraphics[width=8cm]{teyssier.eps}
%\includegraphics[width=8cm]{governato12a.eps}
%\includegraphics[width=8cm]{governato12b.eps}
%\includegraphics[width=8cm]{teyssier}
%\end{center}
\caption{Effect of baryons on density profiles. The top left panel represents  the evolution of a density profile in Ref.~\cite{Governato2012}'s hydrodynamic simulations. The top right panel compares their inner logarithmic slopes (at 0.5 kpc), for galaxies of different stellar mass (red crosses), with THINGS galaxies (open squares). The solid line represents the result of a previous DM-only N-body simulation  \citep[reproduced from][]{Governato2012}.  The bottom left panel displays the density profile evolution in the hydrodynamic simulations of  \cite{Teyssier2013}.
The bottom right panel shows the density profile evolution in the \cite{Onorbe2015} hydrodynamic simulations  for three different dwarfs: early (all stars form in early times), medium, late.
%
%Baryonic effects on CDM halo profiles in cosmological simulations, from Governato et al. (2012). (Left) The upper, dot-dash curve shows the cuspy dark matter
%density profile resulting from from a collisionless N-body simulation. Other curves show the evolution of the dark matter profile in a simulation from the same initial conditions
%that includes gas dynamics, star formation, and efficient feedback. By $z = 0$ (solid curve) the perturbations from the
%fluctuating baryonic potential have flattened the inner profile to a nearly constant density core. (Right) Logarithmic slope of the dark matter profile  measured at $0.5 kpc$, as a function of galaxy stellar mass. %Crosses show results from multiple hydrodynamic simulations. Squares show measurements from rotation curves of observed galaxies. The black curve shows the expectation for pure dark matter
%simulations, computed from NFW profilles with the appropriate concentration. For $M_*> 10^7 M_{\odot}$, baryonic effeects reduce the halo profile slopes to agree with observations.Evolution of the dark %matter density profile over the 2 Gyr of evolution for the control run with cooling, star formation and stellar feedback. We see the formation of a large core. We
%also show for comparison the analytical fit (dashed line) based on a pseudo-isothermal profile (see text for details)
}
\label{fig:DenHydro}
\end{figure}

%Indeed, simulations that produce cores generally adopt a high density threshold for star formation  \cite[$n_H > 100/cm^3$, e.g.][]{Governato2010,Teyssier2013} that results in micro-bursts of star formation
%concentrated in highly compact gas clouds that can be rapidly dispersed by feedback.  As they stress, their model of star formation is totally different from that of e.g. \cite{Governato2010,Teyssier2013}, and %their simulations thus allow star formation to occur throughout the rotationally-supported gaseous disc of a galaxy, limiting the sudden
%fluctuations in the gravitational potential on small scales.
{
Those  results are in agreement with that of Ref~\cite{GonzalezSamaniego2014}. The latter group simulated 7 high-resolution dwarfs, living in $1-2 \times 10^{10} M_{\odot}$ mass halos,  with different assembly history.
They found no case of flattening of the inner core. Their lowest inner slope (at 0.01--0.02~$R_{\rm vir}$) was
$-$0.8 and  corresponded to a dwarf formed in a halo with a very extended assembly history, which also implies  a more
extended star formation rate (SFR) history. {
%\color{red}
In addition, \cite{Schaller2015} got realistic galaxies in simulations, but formed no~cores.}
}

{
%\color{red}
 Besides the contradicting results on core formation obtained in different high resolution hydrodynamical simulations,
the results of \citep[][]{Governato2010,Pontzen2011} (and simulations using their same methods and assumptions)  have been
criticised on several fronts:
{
\begin{enumerate}[leftmargin=*,labelsep=5mm]
\item the energetics of the core formation \citep[][]{Penarrubia2012} (galaxies with $M_*<10^7 M_{\odot}$ have too few stars to generate the requested energy to flatten the cusp) and the required baryonic mass, marginally  exceeding the baryon content of the dSphs \citep{GarrisonKimmel2013}. Figure~\ref{fig:CCenergy} illustrates that problem in its left panel, from Penarrubia \cite{Penarrubia2012}, while the right panel is reproduced from Maxwell \cite{maxwell}'s study that arrives at opposite results.
\\Moreover, while the solution to the CC problem with the SNFF model needs a large number of SNs, and thus a large star formation efficiency (SFE), the solution of the TBTF problem, places an opposite demand on the star formation efficiency (SFE);
\begin{figure}[H]
\begin{center}
\includegraphics[width=6cm]{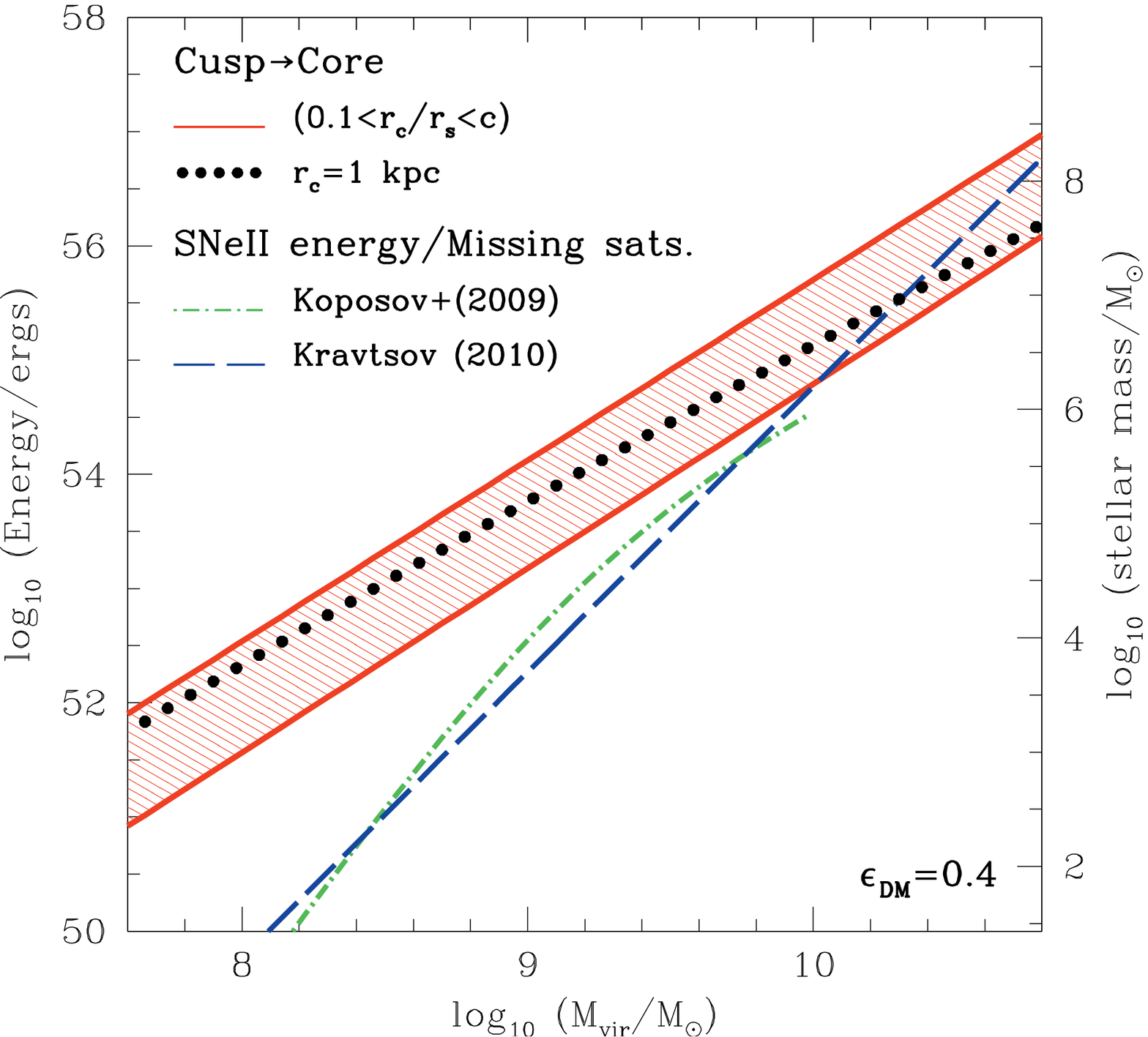}
\includegraphics[width=6cm]{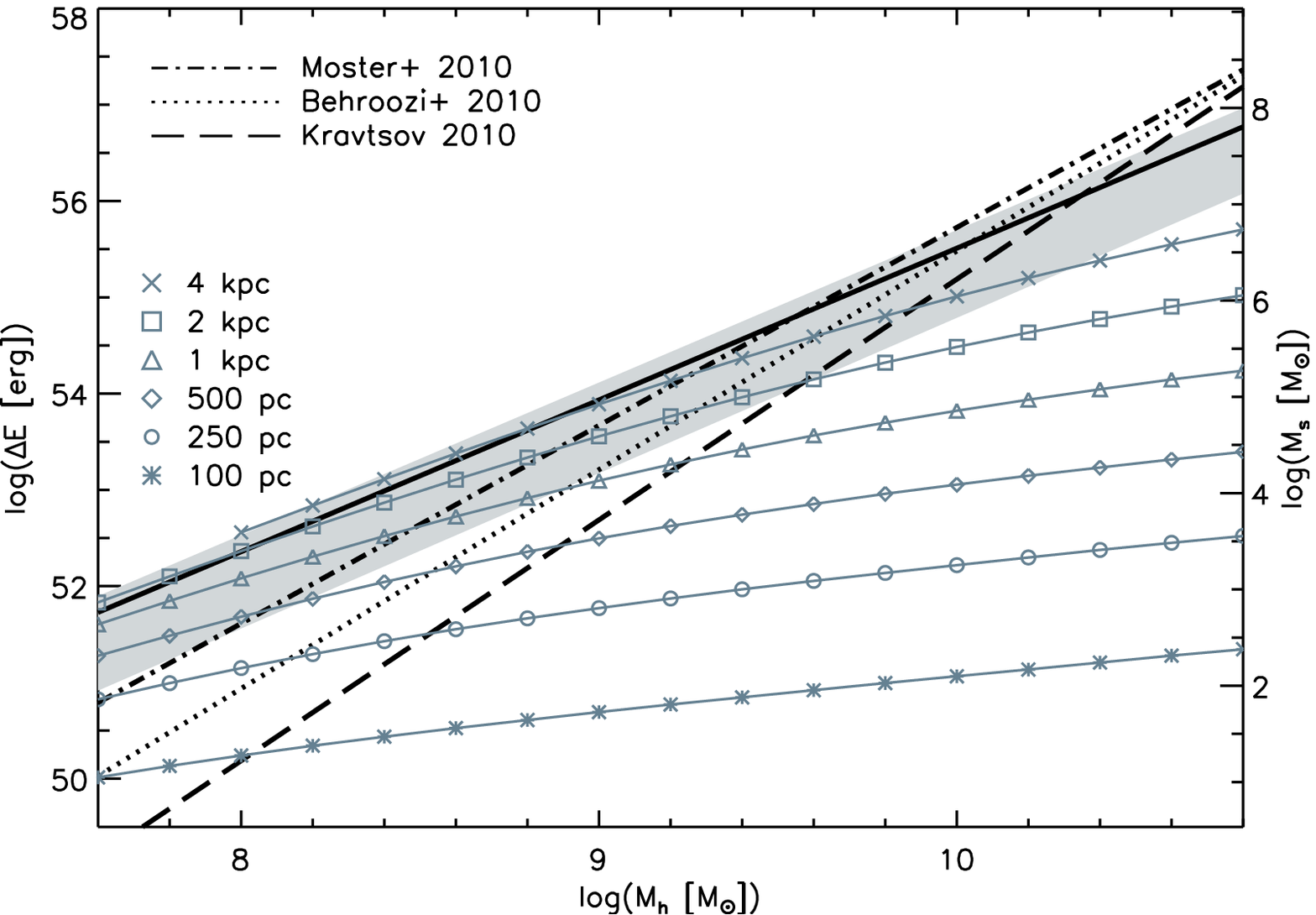}
\end{center}
\vspace{-6pt}
\caption{Left panel: SN energetic output with halo mass. The red shaded area represents the minimum energy of SN explosions, $\Delta E$, needed to generate a core of size  $0.1 <r_c/r_s < c$, $c$, and $r_s$ being the concentration parameter and the scale radius of the NFW profile of virial mass $M_{\rm vir}$, respectively. The~dotted black line corresponds to a core of size $r_c=1$ kpc.
%The dot-dashed green line, and the dashed blue line, represents the energy in SNII explosions.
% needed to solve the missing satellite problem
{%%\bf Re-write
The right vertical axis displays the luminosity, in stellar mass, derived from a star formation
efficiency $F_* = F_*(Mvir)$ constrained by the number of luminous satellites
in our Galaxy. Luminosities are converted into SNeII energy using Equation (6) of Ref.~\cite{Penarrubia2012}
and adopting a strong energy coupling $\epsilon_{DM} = 0.4$. The outputs of energy found for SN explosion compatible with the ``missing satellite'' problem in two different studies~\cite{koposov,Kravtsov2010}. The tension
between the ``core/cusp'' and ``missing satellite'' problems becomes obvious in haloes with
$M_{vir} < 10 M_\odot$ (panel reproduced from~\citep[][]{Penarrubia2012}).
%\\
%
%\textcolor{white}{\bf Figure 8.}
Right panel:
Estimates from Ref.~\cite{maxwell} of the energy $\Delta E$ required to convert a dark matter cusp into a core. Similarly to the left panel, the left axis shows SN energy outputs. However, the right axis shows the stellar mass corresponding to
$\Delta E$ assuming 100\% efficiency ($\epsilon_{DM} = 1$). The energy required to obtain the pseudo-isothermal density profiles, at given fixed core size, yields the solid lines with symbols. The solid black line
shows how $\Delta E$ scales with halo mass $M_h$ while the limit radius of redistribution of cusp mass $r_m$ scale as a fixed fraction of the halo radius $r_h$. The grey area corresponds to the energy estimate of
Penarrubia~et~al.~\cite{Penarrubia2012}, shaded in red in the left panel. The dotted, dashed, and dot-dashed lines display the $M_s-M_h$ relations of Behroozi et al. \cite{Behroozi2010}, Kravtsov
\cite{Kravtsov2010}, and Moster et al. \cite{Moster2010} respectively (panel~reproduced~from~\citep[][]{maxwell}}).
}
\label{fig:CCenergy}
\end{figure}

\item too high a value of energy coupling, $\epsilon_{\rm SN} \simeq 0.4$, compared to  0.05, a value deduced by \cite{Revaz2012};
\item a very high star formation threshold \citep{Sawala2014b,Oman2015} required to obtain the results of \cite{Governato2010};
\item they present difficulties in solving the TBTF problem \citep{Ferrero2012,GarrisonKimmel2013,Papastergis2015};
\item they lack resolution to follow the feedback processes which should transform the cusp into a core
\citep{Choi2014,Laporte2015a,Laporte2015b,Marinacci2014}.
\end{enumerate}
}}

Refs.~\cite{Oh2011a,oh} compared the average slope of THINGS dwarves with the simulations by \cite{Governato2010},
and~\cite{Governato2012} made a similar comparison for larger objects, and found a correlation among $M_{\ast}$
and the inner slope for galaxies having $M_{\ast}>10^6 M_{\odot}$%
\footnote{See footnote \ref{ftn:coreMassLimit}.}.%to be replaced in case of impossibility of footnote by "(see remarks in Section \ref{ftn:coreMassLimit}, p. \pageref{ftn:coreMassLimit} on SN feedback core formation limitations)"

Conversely, for $M_*<10^{6} M_{\odot}$, hydrodynamic simulations predict cuspy profiles . This result is in conflict with Ref.~\cite{Walker2011}'s
% Walker \& Pen$\tilde{a}$rrubia (2011)
results for Fornax and Sculptor inner structure. They showed that both galaxies are compatible with a core, using the slope of the mass profile $\Gamma \equiv \frac{d \log M}{d \log r}<3-\alpha$ as it gives a more reliable inner slope of DM haloes. Ref.~\cite{Madau2014}
%Madau, Shen \& Governato (2014?)
 claimed simulations in agreement with \cite{Walker2011}%
%Walker \& Pen$\tilde{a}$rrubia (2011)
, injecting however 50\% more energy from SNs.

{%\bf
Gnedin and Zhao \cite{Gnedin2002}, considering SN feedback in a Semi-Analytic model based on peculiar assumptions, claimed it cannot produce cores. Without discussing their assumptions, even supposing that they were considering all the details of stellar feedback, their objections are lessened as they did not consider the role of the baryonic clumps on core formation as done in Ref.~\cite{DelPopolo2009}.}

The only SNFF simulations forming cores in dwarf galaxies with masses $<10^6 M_{\odot}$ used the GIZMO code in P-SPH mode \cite{Onorbe2015}. This is much more naturally obtained in the DFBC scenario,
%$\tilde{O}$norbe et al. (2015),
  that we are going to discuss in the next section.

\label{sub:DFclumps}\subsubsection{Gas Clumps Merging}

As already reported, the other mechanism able to transform cusps into cores is that proposed by Refs.~\cite{ElZant2001,ElZant2004}.
%, based on merging gas clumps of $10^{5} M_{\odot}$ in the case of dwarves, and
%$10^{8} M_{\odot}$ in the case of spirals.\footnote{For precision's sake, the concept that large clouds could heat
%stellar systems was proposed by \cite{Spitzer1951}.}
%\footnote{In the \cite{Nipoti2015} simulations, the clump mass was $10^5-10^6 M_{\odot}$, and the total mass
%$10^9 M_{\odot}$.}
There are three ways, previously discussed, by which DM and baryons can interact. In the case when baryons form clumps of masses $\simeq\!0.01\%$ of the system mass, they can transfer --- through their motion --- their orbital energy to DM. As a result, and similarly to the SN feedback scenario, DM particles will move towards outer orbits, flattening the inner DM profile.
The process is more efficient when it occurs earliest, on smallest haloes.

Ref.~\cite{ElZant2001} showed how such mechanism works in galaxies, and \cite{ElZant2004} how it works in clusters. Ref.~\cite{Nipoti2004} used a similar model to study the evolution
of the cluster C0337-2522, finding that after the formation of the brightest cluster galaxy (BCG), the inner DM profile has a baryonically induced inner slope,  $0.49<\alpha<0.90$, smaller than the NFW profile.

Ref.~\citep{RomanoDiaz2008} used N-body simulations, with a hybrid N-body/SPH code, to study the evolution of galaxies in the DFBC scenario. They compared their results between the case of DM-only systems and of  mixed systems containing DM and baryons. They found that baryons subhalos heated up the cusp, forming a $\simeq$1 kpc core.

These results were also confirmed in the simulations of Refs.~\citep{Cole2011,Inoue2011,Nipoti2015}.

The scenario may be summarised as follows. Initially, the proto-structure contains DM and diffuse gas in the linear phase. DM goes non-linear first and  forms the
potential well in which baryons will~fall.

Clumpy structures  form from the instability of accreting gas (e.g. \citep[][]{Noguchi1998,Noguchi1999,Immeli2004a,Immeli2004b,Bournaud2007,Agertz2009,Aumer2010,Ceverino2010,
Ceverino2012}) connected to the arising of a very gas-rich disc.

The rotating disc turns unstable and forms clumps when its surface density, $\Sigma$, becomes too large, namely when $Q \simeq \sigma \Omega/(\pi G \Sigma)<1$ \citep{Toomre1964}, with $\Omega$ the disc angular velocity, and $\sigma$ its orthogonal 1-D velocity dispersion.

The largest clumps reach radii of 1 kpc (e.g., \citep[][]{Krumholz2010}) and  masses of a few percent of the disc mass. Galaxies containing baryon mass of $10^{10}$--$10^{11} M_{\odot}$  typically form clumps of mass $\simeq$$10^8$--$10^9 M_{\odot}$~(see~\mbox{\citep[][]{Agertz2009,Ceverino2010,Ceverino2012}).}

Those long lived clumps ($\simeq$$2 \times 10^8$ Myr) remain in Jeans equilibrium \cite{Ceverino2010}, and are rotationally supported.
Ref.~\cite{Krumholz2010} showed that the clumps to survive when, in agreement with the Kennicutt-Schmidt law, the gas is converted into stars at a rate of a few percent, similarly to local star-forming systems.
The gas clump remains bound, converting into stars,  and thus migrates to the galaxy centre.
%Baryons subject to radiative processes form clumps, which collapse to the centre of the halo while forming stars.
During the collapse phase, baryons are compressed (adiabatic contraction, Refs.~\citep[][]{blumenthal,gnedin}, e.g., at $z \simeq 5$ in the $10^9 M_{\odot}$ galaxy shown in Figure \ref{fig:DFevolDen}), making the DM profile more
cuspy. As dynamical friction (DF) between baryons and DM transfers energy and angular momentum to the DM component, the clumps migrate to the galactic centre. The cusp then heats up, and forms a~core.

{At later stages (e.g., around $z=2$), supernovae explosions provide other events where gas expulsion from the supernova
 explosion decreases the surrounding stellar density. However, as soon as the smallest clumps form stars with a small part of their mass, they get destroyed by such feedback\footnote{The process of star formation is not efficient.}
}\!\!.

Despite some considering that SNFF and DFBC scenarios represent different implementation of the same idea, based on some common features (e.g., gravitational interaction yielding clumps to DM transfer of energy), they essentially differ for two main reasons: Firstly, the epoch of effective profile flattening are markedly different. While the flattening provided by the DFBC (see Figure \ref{fig:DFevolDen}), starts at higher redshifts ($z<5$), at $z \simeq 3$ the DM density profile, in the pure SNFF scenario, the flattening remains similar to the NFW profile \citep{Pontzen2011,Maccio2012,Onorbe2015}.

Secondly, the energy sources moving the clumps come from opposite natures: in the SNFF scenario, the  energy of supernovae is driving the clump \citep{Mashchenko2006,Mashchenko2008}, while the DFBC clumps just ``passively'' infall to the galactic centre, in the sense of
Ref.~\citep{Mashchenko2006,Mashchenko2008}'s definition.

%Clearly in our SAM, supernovae have also a role in flattening the cusp (a few \%), but this is not comparable to that
%produced by baryonic clumps infall which start to flatten the density profile at larger z.

Refs.~\citep{DPPace2016,DelPopolo:2016skd} compared the two scenarios with high resolution data, from Refs.~\citep{Adams2014,Simon2005}, LITTLE THINGS \citep{Oh2015}, THINGS dwarfs \citep{Oh2011a,oh}, Sculptor, Fornax and the Milky Way, examining their respective predictions for the slope-stellar mass, and the slope-circular velocity relationships.
%Del Popolo \& Pace (2016).
They~found the DFBC scenario to perform slightly better than the SNFF, in addition to predicting, in DFBC and differently from SNFF, the emergence of cores at smaller stellar masses than $10^6 M_{\odot}$. {%\bf
However, even the DFBC mechanism cannot produce cores in very small dwarfs ($M_* \le 10^4 M_{\odot}$) in agreement with Ref.~\cite{Weidner:2013pnb} results.}

% Adams et al. (Astrophys. J. 789, 63, 2014), Simon et al. (Astrophys. J. 621, 757, 2005), LITTLE THINGS (Oh et al. in Astron. J. 149, 180, 2015), THINGS dwarves (Oh et al. in Astron. J. 141, 193, 2011a; Oh et %al. in Astron. J. 142, 224, 2011b), THINGS spirals (Oh et al. in Astron. J. 149, 180, 2015),
%Sculptor, Fornax and the MilkyWay, finding that the DFBC scenario performs slightly better than the SNFF. Moreover the DFBC scenario differently from the SNFF scenario predicts cores at $M_* \simeq 10^6 %M_{\odot}$.

Finally, recall that cusp can reform in galaxies with a bulge \citep{DelPopolo2014a},
We recall that in galaxies having a bulge, the cusp can reform as shown by \citep{DelPopolo2014a}, even in dwarf galaxies \citep{Laporte2015a}.
% Del Popolo \& Hiotelis (2014),
%Laporte \& Penarrubia (2014, 1409.3848).

\subsection{Cosmological Solutions}

As previously discussed, the SSP$\Lambda$CDM model could perfectly well hint at the CDM's paradigm failure.
In such a case, the nature of dark matter should be changed , or that of gravity modified,  in~models, both of which have already been widely checked with various degrees of success.

The simplest possibility grants DM with a small velocity dispersion ($\sigma\simeq 100$ m/s nowadays, also~denoted by a small DM relic mass) \citep[][]{Colin2000,som_dol}%Colin et al. 2000; Sommer-Larsen \& Dolgov 2001
, and is usually referred to as warm dark matter (WDM).
As past values velocity dispersion should be higher, such smear could affect small scale structures. This idea leads to an effect resembling the baryonic solutions discussed above: as DM particles retain higher velocity than in the CDM paradigm, small scales cluster less, leading both to flatter profiles and fewer low mass haloes.
Many simulations checked WDM structure formation \linebreak (e.g., \citep[][]{PolisenkyRicotti2011,Lovell,Maccio2012,Angulo}).
%Polisenky \& Ricotti 2011; Lovel et al. 2012; Maccio' et al. 2012; Angulo et al. 2013 (ref in review weinberg)).

Although tuning the WDM particle mass to the scale of the halos considered can solve several of the CDM problems, it is not able to get the correct rotation curves for all galaxies or in the entire mass range for which CDM has problems \citep{kuzio2010}. {%\bf
Both pure CDM and WDM models were explicitly tested against disk galaxies observed rotation curves by Ref.~\cite{Wu:2014eva}, finding no match, however, taking baryons into account, hydrodynamical simulations find the correct rotation curves (see the discussion in Section~\ref{sec:ccp} above).}

Moreover, WDM produces too few  subhaloes compared to, say strong-lensing subhalo fraction, as shown by the $m=2$ keV thermal relic of Ref.~\citep{PolisenkyRicotti2011}, and by several other authors (e.g., \citep[][]{Dalal,Fadely,Fadely1}).
%Dalal \& Kochanek 2002; Fadely \& Keeton 2011, 2012
A 1 kpc core requires a 0.1 keV thermal candidate, while large scale structure imply $m \simeq$ 1--2 keV, corresponding to cores of 10--20 pc (see Figure~\ref{fig:WDM}, and Ref. \citep[][]{Maccio2012b}). WDM too sharp spectrum fall off  \linebreak lead \citep{Schneider}
%Schneider et al. (2014, 1309.5960)
to conclude it does not improve on $\Lambda$CDM.
To make it worse, WDM alters the Lyman-$\alpha$ forest structure \citep{Nayara}
%Narayanan et al. 2000
and thus cannot consistently solve the small scale problems and the observed structure of the Lyman-$\alpha$ forest.

\begin{figure}[t]
\centering
\resizebox{7cm}{!}{\includegraphics{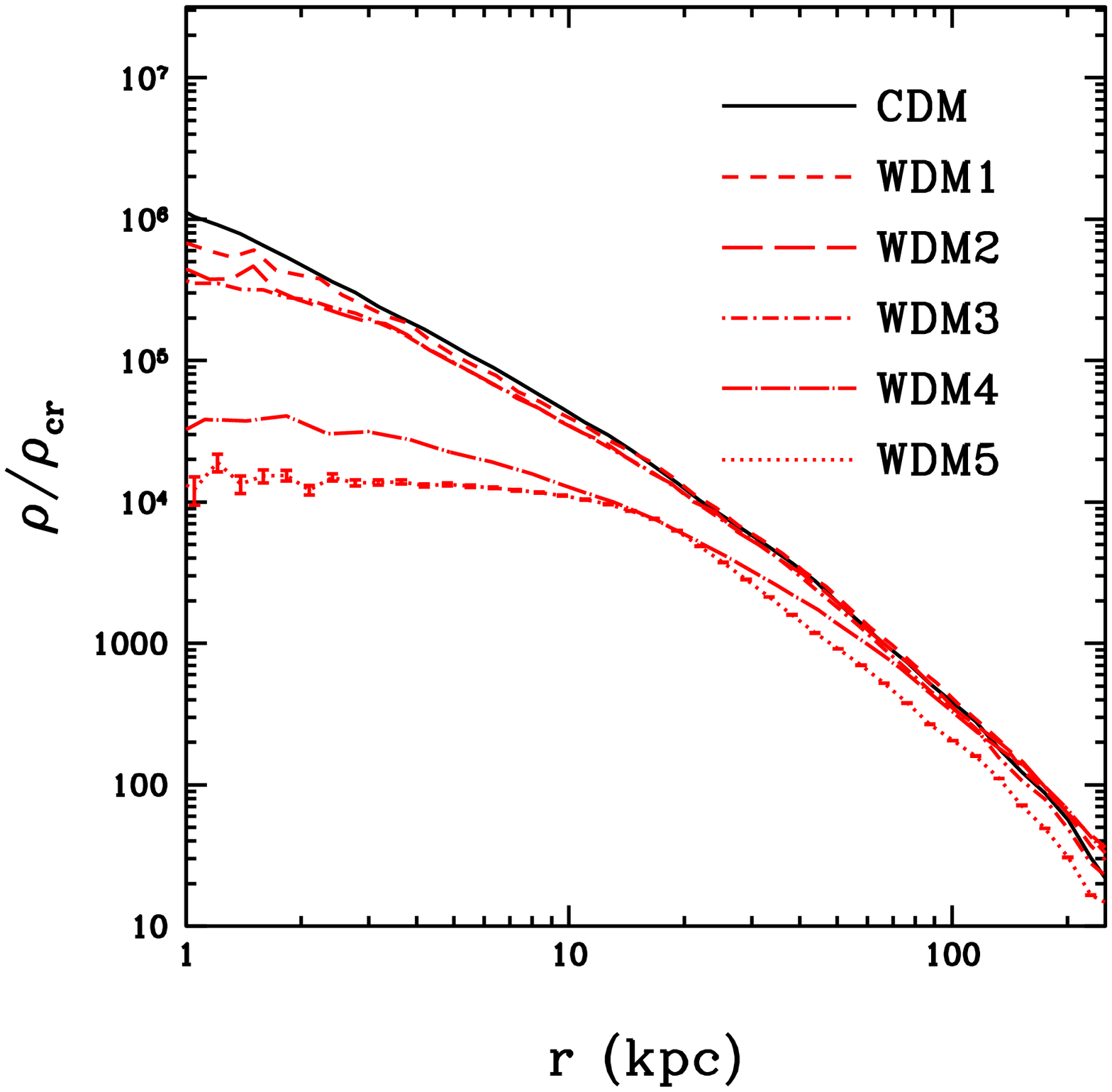}}
\resizebox{6.6cm}{!}{\includegraphics{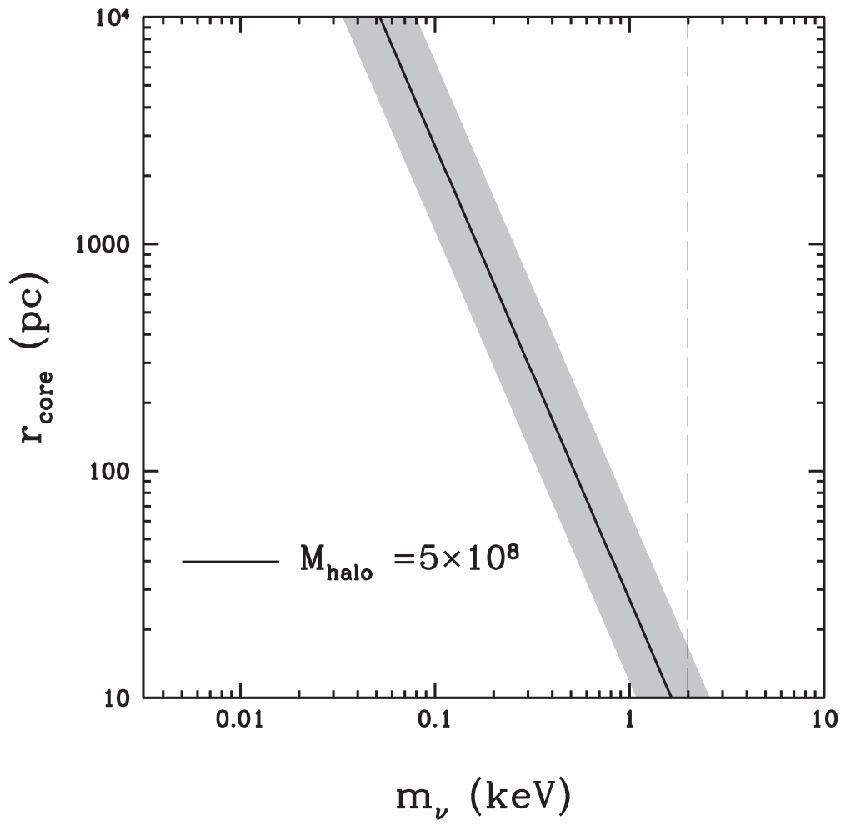}}
\caption{Left panel: comparison of a CDM density profile (solid black line) with five WDM models' profile ranging from $m=2 keV$ (WDM1) to 0.05 keV (WDM5).
%\\
%Expected core size for the typical dark matter mass of Milky Way satellites as a function of the WDM mass $m_\nu$ .
%\textcolor{white}{\bf Figure 9.}
Right panel: core radius obtained in terms of the WDM particle mass. The shaded band corresponds to  allowed cosmological values $0.15 < \Omega_m < 0.6$.
The vertical line represents the upper limit WDM mass, constrained from large scale observations  (reproduced from \citep[][]{Maccio2012b}).
 }\label{fig:WDM}
\end{figure}

The next simple possibility endows DM with self-interaction (Self-interacting DM, \linebreak hereafter SIDM, \citep[][]{sperg_ste}),
%Spergel \& Steinhardt 2000;
%Yoshida et al. 2000; Dave et al. 2001),
with cross-section of same order as for nucleon-nucleon ($\simeq$(m/g)$^{-1}$cm$^2$)\footnote{1 cm$^2$/g $\simeq$ 1 barn/GeV}. Redistribution of angular momentum and energy results from elastic scattering in the galaxies inner region, reduces tri-axiality and forms a Burkert profile core \citep{burkert}.
%Burkert 2000
 Some cosmological simulations~\citep{Newman2013a,newman2,rocha,peter}
%Newman et al. 2013a,b; Rocha et al. 2013; Rocha et al. 2013; Peter et al, 2013,
running with  0.1--0.5 cm$^2$/g cross sections, consistent with clusters of galaxy merging  observations,  \citep{clowe,randall,dawson},
%Clowe et al. 2006; Randall et al. 2008; Dawson et al. 2012
 claim that this scattering mechanism  solves the CC problem in dwarfs, MW sized galaxies, and clusters of galaxies. The result of \citep{rocha}'s simulations are presented in  Figure~\ref{fig:SIDM} for halo mass ranging from galaxies to clusters and two values of $\sigma/m$.
As the cored SIDM subhaloes feel more tidal stripping and disruption than CDM subhaloes, it solves the CC problem and improves on WDM \citep{rocha,peter}
%Rocha et al. 2013; Peter et al. 2013)
since it leaves enough subhaloes (see however \citep[][]{kuzio2010} for a different point of view).
%Kuzio de Naray et al. 2010
%0912.3518
Note also that SIDM have some natural appeal since particle models of the ``hidden sector'' produce them, e.g. \citep[][]{sperg_ste,Yoshida2000,Dave2001}.

\begin{figure}[t]
\centering
\resizebox{11cm}{!}{\includegraphics{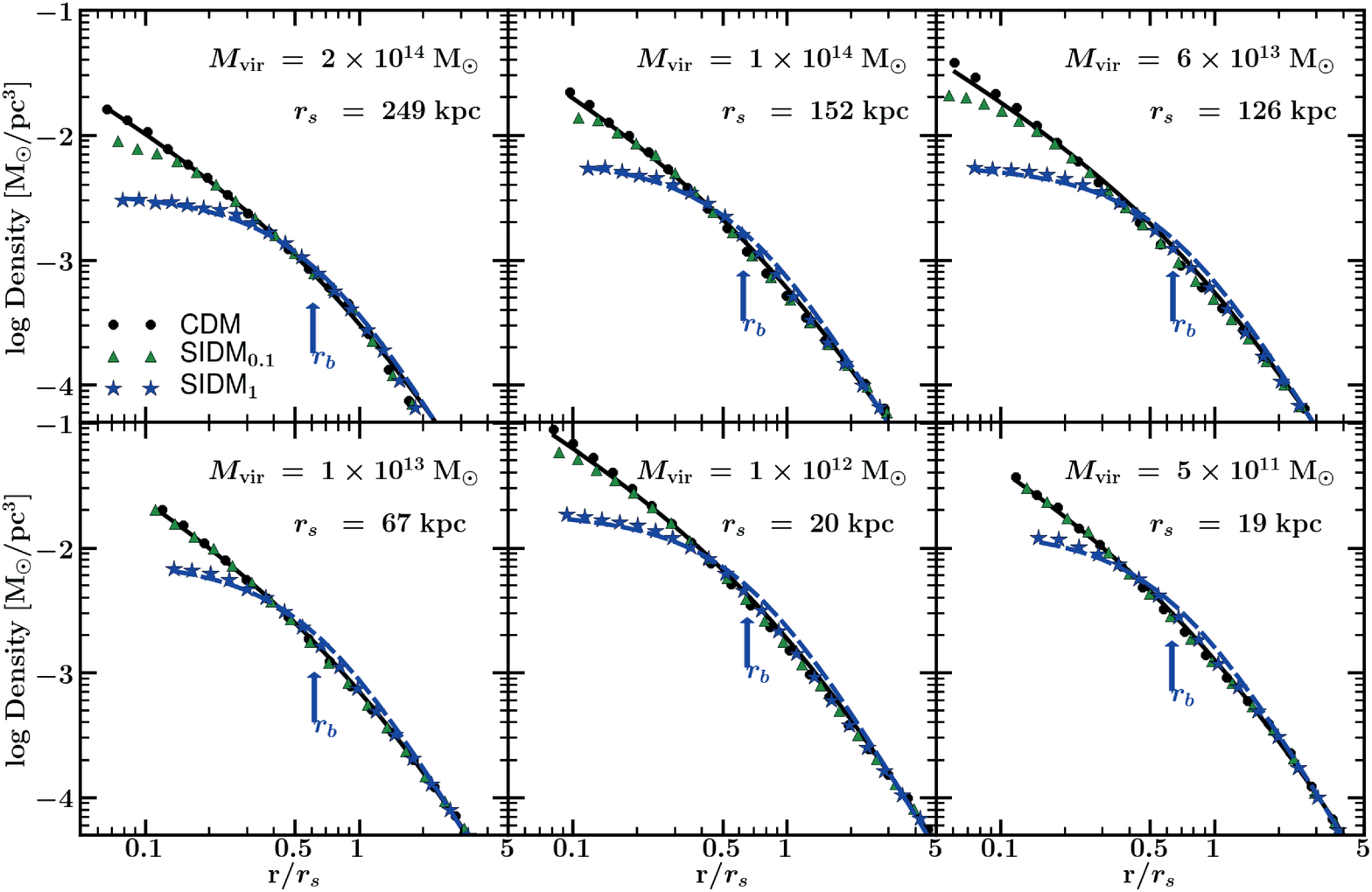}}
\caption{Comparison of the NFW (black line) and Burkert (blue line) density profiles with simulations of SIDM universes, using the two cross section over DM particle masses  \mbox{$\sigma/m=1$ and $0.1$}, denoted~$SIDM_{1}$ (blue stars) and $SIDM_{0.1}$ (green triangles) respectively. The arrow indicated the location of the Burkert's profile core radius (figure reproduced from \citep[][]{rocha}).
%Density profiles for our six example halos from our SIDM1 (blue stars) and $SIDM0.1$ (green triangles) simulations and their CDM counterparts.
%With self-interactions turned on, halo central densities decrease, forming cored density profiles. Solid lines are for the best NFW(black) and Burkert (blue) fits,
%with the points representing the density at each radial bin found by AHF. The arrow indicates the location of the Burkert core radius rb. rs is the NFW scale
%radius of the corresponding CDM halo density profile (black solid line). Burkert profiles provide a reasonable fit to our SIDM1 halos only because $rb \simeq rs$
%for $\sigma/m = 1 cm^2/g$, so a cored profile with a single scale radius works. As discussed in sect 7 this is not the case for $\sigma/m = 0.1 cm^2/g$ and thus Burkert
%profiles are not a good fit to our $SIDM0.1$ halos.
}\label{fig:SIDM}
\end{figure}

%SIDM (Spergel & Steinhardt 2000; Yoshida et al. 2000; Dave et al. 2001)(Light versions of WIMPZILLAS, and Q-balls): (Self-Interacting)
%Interaction: QCD but no EM;  Strength comparable to neutron-neutron; Significant self-scatering cross section
%Scattering strips the halos from small clumps of dark matter orbiting larger structures, making them vulnerable to tidal stripping and reducing their number.
%Difficulty: a. spherical clusters; b. against lensing. C. ALSO: Kuzio de Naray et al. (2010): high-resolution rotation curves of nine LSB galaxies -> the minimum core size in WDM models is predicted %to decrease with halo mass, WHILE the inferred core radii increase with halo mass and also cannot be explained with a single value of the primordial phase space density

SIDM has been declined in several variations with altered properties: negative scattering leads to repulsive DM (RDM) \citep[][]{goodman},
%Goodman 2000
 a condensate of massive bosons, which superfluid behaviour in the central part of DM haloes
%(Bose-Einstein condensate of DM particles, similar to axion but with a short range repulsive potential). The inner parts of dark matter halos would behave like a superfluid and be less cuspy.
smoothes down the cuspy profiles \citep{harko}.
%Harko 2011, 1105.2996
Such non-relativistic Bose condensate was recently simulated for structure formation \citep{schive},
%Schive, Chiue \& Broadhurst (2014)
where large scale structures could not distinguish CDM universes from their model, while flat density profiles and reduced small scale substructures resulted from the opposition, at small scales, between gravity and the uncertainty principle. A further implementation of this model using scalar field condensation, dubbed Scalar Field Dark Matter (SFDM), produced galaxies flat inner profiles \citep{robles}.
%Robles \& Matos 2012, 1201.3032

Wave-particle duality was summoned for Fuzzy DM (FDM) \citep[][]{hu},
%Hu et al. 2000
ultra-light ($m \simeq 10^{-22}$ eV) scalar particles which galactic core sized Compton wavelength cannot be ``squeezed'' further, resulting in flatter profiles and less substructure.

The last two aspects of SIDM are very common when related to indirect DM detection:  Self-Annihilating DM  (SADM) \citep[][]{kap}
%Kaplinghat et al. 2000
proposes that the self-interaction results in annihilation, with cross section-velocity $\sigma v \simeq$ 10$^{-29}$ (m/GeV) cm$^2$. In dense regions, the annihilation leads to possibly detectable levels of radiation. Annihilation reduces the structure's particle number in the centre, thus reducing gravity and consequently expanding particles orbits and flattening the density profile.

The second aspect, Decaying DM  (DDM) \citep[][]{cen1},
%Cen 2001
considers DM to decay into relativistic particles, also leading to radiation detection. The gravitational effect is similar to SADM in structure, reducing significantly the core's density of galaxies while large scale structures remain unaltered.

\label{sub:MTG}\subsection{Modified Theories of Gravity}

The preceding alterations of DM spoil the simplicity of the CDM paradigm, a further possible solution to the small scale problems legitimately questions the existence of DM and proposes changing  gravity itself: this leads to the branch of modified theories of gravity (MG). Although MG is an old issue, the discovery of the universe's accelerated expansion \citep{riess,perlmutter} literally exploded interest in it. Most of the more recent alternatives to GR  are cosmologically motivated, attempting to replace or supplement the concordance cosmology postulates of ``inflation'', ``dark matter'' and ``dark energy''~\cite{Pace:2014taa}. They all build on the premises that, although agreeing with GR locally in time and space, gravity may be quite different in the early universe or at large scales.

In this context, instead of interpreting the ``anomalous'' rotation curves of spiral galaxies nowadays as the trace of missing mass (DM), they reveal a lack in the gravity theory.

%The discovery of unexpected rotation curves for galaxies took everyone by surprise. Could there be more mass in the universe than we are aware of, or is the theory of gravity itself wrong? The %consensus now is that the missing mass is "cold dark matter", but that consensus was only reached after trying alternatives to general relativity and some physicists still believe that alternative %models of gravity might hold the answer.

From the discovery of the universe's accelerated expansion by the supernova surveys,  Einstein's~cosmological constant was rapidly reinstated, and quintessence was proposed to overcome the problems that such constant entails. Alternatives to GR also attempted to explain such results.

%Another observation that sparked recent interest in alternatives to General Relativity is the Pioneer anomaly. It was quickly discovered that alternatives to GR could explain this anomaly. This is now %believed to be accounted for by non-uniform thermal radiation

Nowadays, the catalogue of MGs theories, cosmologically motivated or not, extends to: $f(R)$, $f(T)$, Modified Newtonian Dynamics (MOND), BIMOND, Tensor-Vector-scalar theory (TeVeS),  Scalar-Tensor-Vector
Gravity Theory (STVG), Gauss-Bonnet MG, Lovelock MG, non-minimal scalar coupling, non-minimal derivative coupling, Galileons, Hordenski, etc.  (for a review, see \citep[][]{clifton} and Figure~\ref{fig:MGtree}, reproduced from their paper).

%Figure 11 plots a diagram of MGs.

\begin{figure}
\begin{center}
\includegraphics[width=12cm]{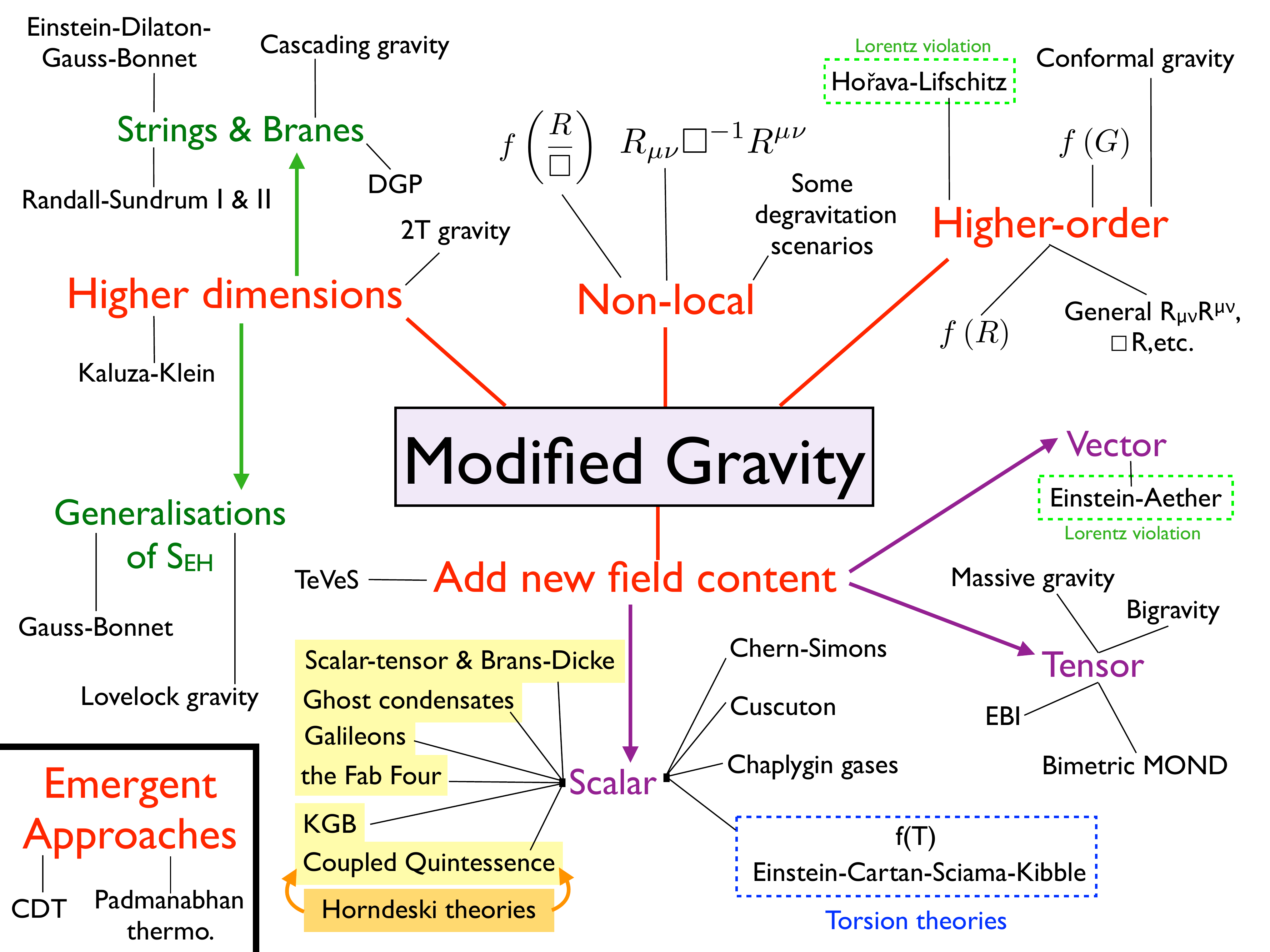}
\end{center}
\caption
{MGs tree diagram, reproduced from \cite{lcdm} %\textcolor{red}{Copyright permission from the original publisher is requested}
 (with permission from Tessa Baker. \href{file:///C:/Users/nino/Downloads/landscape_TBaker.pdf}{See updated version here}).
}\label{fig:MGtree}
\end{figure}

MOND demonstrated particular effectiveness in solving the SSP$\Lambda$CDM.
An {%\bf
originally} ad hoc modification of Newton's gravitation law was proposed in 1983 by Milgrom \citep{Milgrom1983a,Milgrom1983b}, for Newton's second law\footnote{However, this change {%\bf
in Newton's law} violates momentum conservation}:
%More precisely, there are two different ``interpretation" of his idea:

%a) modification of Newton's second law of motion, which is incompatible with momentum conservation;

\begin{equation}
F=m\mu (a/a_0)a
%=GMm/r^2
 \end{equation}

%b) modification of the law of gravity:
%\begin{equation}
%\mu (a/a_0)a=GM/r^2
 %\end{equation}
There, with the universal constant $a_0 \simeq 10^{-10}$  m/s$^2$, the gravitational force results nonlinearly in the acceleration, $a$, as the functional form $\mu (a/a_0)$ tends to 1 for high values of the acceleration, while~small accelerations modify Newton's law with $\mu\sim a/a_0$.
% This interpretation again contradicts momentum conservation.

The analytical form of $\mu(a/a0)$ in MOND remains free to be determined from observations. RCs-fitting leads to $\mu(x) = x/sqrt(1+x^2)$, while the so called interpolation function reads
 \mbox{$\mu(x) = x/(1+x )$.}

In both forms, $a>>a_0$ recovers Newton's second law, while the case $a<<a_0$ yields a force $F$ proportional to the velocity squared:
\begin{equation}
F=m(a/a_0) a=ma^2/a_0=GmM/r^2,
\end{equation}
such that
\begin{equation}
a=\sqrt{GMa_0}/r.
\end{equation}

For a test particle in circular motion around the galactic centre, for small acceleration, far away from the centre,
\begin{equation}
a=v^2/r=\sqrt{GMa_0}/r,
\end{equation}
the rotation velocity reaches a plateau $V_f=\sqrt[4]{GMa_0}$ and one re-obtains the baryonic Tully-Fisher relation
$V_f^4=Ga_0 M_b$.

The success of MOND extends beyond fitting the RCs of spirals, and reproducing the Tully-Fisher relation: it provides explanations for several galactic phenomena, from dwarfs to ellipticals (see~\mbox{\citep[][]{sanders,famaey2012}}),
% 12, 15
to Freeman's law \citep{freeman}, that sets
%16
 the upper limit for spirals surface brightness, to Fish's law \citep{fish}, determining
%(17
 ellipticals characteristic surface brightness, and to the Faber-Jackson relation. {%\bf
 MOND is actually the simplest way to reproduce observed scaling relations, such as the relation between the rotation curve's shape and the baryonic surface density (see Figure 15 of Ref.~\citep[][]{famaey2012}, relevant to the diversity of rotation curve shapes at a given $V_{max}$ scale), with the stellar and dynamical surface densities in the central regions of disk galaxies (recently discussed in \citep[][]{Lelli:2016uea,Milgrom:2016ogb}), or the small scatter of the BTFR discussed in Ref.~\cite{Lelli:2015wst}. These tight relations appear as less natural consequences of baryonic solutions to the CC problem than of MOND, from a formal point of view.}

The RCs of two different dwarf galaxies, UGC11583, and ESO138-G014 are plotted in Figure~\ref{fig:MOND}, together with their MOND fit. Although MOND usually fits galactic RCs well, some cases, such as~ESO138-G014 here, escape its grasp. {%\bf
This should be taken with caution as, according to Ref.~\cite{Famaey:2013yua}, there are misunderstandings on MOND on the problems they quote in their section 3. See also Ref.~\cite{Haghi:2016jdv} for a different point of view.}

\begin{figure}[t]
\centering
%\hspace{-1.5cm}
\resizebox{7cm}{!}{\includegraphics{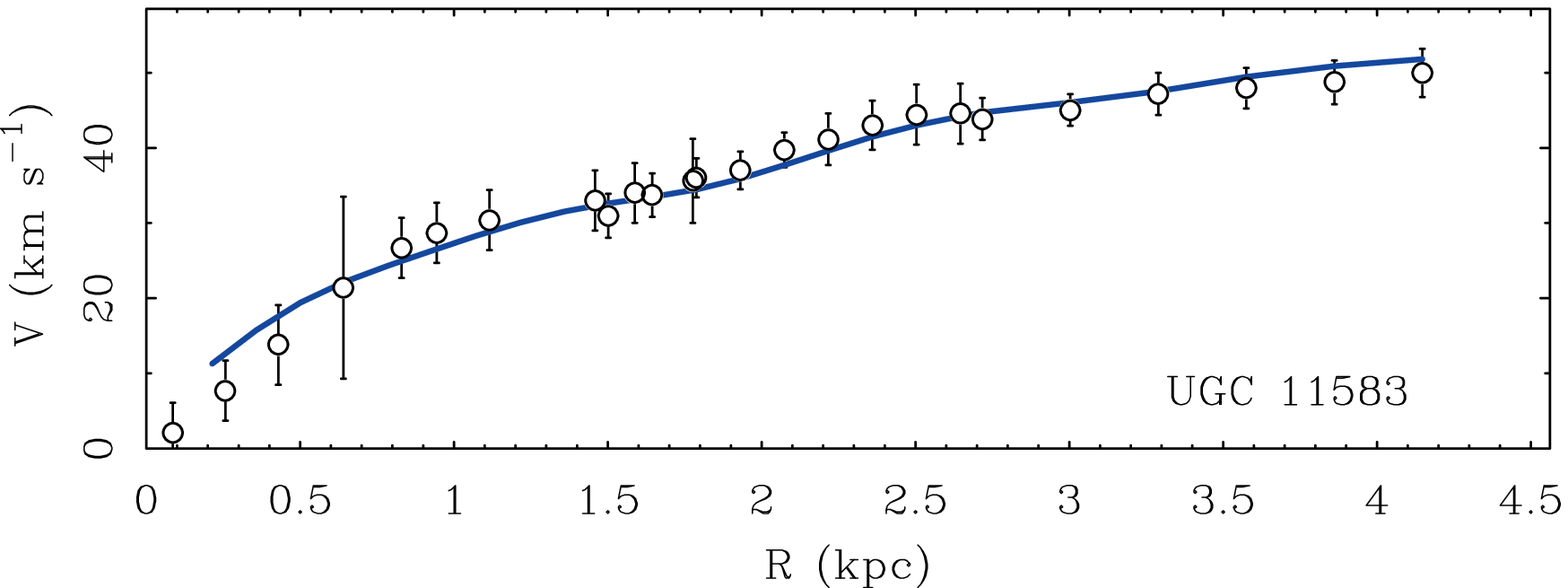}}\hspace{-2.3cm}%\hspace{-.03cm}%
\resizebox{9cm}{!}{\includegraphics{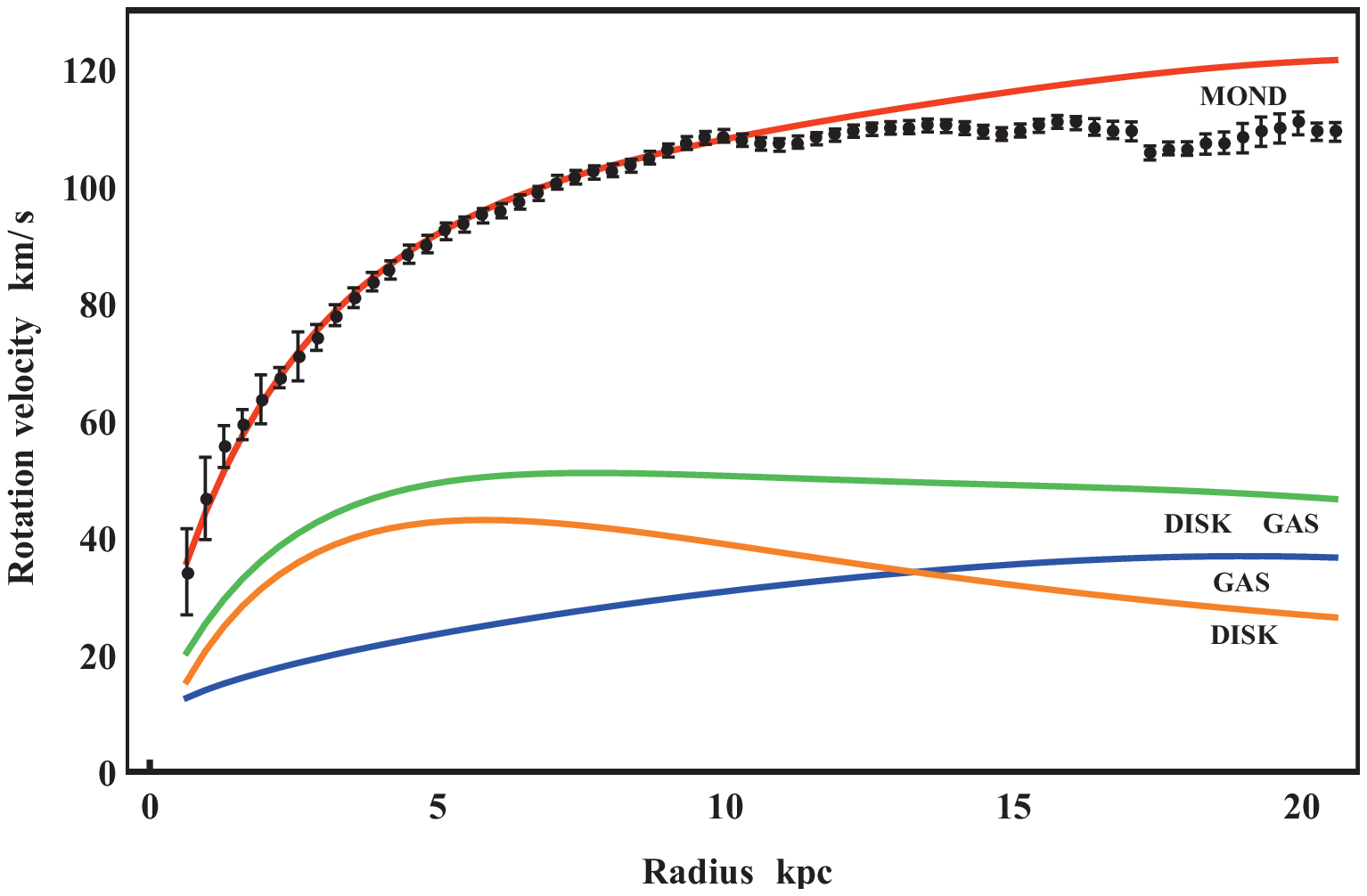}}
\caption{Left panel: Example of MOND fit (blue lines) to a dwarf galaxy  (UGC11583, reproduced from \citep[][]{famaey2012}, Figure 25).
%\\
%
%\textcolor{white}{\bf Figure~12.}
Right panel: RC comparison between observed
%The observed rotation curve of
 ESO138-G014 (black dots with error-bars) and MOND prediction (red line) (reproduced from  \citep[][]{hashim}).
% with its separation of disk and gas contributions (orange line: disk; blue line: gas estimation from the
%code with additional factor).
}\label{fig:MOND}
\end{figure}

% \begin{figure}
% %\begin{center}
% \phantom{e}\hspace{-1.5cm}
% \includegraphics[width=18cm]{power.eps}\hspace{2cm}
% %\end{center}
% \caption[]{
% {\bf
% The class of models reducing to MOdified Newtonian Dynamics (MOND) in the weak field limit does an excellent job fitting the rotation curves of galaxies, predicting the relation between baryonic mass and velocity in gas-dominated galaxies, and explaining the properties of the local group. Several of the initial challenges facing MOND have been overcome, while others remain. Here I point out the most severe challenge facing MOND.
% }\label{fig:power}
% }
% \end{figure}

At cluster scales, nonetheless, MOND proves much less successful (however, see \citep[][]{famaey2012} for a different point of view). In addition, MOND {%\bf
historically} being a mere Newtonian fit, to be considered a full theory requires, also to be applicable on cosmological scales, a relativistic version. One such TeVeS theory was proposed by Sanders and Bekenstein \citep{sanders1,bekenstein,sanders2}.

% \begin{figure}
% %\begin{center}
% \phantom{e}\hspace{-1.5cm}
% \includegraphics[width=18cm]{redshift.eps}\hspace{2cm}
% %\end{center}
% \caption[]{
% {\bf
% Here we report observation of the gravitational redshift of light coming from galaxies in clusters at the 99 per cent confidence level, based upon archival data. The measurement agrees with the predictions of general relativity and its modification created to explain cosmic acceleration without the need for dark energy (f(R) theory), but is inconsistent with alternative models designed to avoid the presence of dark matter.
% }\label{fig:redshift}
% }
% \end{figure}

Since the successes and problems of MOND and the $\Lambda$CDM model appear on complementary scales (galactic vs large scales), Khouri  \citep{khouri} proposed combining both theories, keeping each on their respective successful scales. A wide survey of MOND's successes and problems, and of its attempted relativistic extensions was presented in \citep{famaey2012}.

In summary, there exist several apparently valid proposals solving the CC problem, as well as for the other small scale problems, but the challenge is to single out the correct one.

\section{The Missing Satellite Problem}\label{sec:MSP}

Galactic %and cluster
mass halos were noticed by {%\bf
Klypin et al. and Moore et al. \protect\cite{Klypin:1999uc,moore1}} to present many more subhaloes predicted by N-body simulations than observed satellite galaxies. The scale invariant CDM primordial fluctuations at small scales leads, through collapse, to a large number of subhaloes, hence~creating this MSP.

The MW counts 9 bright dSphs,
Sagittarius, the LMC and the SMC, thus much less than the 500 satellites, obtained in simulations, with larger circular velocities than Draco and Ursa-Minor (i.e.,~bound~masses $>\!10^8 M_{\odot}$ and tidally limited sizes $>$ kpc, see Refs. \citep[][]{boyl,boyl1} and Figure \ref{fig:MSP}, top~left~panel).
%Boylan-Kolchin, Bullock, and Kaplinghat 2012

\begin{figure}[ht]
\centering
\resizebox{5.5cm}{!}{\includegraphics{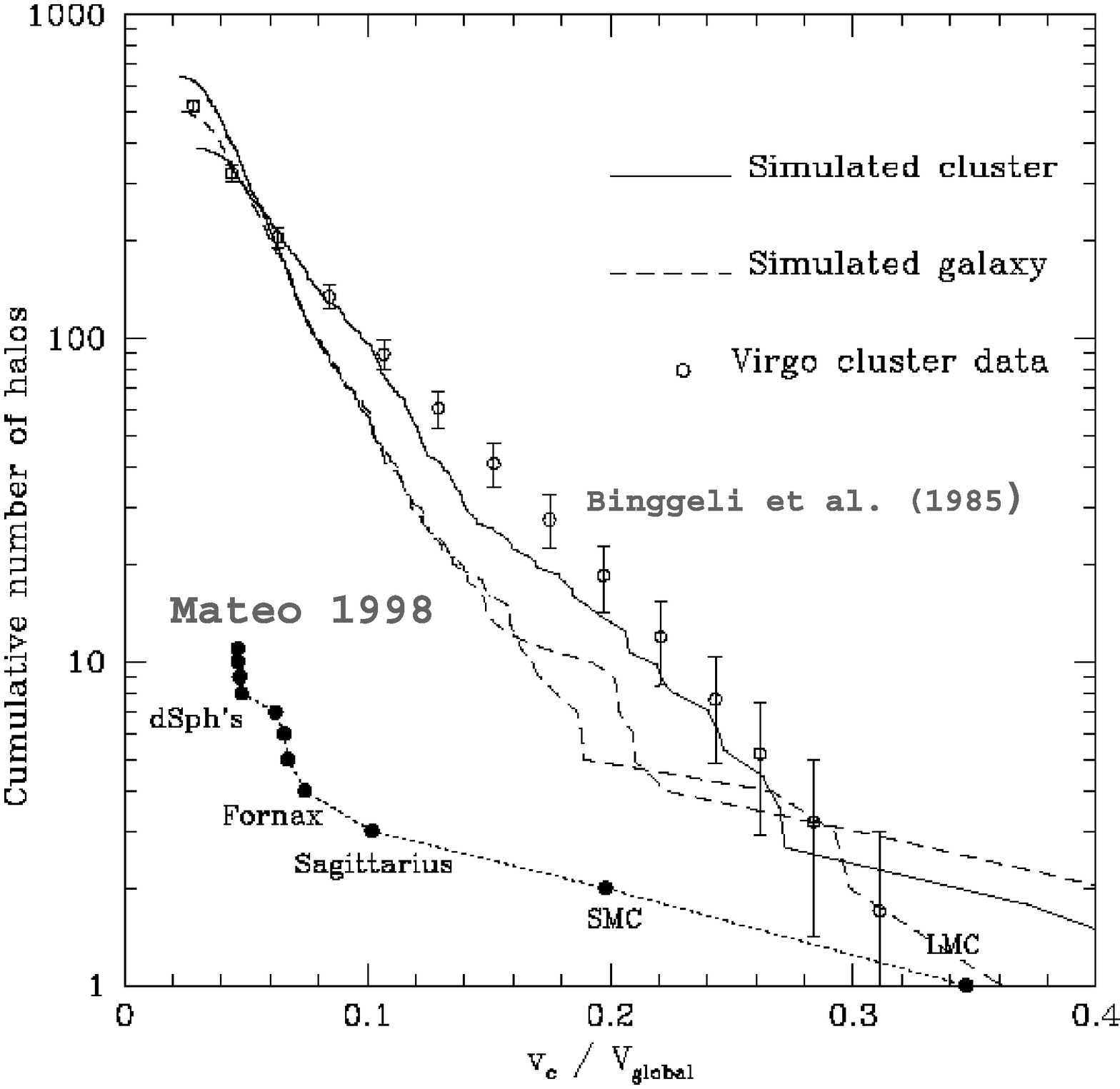}}
\resizebox{5.5cm}{!}{\includegraphics{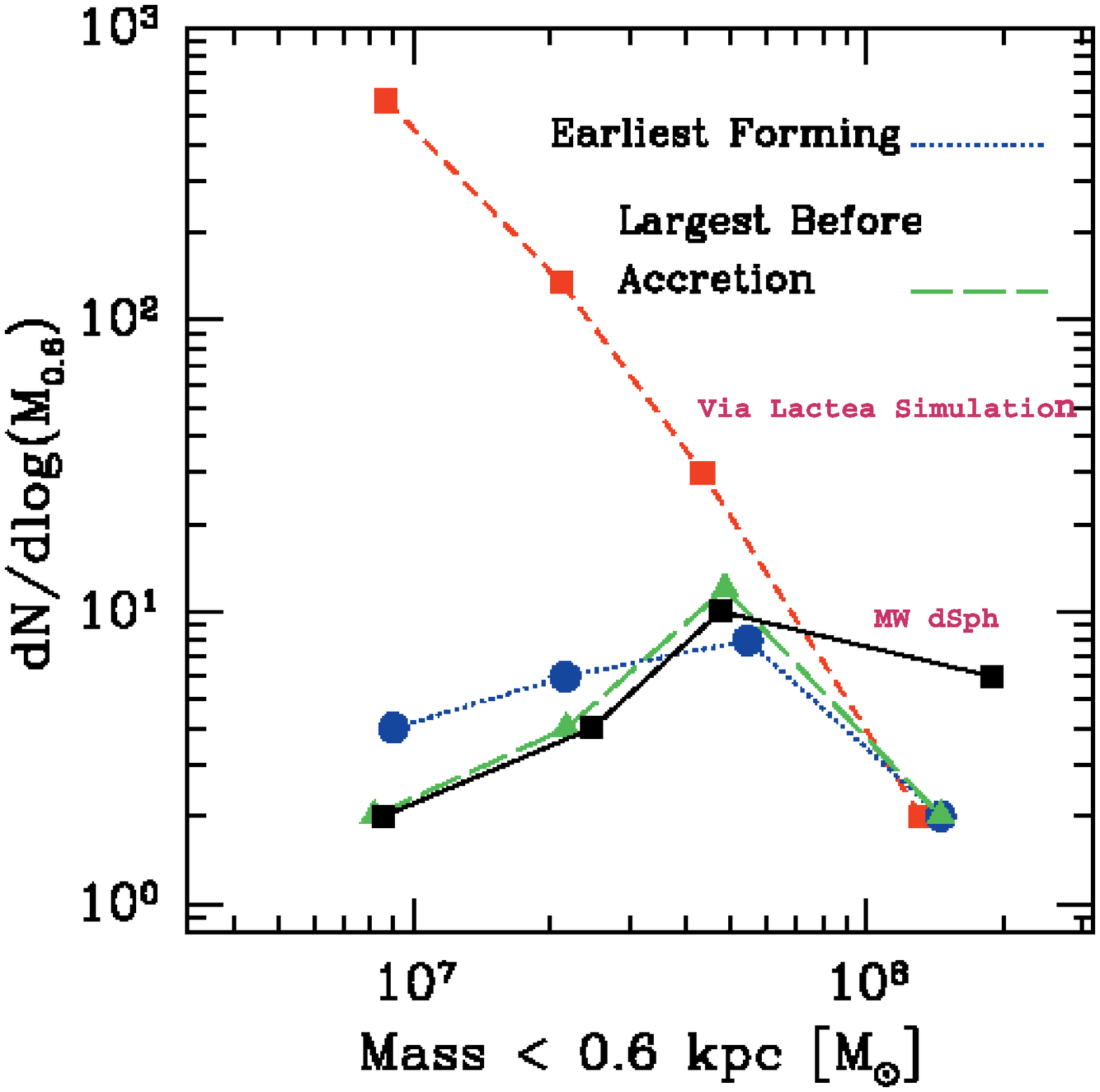}}
\resizebox{6cm}{!}{\includegraphics{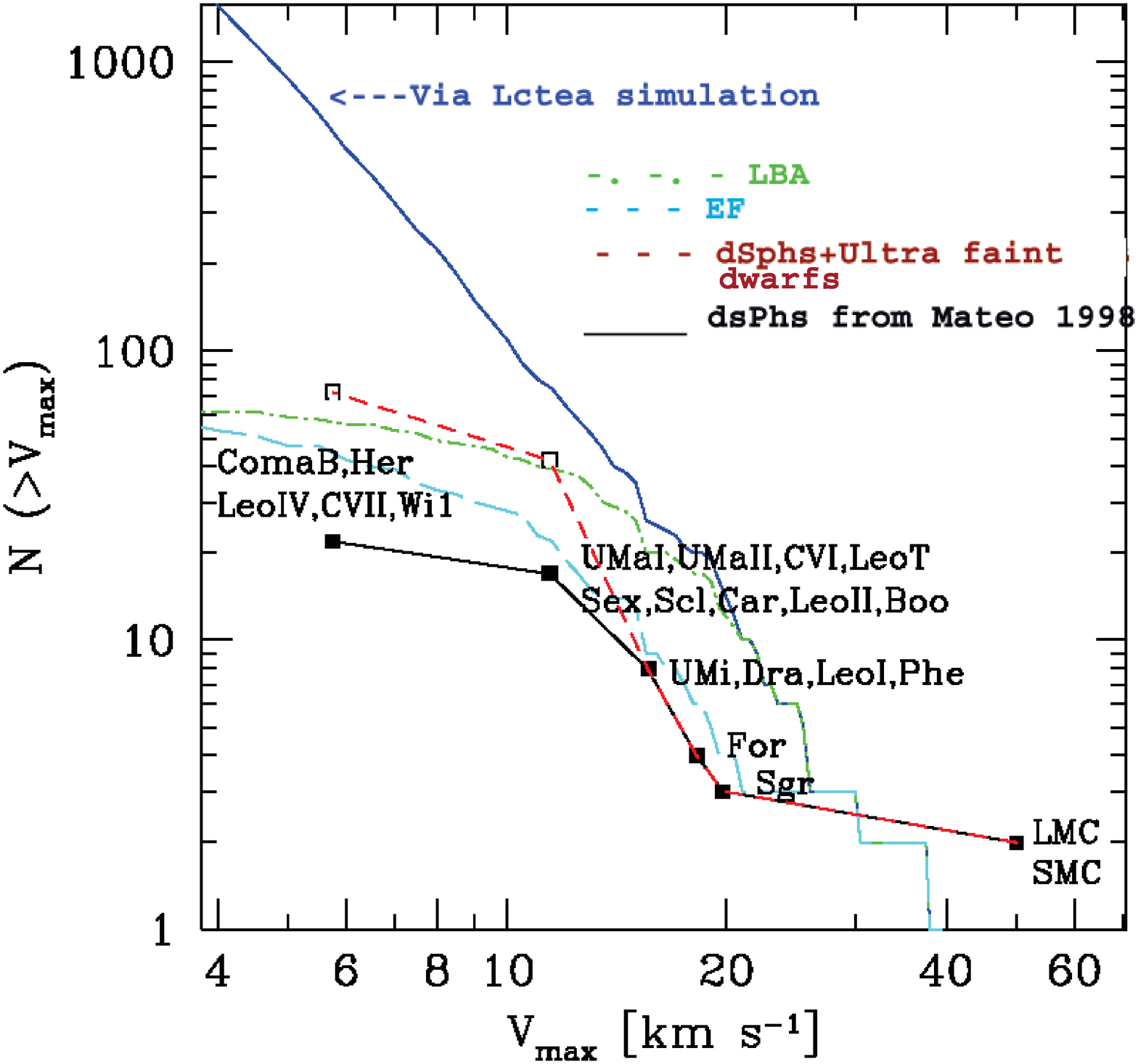}}
\resizebox{5.5cm}{!}{\includegraphics{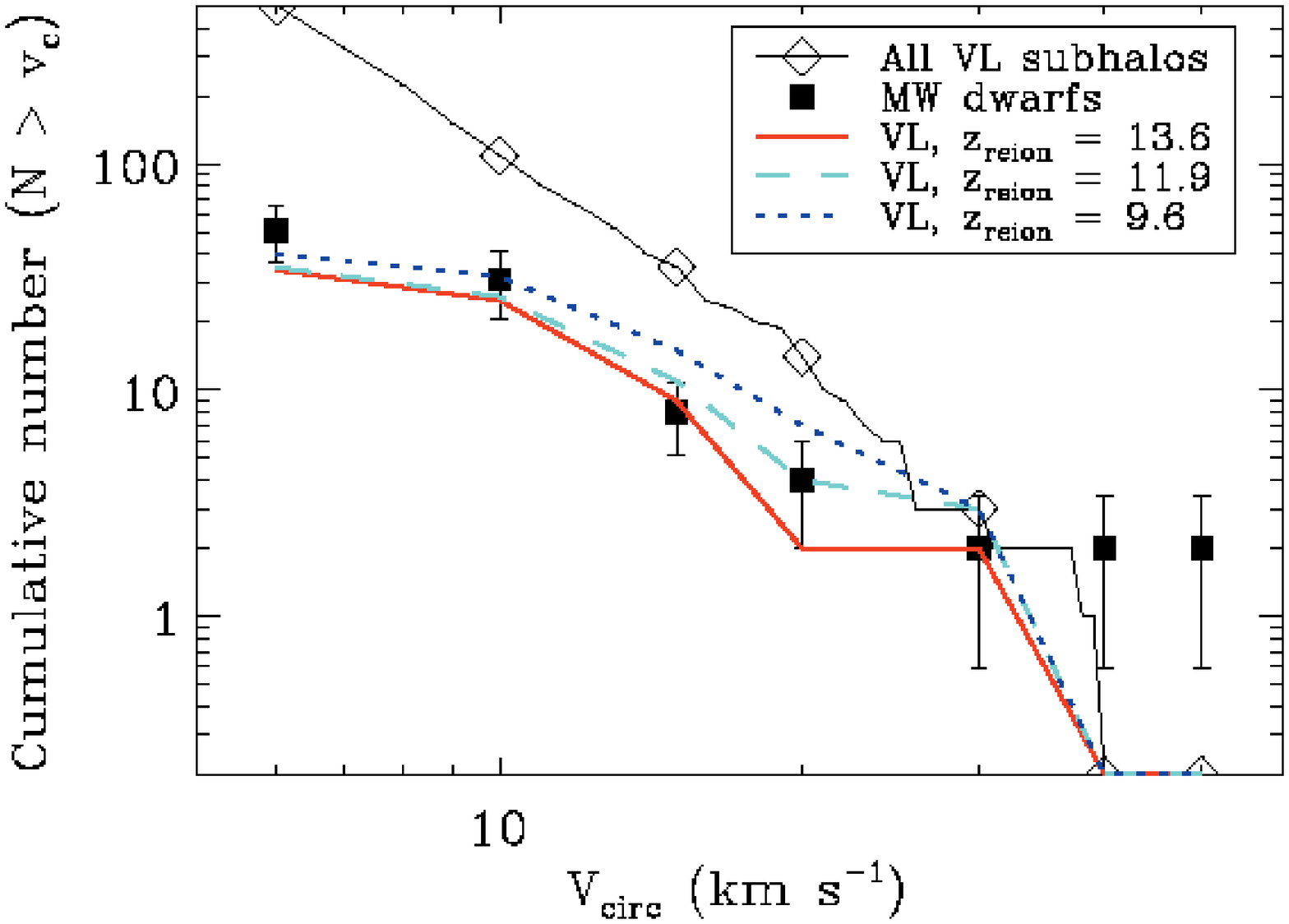}}
\caption{\protect{Top left panel:
%\textcolor{white}{eelanel}
comparison,} (from \protect\citep[][]{moore1}) of the simulated cumulative clusters and dSphs number
with observed clusters \protect\cite{binggeli} and dSphs
% simulations for dSphs and clusters with observations from
\protect\cite{mateo1}.
%\\
%\textcolor{white}{\bf Figure~13.~}
\protect{Top right panel:
%\textcolor{white}{elanel}
comparison} of the abundances for the Via Lactea simulation (red dashed line) with the MW dSphs in the EF (blue dotted line), and LBA (green dashed line) scenarios (from \protect\citep[][]{strigari}).
%\\ \textcolor{white}{\bf Figure~13.~}
Bottom left panel:
%\textcolor{white}{anel}
comparison of the abundances for the Via Lactea simulation (blue solid line) with the MW dSphs (black line) , MW dSphs+Ultra-faint dwarfs (brown dashed line), and the EF (blue dashed line) , and LBA (dot-dashed green line) scenarii (from \protect\citep[][]{madau}).
%\\ \textcolor{white}{\bf Figure~13.~}
Bottom right panel: comparison of the abundances for the Via Lactea Simulation (black solid line) with the MW dSphs+Ultra-faint dwarfs (black dots), and with the Via Lactea subhaloes at redshifts 9.6 (dashed dark blue line), 11.9 (dashed light blue line), and 13.6 (solid red line), respectively (from \protect\citep[][]{simon_geha}).
}\label{fig:MSP}
\end{figure}

The problem was confirmed in subsequent cosmological simulations (Aquarius, Via Lactea, and~GHALO simulations: \citep[][]{springelll,Stadel2009,diemand1}).
%Springel et al. 2008; Stadel et al. 2009; Diemand et al. 2007b)
In the end, every cosmological simulations predicts Milky Way-like galaxies surrounded with at least one order of magnitude more small subhalos (dwarf galaxies) than observed (e.g. Via Lactea simulation \protect\citep[][]{diemand1}).

The ultra faint dwarf satellites discovery (UFDs) \citep[][]{willman,belokurov,zucker,sakamoto,irwin}
%Willman et al. 2005; Belokurov 2006; Zucker 2006; Sakamoto \& Hasegawa 2006; Irwin et al. 2007
 alleviated the problem without solving it: adding them to known MW satellites reduces the discrepancy (see surveys like SkyMapper, DES, PanSTARSS, and LSST \citep[]{tollerud}).

%->234 million particles in (90 Mpc/h)³ multimass simulation, mp=20900 Msun

The key idea of the solution lies in the distinction between visible satellites and the entire population: if only a subset of the population is visible, the observed vs predicted satellites discrepancy can be reduced.
%some of the satellites
Thus, various suppression mechanisms for the visible population have been proposed:
\begin{enumerate}[leftmargin=*,labelsep=5mm]
\item Tidal stripping from  the satellites' parent: presently observed satellites  had the largest masses before accretion (LBA), large enough to retain visible stars, resisting tides when accreted by their parent \protect\cite{diemand1}.
\item Re-ionisation stripping satellites gas, thus star formation, hence suppressing visible satellites formation \protect\cite{bullock,moore2}: presently observed, earliest forming satellites (EF see Refs. \protect\citep[][]{bullock,moore1} and~Figure \ref{fig:MSP}, top right, and bottom left) are visible because they acquired gas, and thus form stars, before re-ionisation.\\
Ref. \protect\cite{simon_geha} compared the cumulative number of dSphs and UFDs  (with $M/L \simeq 1000$, from SDSS data) with the cumulative number of satellites obtained in the Via Lactea simulation, assuming a $z= 9-14$ reionisation epoch,
% protostructures could not attract enough baryonic matter to create a visible dSPh.
almost solving the problem (see {Figure \ref{fig:MSP}}, bottom right panel).
\item Photo-ionisation from stellar and supernova feedback  (e.g., Refs. \protect\citep[][]{dekel1,mori,koposov}), and generally stripping gas by ram pressure  (e.g., Ref. \protect\citep[][]{mayer}). Because of the photo-ionisation threshold, UFDs'
\noindent baryons to stars conversion efficiency lies in the range 0.1\%--1\%, thus making it is not clear whether they are "fossils" from reionisation epoch \cite{bovill}.
\item Transfer of angular momentum from baryons to DM through dynamical friction \cite{DelPopolo2009,DPK2009}, which~turns also cuspy profiles to cores. The number of visible satellites decreases since tidal stripping acts more on cored density profiles.
\end{enumerate}

In conclusion, adding baryon physics to the usual dissipationless DM model solves the MSP, {%\bf
for the majority of the Cosmology community,} as well as the other SSP$\Lambda$CDM, as shown by recent hydrodynamic simulations (e.g., \citep[][]{zolo,Sawala:2015cdf,zhu})
%Zolotov et al. 2012; Sawala et al. 2015 (1511.01098); Zhu et al. 2015 (1506. 05537)
 or semi-analytic models \cite{dpet14}.
%Del Popolo et al. 2014

\section{The Too Big to Fail Problem} \label{sec:TBTF}

%\begin{figure}
%\resizebox{14.65cm}{!}{\includegraphics{tbtf.eps}}
%\caption{The Too Big To Fail problem (from Boylan-Kolchin et al. (2012)\protect\cite{boylan2011}).}
%\label{spectra}
%\end{figure}

%\begin{figure}
%\resizebox{14.65cm}{!}{\includegraphics{brooks.eps}}
%\caption{A solution to the Too Big To fail problem through baryon physics (from Brooks et al.\protect\cite{brooks}).}
%\label{spectra}
%\end{figure}

On closer inspection, eliminating visible satellites from the faint end of the distribution does not exhaust the model vs observation discrepancies in satellites (MSP): the most massive (luminous) satellites also pose problem.
%The MSP recently has shown another feature,
Boylan-Kolchin \cite{boyl,boyl1}
%Boylan-Kolchin, Bullock, and Kaplinghat 2011, 2012
discovered, in the Aquarius and the Via Lactea simulations, a population of $\simeq$10 subhaloes that were too massive and dense, by a factor $\simeq$5, to host even  the brightest satellites of the MW, and dubbed it the ``Too Big to Fail'' (TBTF) problem\footnote{``Too Big to Fail'' refers to simulation satellites being too big for MW satellites, while no mechanism would lead them to fail being visible.}. In general, $\Lambda$CDM simulations of the MW predict at least 10 subhaloes with $V_{\rm max} >25$ km/s, while $12 < V_{\rm max} < 25$ km/s for all the dSphs.

This is shown in Figure \ref{fig:TBTF}. Both the left, and right panels display RCs with $V_{\rm max}<24$ km/s while simulations have much larger values.
Rotation curves obtained from NFW-shaped subhaloes with $V_{\rm max}=(12, 18, 24, 40)$ km/s, with a 1 $\sigma$ scatter, taken from Aquarius simulations, present the simulation side of the TBTF problem in the left panel. They are confronted, for this purpose, with the bright dSphs, the dots with error-bars.
The results show that $V_{\rm max}< 18~km/s$ for most of the dSphs, all~have $V_{\rm max}$ under 24 km/s and only Draco is consistent with a
 subhalo modelled with \mbox{$V_{\rm max}\simeq 40$ km/s}. The right panel plots  $V_{\rm max}$ vs the visual magnitude, $M_V$, and confirms that the simulations produce much more massive subhaloes than the observed dSphs.

%
%The second problem, named the Too Big To Fail problem, is connected to the previous one. The problem arose from analyses of the Aquarius and Via Lactea simulations.
%Each simulated halo had  $\simeq 10$ sub-halos that were too massive and
%dense with respect to MW dSphs that they would appear to be too big to fail to form lots of stars. The TBTF problem is
%that none of the observed satellites of the Milky Way or Andromeda have stars moving as fast
%as would be expected in these densest sub-halos\protect\cite{boylan,boylan1} (see fig. 50).
%

%\begin{figure}
%\resizebox{8.6cm}{!}{\includegraphics{madau.eps}}
%\resizebox{8.6cm}{!}{\includegraphics{simon.eps}}
%\caption{Left panel: Comparison of Via Lactea Simulation with the MW dSphs, MW dSphs+ Ultra-faint dwarfs, the EF , and LBA scenario (from Madau et al. %2008\protect\cite{madau}). Right panel: comparison of Via Lactea Simulation with the MW dSphs+ Ultra-faint dwarfs, and Via Lactea subhaloes at redshifts 9.6, %11.9, 13.6 (from Simon \& Geha 2007\protect\cite{simon_geha}).}
%\end{figure}

\begin{figure}[t]
\centering
\resizebox{6.75cm}{!}{\includegraphics{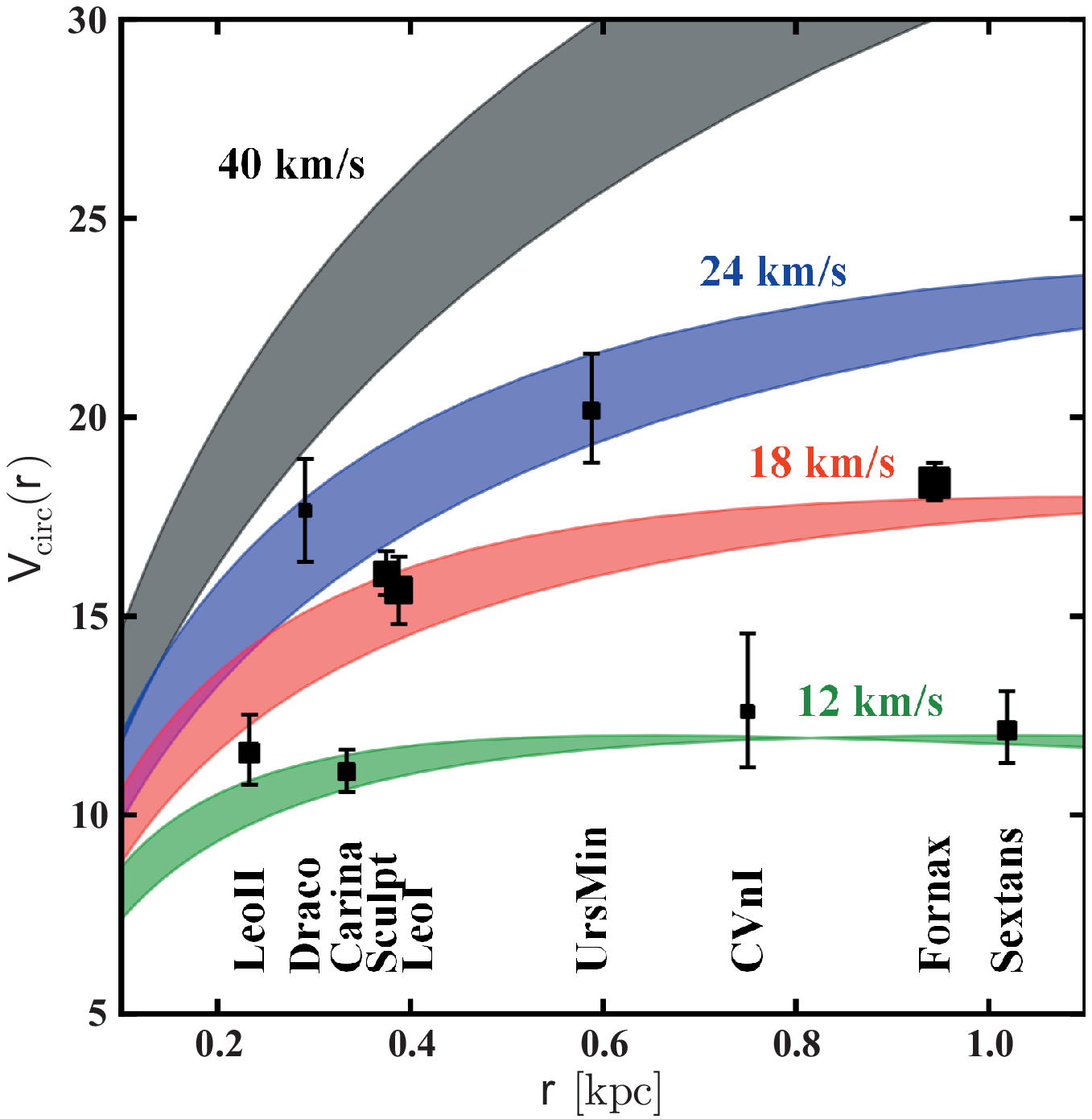}}
\resizebox{7cm}{!}{\includegraphics{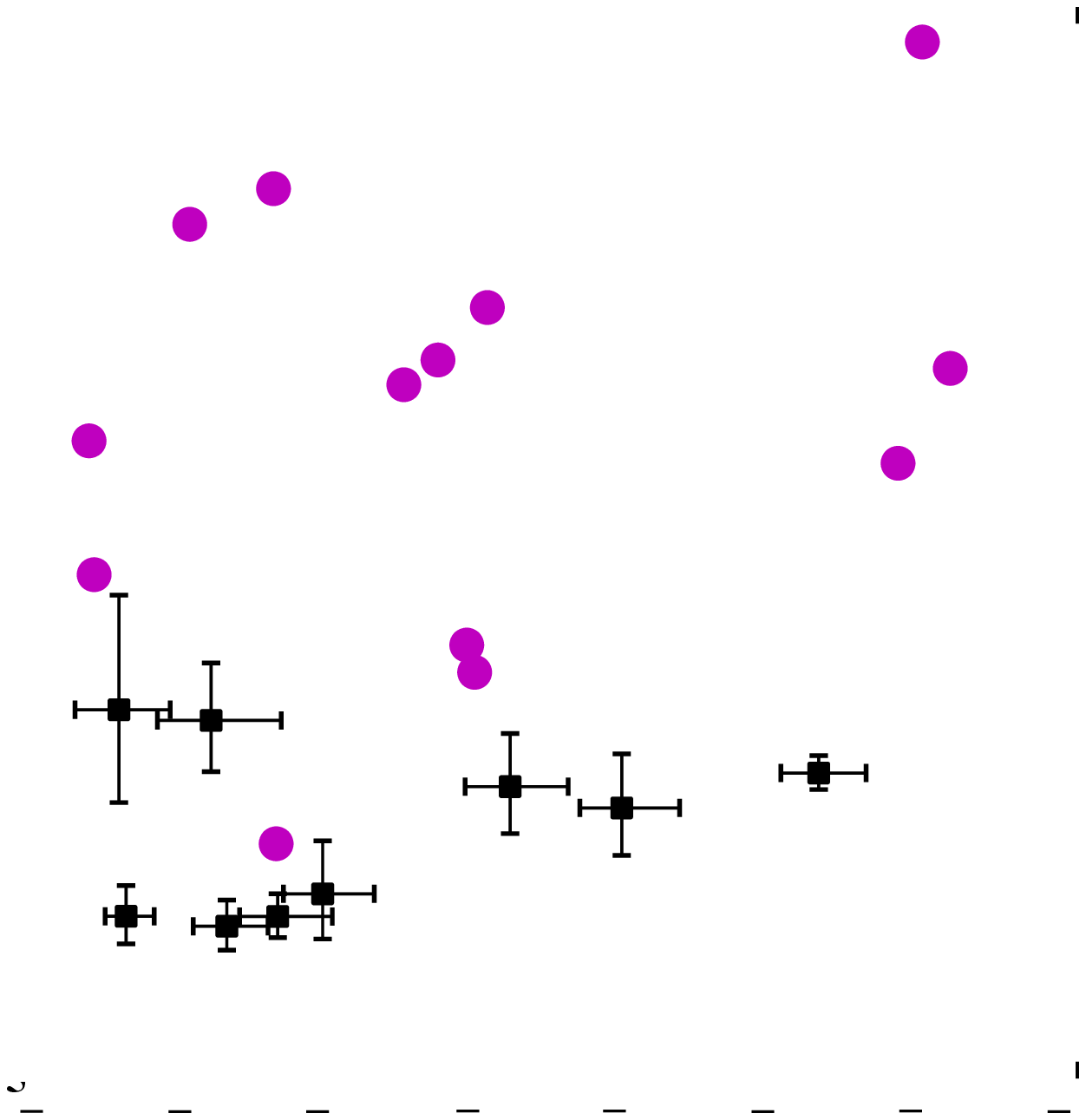}}
\caption{Left panel:
rotation curves for NFW subhaloes with $V_{\rm max}=(12, 18, 24, 40)$ km/s $\pm 1 \sigma$ from Aquarius simulations, confronted with the bright dSphs, represented by the dots with error-bars.
%\textcolor{white}{\bf Figure~14.~}
Right~panel:  $V_{\rm max}$ vs. $M_V$, the visual magnitude, for the same bright dSphs (black dots with error-bars) and the most massive subhaloes of the Aquarius simulations (magenta dots). The observed dSphs are much less massive than the simulations (panels reproduced from \citep[][]{boyl1}).
%
%Observed Vcirc values of the nine bright dSphs (symbols, with sizes proportional to log LV ), along with rotation
%curves corresponding to NFW subhalos with $Vmax =(12, 18, 24, 40) km/s$. The shading indicates the 1 $\sigma$ scatter in
%rmax at fixed Vmax taken from the Aquarius simulations. All of the bright dSphs are consistent with subhalos having $Vmax <
%24 km/s$, and most require $Vmax < 18 km/s$. Only Draco, the
%least luminous dSph in our sample, is consistent (within 2 $\sigma$) with
%a massive CDM subhalo of $\simeq 40 km/ s$ at $z = 0$.
%
%Right: Values of Vmax computed in Section 4.1 for the nine luminous Milky Way dwarf spheroidals (square symbols with errors), along with $Vmax$ $(z = 0)$ values of the
%subhalos with $MV < −8$ (magnitudes are assigned by abundance matching) from the halo that best reproduces the luminosity function
%n the left panel (Aq - E). While numerical simulations combined with abundance matching reproduces the luminosity function of MW
%satellites, the structure of the dwarf spheroidals hosts’ in this model does not match observations: the simulated subhalos are much more
%massive (have larger values of Vmax) than the dSphs.
}\label{fig:TBTF}
\end{figure}

This issue can be compared with the CC problem for haloes: the $\Lambda$CDM seems to produce too much mass in haloes and subhaloes.
Thus, similarly to the CC problem, two main classes of solutions have been proposed: cosmological and astrophysical. As for the CC problem, cosmological solutions modify the perturbation spectrum or the nature of dark matter particles. Astrophysical solutions are driven by baryon physics, and consider, in analogy with solutions to the CC problem:
\begin{enumerate}[leftmargin=*,labelsep=5mm]
\item The shape of satellites inner densities, shifting from cuspy to cored \citep[][]{zolo,brooks} (hereafter Z12 and B13 respectively)
%Zolotov et al. 2012 (Z12); Brooks et al. 2013 (B13)
, thus making them more susceptible to tidal stripping and even to tidal
destruction~\cite{strigari,pen10}. This picture would see the present-day dwarf galaxies, more massive in the past, transformed and reduced
 strong tidal stripping  (e.g., \citep[][]{Kravtsov2004});
%Kravtsov, Gnedin \& Klypin 2004).

\item The suppression of star formation from
\begin{enumerate}[leftmargin=20pt,labelsep=7pt]
\item Supernova feedback (SF)
\item Photo-ionisation \cite{okamoto,brooks}, and
\item Reionisation. This can prevent small mass DM haloes gas accretion, ``quenching'' star formation after $z \simeq 10$ \cite{bullock,ricgne,moore2}.
%Bullock, Kravtsov, \& Weinberg 2000; Ricotti \& Gnedin 2005; Moore et al. 2006).
\end{enumerate}
This would suppress dwarfs formation or could make them invisible;

\item The dynamical effects of a baryonic disc \cite{zolo,brooks}. Satellites crossing such disc experience disk shocking, strong tidal effects, even more so for cored inner profiles.
\end{enumerate}

Alternative solutions to the TBTF problem have been proposed: the TBTF excess of massive subhalos in simulations on the MW could vanish if Einasto's are the correct satellite density profiles, or if the correct measurement of MW's virial mass reduces from  $\simeq\!10^{12} M_{\odot}$ to $\simeq\!8 \times 10^{11} M_{\odot}$ \protect\cite{vera_cintio,vera_cintio1}.

The above solutions to the TBTF problem are focused on the MW satellites.
%As reported, one of the solutions to the TBTF problem is connected to baryon physics, as shown by
However, the TBTF problem also concerns the Local Group, and Local Volume galaxies \cite{GarrisonKimmel2014,Ferrero2012,Papastergis2015}, galaxies that are too massive to be modified by (reionisation) feedback \cite{boyl1,Penarrubia2012,GarrisonKimmel2013,BrookCintio1,BrookCintio2,Pawlowski2015,PapastergisShankar}, contrary to the baryonic solution of Refs.~\cite{zolo,brooks,dpet14,sawafre}.
%%Zolotov et al. (2012), Brooks et al. (2013), Del Popolo et al. (2014), Sawala et al. (2015)(Sawala, T., Frenk, C. S., Fattahi, A., et al. 2015, MNRAS, 448, 2941)
Nevertheless, such limitations of feedback were challenged for a few  galaxies by \cite{Madau2014,maxwell,chan},
%Madau et al. 2014; Maxwell et al. 2015; Chan et al. 2015 (ref in Nihao V)),
and for an entire galaxy population by \cite{Dutton2015}.

%
%Inclusion of baryonic physics can create shallower slopes of the dark matter densities in the centers of low-mass galaxies reducing or solving the discrepancy between cuspy profile predicted in N-body %simulations and flat ones seen in observation. One possible solution is that showed by Brooks et al. (2012)\protect\cite{brooks} using a suggestion of Zolotov et al. (2012)\protect\cite{zolotov}. The last %author proposed a correction to the velocity in 1 kpc
%\begin{equation}
%\Delta (v_{\rm 1kpc})= 0.2 v_{\rm infall}-0.26 {\rm km/s}; \hspace{0.5cm} 20 {\rm km/s}<v_{infall}< 50 {\rm km/s}
%\end{equation}
%that must be applied to the central parts of low-mass galaxies, in order to  take into account tidal stripping enhancement due to baryons, and the effect of supernova feedback flattening the cusp in the haloes %(see fig. 51).
%
%Similar result is obtained in the dark baryons-DM interaction through dynamical friction (see the section ``A Unified baryonic solution to the $\Lambda$CDM small scale problems").
%

\section{A Unified Baryonic Solution to the $\Lambda$CDM Small Scale Problems}\label{sec:UsolSSP}

So far, each SSP$\Lambda$CDM was solved separately through different recipes. Unified solutions were however proposed, after understanding their inter-relations.
%While it is not complicated to separately solve the MSP problem, and the TBTF problem with the recipes discussed above, a simultaneous solution of both problems in models of galaxy
%formation based on DM-only simulations of the CDM model (Boylan-Kolchin, Bullock \&
%Kaplinghat 2012),5 is much more complicated.
%The quoted problems can be solved once we take into account the baryons physics. The quoted problems are connected.
For example, transforming cuspy density profiles into cored distributions, through the SNFF or the DFBC scenario,
%SF feedback , or dynamical friction,
also affects the parent halo distribution and number of substructure/satellites as
tidal effects of a parent halo on a satellite depend fundamentally on
its shape (e.g., \citep[][]{Mashchenko2006,Mashchenko2008,pen10}):
%Mashchenko et al. 2006, 2008; Pe~narrubia et al. 2010)
cuspy satellite structure will not suffer big changes when entering the main halo, while the parent's tidal field can easily strip a cored profile from its gas, in~some cases down to its destruction \cite{pen10}.
%Penarrubia et al. 2010).

%In fact, cored profiles are more subject to be destroyed by tidal stripping. Ejection of low angular momentum gas by (e.g.) SF feedback can solve the angular momentum problem, as previously discussed.

A recent sketch of a baryonic solution to the SSP$\Lambda$CDM was proposed \protect\cite{zolo,brooks}, based on SF explosions \protect\cite{Navarro1996a,Gelato1999,Read2005,Mashchenko2006,Mashchenko2008, Governato2010} removing angular momentum and gas from the proto-galaxy,
%a)
thus~flattening the density profile.
%; b) giving rise to the correct angular momentum distribution in discs.
Indeed, the discrepancy between observed and predicted number, and density, of satellites (see \protect\citep[][]{brooks}) can significantly reduce, by applying a correction to satellites circular velocity at 1 kpc, $v_{1kpc}$ \protect\cite{zolo}, to large N-body simulations results (e.g., Via Lactea II, VL2 hereafter).

A similar result is obtained in the DFBC \protect\cite{ElZant2001,ElZant2004,DelPopolo2009}.
%, and the angular momentum discrepancy is not present (Del Popolo (2009)\protect\cite{delpopolo2}).
Indeed, the Del Popolo model \protect\cite{DelPopolo2009} provided a similar correction to $v_{1kpc}$ than Zolotov et al. \protect\cite{zolo}, which, applied to the Via Lactea satellites, as in Brooks et al. \protect\cite{brooks}, simultaneously solves the MSP and the TBTF problems \protect\cite{dpet14,dpet14a}.
The~model of Ref.~\cite{dpet14a} is in fact more complete: it couples satellites interaction with the halo to dynamical friction energy and angular momentum exchange from baryons clumps to DM, which flattens the density profiles. The effects of tidal stripping and heating on
the satellites were modelled with a combination of the procedures from Refs.~\cite{TB01,pen10}.
%Taylor \& Babul (2001) (TB01) with that from P10.

%
%Inclusion of baryonic physics can create shallower slopes of the dark matter densities in the centers of low-mass galaxies reducing or solving the discrepancy between cuspy profile predicted in N-body %simulations and flat ones seen in observation. One possible solution is that showed by Brooks et al. (2012)\protect\cite{brooks} using a suggestion of Zolotov et al. (2012)\protect\cite{zolotov}. The last %author proposed a correction to the velocity in 1 kpc
%\begin{equation}
%\Delta (v_{\rm 1kpc})= 0.2 v_{\rm infall}-0.26 {\rm km/s}; \hspace{0.5cm} 20 {\rm km/s}<v_{infall}< 50 {\rm km/s}
%\end{equation}
%that must be applied to the central parts of low-mass galaxies, in order to  take into account tidal stripping enhancement due to baryons, and the effect of supernova feedback flattening the cusp in the haloes %(see fig. 51).
%
%Similar result is obtained in the dark baryons-DM interaction through dynamical friction (see the section ``A Unified baryonic solution to the $\Lambda$CDM small scale problems").
%

In summary, the method proceeded in two main phases. The first phase calculated the satellite density profile flattening from baryonic physics (in particular, the subhaloes' central mass lowering), considering it isolated, without interactions with the host halo, in the same fashion as in Ref.~\cite{DelPopolo2009} (see also \citep[][]{HioDP06,HioDP13}, to have a semi-analytical description of halos growth).
%Hiotelis \& Del Popolo 2006, 2013,

The flattening was translated into the difference in circular velocity, at 1 kpc, between the DM-only (hereafter DMO) satellites equivalent to those considered and containing also baryons (hereafter DMB
satellites),  $\Delta v_{{\rm c,1kpc}}=v_{{\rm c,DMO}}-v_{{\rm c,DMB}}$.

Figure \ref{fig:DV-V_DMO-DMB} displays the resulting $\Delta v_{{\rm c,1kpc}}$ at $z=0$ and the fitted dashed line through the output points of the model,
given by
\begin{eqnarray}
\Delta v_{\rm c,1kpc} & = & {\rm 0.3v_{infall}-0.3km/s}%\nonumber \\
 \quad \quad \quad \quad% &  &
 {\rm 10\,km/s<v_{infall}<50\,km/s}\label{eq:mia}
\end{eqnarray}

\begin{figure}[t]
\centering
%\hspace{-1.0cm}
%\begin{center}
\includegraphics[width=11cm]{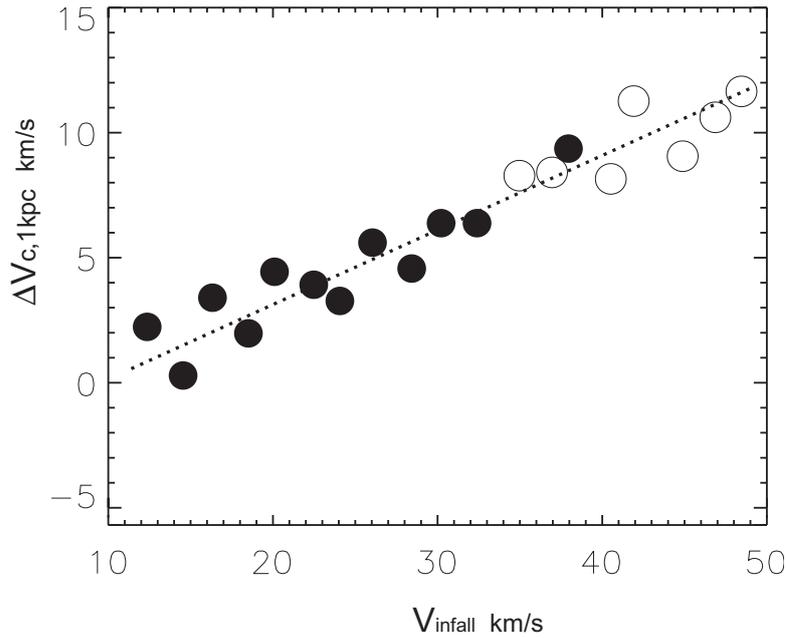}
%\end{center}
%\psfig{file=DV4.eps,width=13.0cm}
%\psfig{file=DV3.eps,width=10.0cm}
\vspace{-36pt}
\caption[]{ Flattening of satellites from inner Baryonic tidal stripping and heating, expressed in terms of difference in central $v_{\rm c}$ at 1 kpc, and $z=0$,
between equivalent DMO, and DMB
satellites as a function of equivalent DMO satellites maximum velocity $v_{\rm max}$ at infall. The model produces, for satellites with $M_b/M_{500}<0.01$, filled circles, while for $M_b/M_{500}>0.01$, the open circles and results in a linear fit (dashed line) (figure reproduced from \citep[][]{dpet14a}).}\label{fig:DV-V_DMO-DMB}
\end{figure}

%In this paper we deliberately chose not to take account of SF and to concentrate on the model of baryonic clumps exchanging energy and angular momentum with DM through DF, since it has clearly been %shown (Del Popolo, 2014c, figure 4) that in the mass (circular velocity) range of the dwarfs studied in the present paper, the former is less efficient in transforming cusps into cores
%than the latter. Del Popolo & Hiotelis (2014) compared also the result of the current model, adding SF, to the SPH simulations of Inoue & Saitoh (2012): the full model
%agrees with the results of Inoue & Saitoh (2012). There, the addition of SF does not alter the outcome significantly. This phase is originally described in DP09.

The second phase no longer considered the satellite as isolated, and subjected it to the host halo's tidal field and accretion.

The merger history and interacting satellites' growth was followed, and the substructure evolution was tracked in Ref.~\cite{dpet14a}'s
%Del Popolo et al. (2014)
model, taking into account the mass loss induced from tidal stripping,
tidal~heating, as well as the disc's stripping enhancement, caused by the host halo.

The destruction rates by tidal stripping induced a second correction to satellites outputted from N-body simulations, requiring the link between satellite mass loss, or remaining, and velocity change (e.g., $V_ {\rm max}$) during infall in the parent. Such link was shown in Figure \ref{fig:V-MwB}: DMB satellites were shown to lose more mass than DMOs because
\begin{enumerate}[leftmargin=*,labelsep=5mm]
\item DMB satellites contain gas, DMOs do not;
\item DMB satellites profiles are flatter than DMO's, thus tidal stripping affects them more (e.g., \citep[][]{pen10}). Similar trend affect baryon-richer DMB (filled circles) compared to baryon-poorer DMB (open circles).
\end{enumerate}
The maximum tidal loss for DMB satellites corresponds to those in the host galaxy disc's vicinity.

\begin{figure}[t]
\centering
%\begin{center}
\includegraphics[width=12cm]{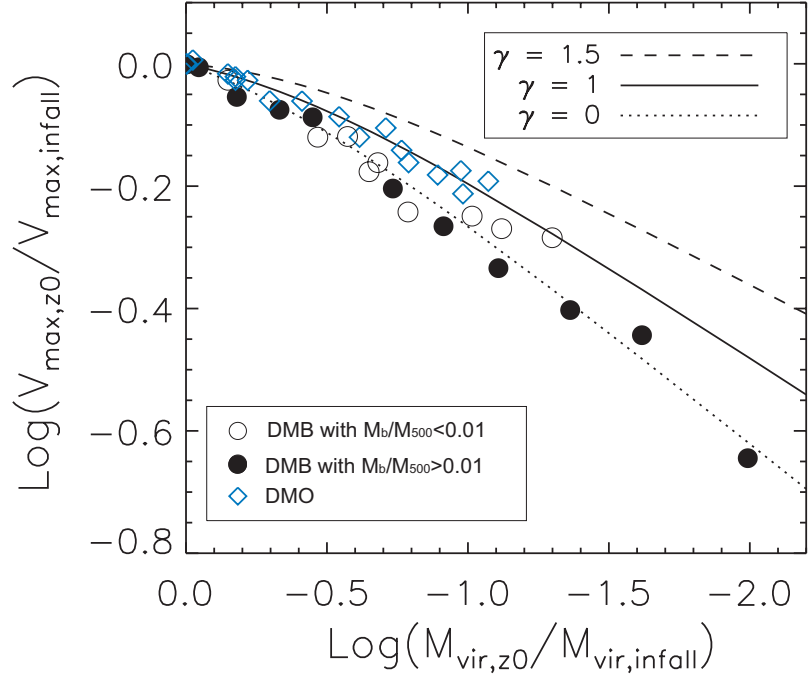}
%\end{center}
\caption
{Reduction of the 1 kpc circular velocity after infall, from $z_{\rm infall}$ to
$z=0$, in terms of corresponding mass reduction. DMB satellites with baryonic fraction $M_b/M_{500}>0.01$ are shown as filled
circles,
 while $M_b/M_{500}<0.01$ yield open circles and DMO satellites are marked as open diamonds. Equation (8) of \cite{pen10} fit for slope
$\gamma=1.5, 1, 0$ are displayed as dashed,
solid, and dotted lines, respectively (figure reproduced from \citep[][]{dpet14a}).
}\label{fig:V-MwB}
\end{figure}

The analytic fits from Equation (8) of \cite{pen10}, linking the $v_{{\rm max}}$ change to tidal stripping mass loss,
\begin{equation}
\frac{v_{{\rm max}}(z=0)}{v_{{\rm infall}}}=\frac{2^{\zeta}x^{\eta}}{(1+x)^{\zeta}},
\label{eq:pe}
\end{equation}
 where $x\equiv mass(z=0)/mass(z=infall)$, was also shown in Figure \ref{fig:V-MwB}, for the exponent values $\zeta=0.40$ and $\eta=0.24$, corresponding to central density profile logarithmic slopes $\gamma=1.5$, represented by a dashed line, $\zeta=0.40$ and $\eta=0.30$, corresponding to $\gamma=1$, with a solid line, and $\zeta=0.40$ and $\eta=0.37$, yielding $\gamma=0$, with a dotted line, respectively.

Then following \cite{brooks}, a destruction criterion (in terms of mass loss) was fixed, determining tidally disrupted satellites from Equation (\ref{eq:pe}): mass losses
\begin{itemize}[leftmargin=*,labelsep=5mm]
\item above 97\% mass (x = 0.03), or
\item above 90\% mass, with $v_{\rm infall} > 30$ km/s and pericentric distance $<$20 kpc,
\end{itemize}
for a given satellite, was set as destroyed.
The photo-heating induced star formation suppression, from the \cite{okamoto} results, produced a third correction. The last step assigned surviving satellites luminosity. Stellar masses of satellites were allocated relating their infall velocity $v_{\rm infall}$ to the stellar mass $M_*$, as~plotted in Figure \ref{fig:Vmax-Mstar}. The fit to the $v_{\rm infall}$-$M_\ast$ data yielded
the relation\footnote{Note that scatter in the
$v_{\rm infall}-M_\ast$ relation proceeds from a halo mass reduction caused by tidal stripping and heating from $z_{\rm infall}$ to $z=0$. }
\begin{equation}
\frac{M_{\ast}}{M_{\odot}}=0.1 (\frac{v_{\rm infall}}{\rm kms^{-1}})^{5.5}.\label{eq:MstarVinfall}
\end{equation}

Finally, the V-band magnitude, $M_{\rm V}$ was related to $M_\ast$ by applying \cite{brooks}'s relation from~\cite{zolo}'s~simulations,
\begin{equation}
\log_{10}(\frac{M_{\ast}}{M_{\odot}})=2.37-0.38 M_{\rm V}.
\end{equation}

\begin{figure}
\centering
%\hspace{-0.5cm}
%\begin{center}
\includegraphics[width=12cm]{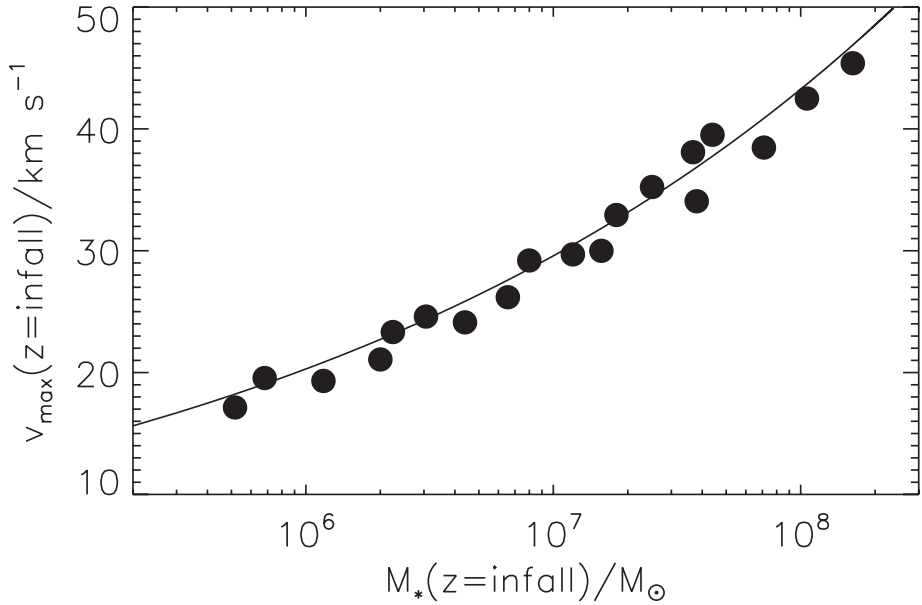}
%\end{center}
%\psfig{file=v_m1.eps,width=13.0cm}
%\psfig{file=DMO_DMB.eps,width=10.0cm}
\caption[]{DMO subhaloes maximum velocity $v_{\rm max}$ as function of their stellar
mass, $M_{\ast}$, at infall time. The stellar mass change with $v_{\rm infall}$ is synthesised by the data fit from Equation~\eqref{eq:MstarVinfall}, shown as a solid line (figure reproduced from \citep[][]{dpet14a}).}\label{fig:Vmax-Mstar}
%}
\end{figure}

Figure \ref{fig:V-M_V} showed the result of the all corrections, with the raw results from the VL2 simulations at $z=0$ displayed on the top panel, to compare with the same satellites corrected with heating, destruction, and velocity corrections, as discussed above, presented on the bottom panel. There,
\begin{itemize}[leftmargin=*,labelsep=5mm]
\item Red filled symbols mark ``observable'' objects produced by the VL2 simulation,
\item Filled black circles mark much fainter satellites than ``observable'', stripped of their stars \linebreak (see~\citep[][]{pen10,brooks}), their mass loss $\ge$90\% still does not grant them destruction from the above criteria,
\item Empty circles indicate totally dark objects
\begin{itemize}[leftmargin=*,labelsep=5mm]
\item Simple empty circles merely lost all their baryons and thus did not form stars as their mass was smaller than the minimum to retain them, while
\item  Empty circles crossed with an ``x'' represent destroyed subhaloes from baryonic effects \mbox{(e.g., baryonic disc, etc).}
\end{itemize}
\end{itemize}
In agreement with \cite{brooks}, 3 satellites with $v_{1kpc}>20$ km/s were obtained.
%However, our central velocities are smaller: the correction to the circular velocity, $\Delta(v_{1kpc})$, is larger in our model compared to Z12 and B13. In addition, in our case, some satellites
%are ``overcorrected'': their corrected velocities are negative.

On top of a reduction of the number of satellites to reach the levels
observed in the MW, noticeable~from Figure \ref{fig:V-M_V}, clearing up the MSP, the model, applying baryonic correction to subhaloes, also reduces their central velocity, solving the TBTF problem. As evidenced in Figure~\ref{fig:Fn}, those corrections from baryonic physics are sufficient to solve the velocity and number counts problems of MW-type satellites produced by the VL2 simulation.

\begin{figure}[t]
\centering
%\hspace{-1.0cm}
%\begin{center}
\includegraphics[width=8cm]{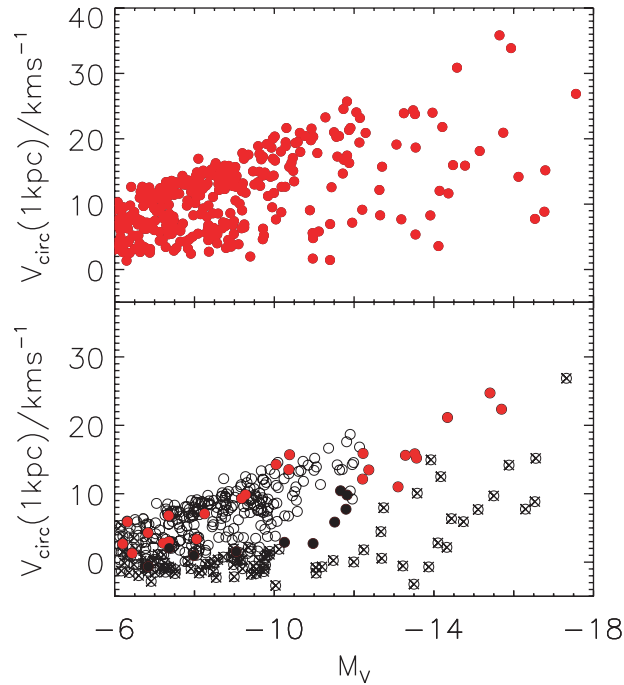}
%\end{center}
%\psfig{file=brooks_new2i.eps,width=13.0cm}
%\psfig{file=fig3_mio.eps,width=10.0cm}
%\psfig{file=fig3.ps,width=8.0cm}
\caption[]{Baryonic effects on VL2 simulation subhaloes, seen through their changes in $v_{1kpc}$ vs. $M_{V}$. VL2 satellites are labelled, as in \cite{brooks}, in the top panel, by their velocities vs. $M_{V}$ at $z=0$. The same, after~baryonic corrections, are presented in the bottom panel.
Satellites stripped of their stars from losing enough mass are represented by filled black circles: at infall, their actual luminosities are much fainter and indicated luminosities are only upper limits. Satellites actually observable at $z=0$ appear as filled red circles. Empty circles mark dark
subhaloes, those with an x being most probably tidally destroyed (figure reproduced from \citep[][]{dpet14a}).
}\label{fig:V-M_V}
\end{figure}
\begin{figure}[t]
%\begin{center}
\centering
\hspace{-3.0cm}
\includegraphics[width=13cm]{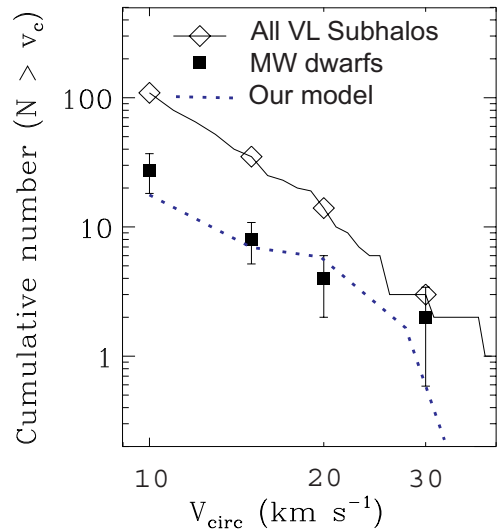}\label{fig:Nsat-Vmw}
%\end{center}
%\psfig{file=cumulative4.eps,width=16.0cm}
%\psfig{file=cumulative1.eps,width=14.0cm}
\caption[]{Corrections from baryonic physics to cumulative number of MW satellites in terms of circular velocity in VL2 simulation.
The classical
MW satellites augmented by the ultra-faint-dwarfs from \cite{simon_geha}
%Simon \& Geha (2007).
are shown as filled squares with error bars.
The Via Lactea subhaloes abundance \cite{diemand1} produces the solid line with diamonds. The model of baryonic corrections applied to VL2 subhaloes yields the abundance shown by the dotted line (figure reproduced from \citep[][]{dpet14a}).}\label{fig:Fn}
\end{figure}

\label{sec:concl}\section{Conclusions}

The discussions on vertical motions of stars near the Galactic plane by \"Opik \cite{Opik} were the first evaluation of our neighbourhood's matter content. One century has passed since, and present evidence points towards a large part of matter in the universe consisting of DM. Despite this mounting evidence, direct or indirect experiments have still not detected DM particles, to date \cite{dp13,dp14a}.

As DM dominated structures, dwarf galaxies, small scale structures, and galaxy satellites are at the front of our understanding of the nature of DM. Moreover, observation of those objects have presented significant discrepancies with predictions of the $\Lambda$CDM model. This review discussed some of those discrepancies such as the CC problem, and issues with satellites.

The understanding of the nature of dark matter, and/or the physics of galaxy formation certainly can improve a great deal from the study of those discrepancies. Noticed more than twenty years ago, we still have no definitive idea of what causes the CC problem. Recently, baryonic effect on small scale structure problems have been the focus of several researches.  Interaction of baryon clumps with DM, from $M >10^5 M_{\odot}$ dwarfs, in the DFBC scenario, can create cores, and MW-type galaxies acquire cuspy profiles, as shown in \cite{DPPace2016}. The SNFF scenario had found similar results, but with cores formed
 from heavier satellites, $M_*>10^6 M_{\odot}$ \cite{Governato2012}. Both scenarios also solve the subhaloes abundance and TBTF problems by reducing the galaxies central DM. However, the baryonic effects debates are still open, as the SNFF scenario presents difficulties in some small scale issues, in particular in some of the MW classical dwarfs \cite{Penarrubia2012,GarrisonKimmel2013}, while isolated galaxies do not clearly present baryonic physics solutions to the TBTF problem.
In particular, the SNFF scenario, with future flat inner profiles dwarfs with $M_*<10^6 M_{\odot}$, would conclude that DM is not cold\footnote{This conclusion is not true in the DFBC scenario.}.
Further investigations of the SIDM model or its variant would in such case be required to understand if they can solve the small scale issues. In any case, new hints on the small scale issues would require the future surveys to discover new satellites with their stars velocity measurements.

Be that as it may, the difficulty to distinguish cusps from cores retains the  debate on dSphs structure open, as discussed in the introduction (e.g., \citep[][]{Strigari2014}). GAIA \cite{Bruijne} and the Subaru Hyper-Supreme-Camera~\cite{Takada} were proposed to provide possible solutions, but both instruments can only hope to solve the problems for larger dwarves, such as Sagittarius \cite{Richardson2014}. The stars velocity line
 of sight components and stars' 2D projection radius that they could provide are plagued with degeneracy between the anisotropy parameter and the density profile, because the determination of the latter is based on Jeans equations. Improvements were proposed measuring just one of three velocity components, and two of three position components \citep[see][]{Battaglia2013}, to obtain proper motions of the dwarves stars \cite{strigari}, and thus the half-light radius density slope, despite the challenge it poses with GAIA \cite{Richardson2014}. The knowledge on dwarves inner structures remains of
fundamental importance, no matter the technical problems.
Yet, the SNFF scenario's tension with the $\Lambda$CDM  model is not fundamental: as shown, the DFBC scenario provides a way to form cores from cusps before gas forms stars, and is more efficient. Thus,  although the SNFF would have serious problems if cored dwarves with $M_* < 10^6 M_{\odot}$ were found, so long as $M_* > 10^5 M_{\odot}$, the DFBC scenario within the $\Lambda$CDM  model would not disagree with observations.

{%\bf
The Baryonic Tully--Fisher Relation and, mostly, the satellite planes problem remain open problems for the $\Lambda$CDM model to tackle, although the first one seems to be on the right path.}

Alternate tests of the $\Lambda$CDM could proceed from checking
\begin{itemize}[leftmargin=*,labelsep=5mm]
\item The number of small subhaloes in galaxies' virial radius,predicted to be large: gravitational lenses flux anomalies barely found agreement with those predictions \cite{MetcalfAmara};
%Metcalf \& Amara 2012).
\item The Galaxy's cold tidal streams perturbations by subhaloes (see \citep[][]{Carlberg}).
%Carlberg \& Grillmair 2013).
\end{itemize}

Obviously, collider, direct, and indirect DM particles detection would solve the problem of DM nature and existence. Unfortunately, no evidence of super-symmetry (SUSY) has appeared so far at the LHC, so WIMPS particles DM are yet elusive\footnote{The diphoton excess might be able to rescue the WIMP hypothesis.}, while a di-photon excess decay at 750 GeV, rumoured at 3.9 $\sigma$ significance is still debated. The claimed DAMA/LIBRA/CoGeNT annual modulation continues to be the only, although controversial, news from direct searches and similarly no incontrovertible indirect evidence of DM has been presented, to date.

\vspace{6pt}

%%%%%%%%%%%%%%%%%%%%%%%%%%%%%%%%%%%%%%%%%%
%% optional
%\supplementary{The following are available online at www.mdpi.com/link, Figure S1: title, Table S1: title, Video S1: title.}

%%%%%%%%%%%%%%%%%%%%%%%%%%%%%%%%%%%%%%%%%%
%\acknowledgments{All sources of funding of the study should be disclosed. Please clearly indicate grants that you have received in support of your research work. Clearly state if you received funds for covering the costs to publish in open access.}

%%%%%%%%%%%%%%%%%%%%%%%%%%%%%%%%%%%%%%%%%%
%\authorcontributions{\hl{Please list each author's contribution in this work.}}

%%%%%%%%%%%%%%%%%%%%%%%%%%%%%%%%%%%%%%%%%%
\conflictofinterests{{The authors declare no conflict of interest.}}

\bibliographystyle{mdpi}

%=====================================
% References, variant A: internal bibliography
%=====================================
\renewcommand\bibname{References}
\bibliography{lite,shortnames}
\end{document}